\documentclass[5p,twocolumn,times]{elsarticle}
\newcommand{\add}[1]{#1}
\newcommand{\addcad}[1]{#1}
\newcommand{\addrevision}[1]{#1}
\newcommand{\addrevisiongraphics}[1]{\fcolorbox{white}{white}{#1}}

\newcommand{\addmorerevision}[1]{\textcolor{black}{#1}}
\newcommand{\addmorerevisiongraphics}[1]{\fcolorbox{white}{white}{#1}}
\newcommand{\addmorerevisiontable}[1]{\color{black}{#1}}

\newcommand{\additionalrevision}[1]{\textcolor{black}{#1}}
\newcommand{\additionalrevisiongraphics}[1]{\fcolorbox{white}{white}{#1}}

\newcommand{\addextrarevision}[1]{\textcolor{black}{#1}}
\newcommand{\addextrarevisiongraphics}[1]{\fcolorbox{white}{white}{#1}}
\newcommand{\addextrarevisiontable}[1]{\color{black}{#1}}

\usepackage[T1]{fontenc}
\usepackage[utf8]{inputenc}
\usepackage{color}
\usepackage[american]{babel}
\usepackage{array}
\usepackage{calc}
\usepackage{amsmath}
\usepackage{amssymb}
\usepackage{graphicx}
\usepackage{wasysym}
\PassOptionsToPackage{normalem}{ulem}
\usepackage{ulem}

\providecommand{\tabularnewline}{\\}
\usepackage[T1]{fontenc}
\usepackage{courier}
\usepackage{array}
\usepackage{color}
\usepackage{upquote}
\usepackage{xcolor}
\usepackage{listings}
\usepackage[most]{tcolorbox}
\usepackage{caption}
\usepackage{graphics}
\usepackage{placeins}
\usepackage{graphicx, epstopdf}
\usepackage[framed, numbered]{matlab-prettifier}
\usepackage{colortbl}
\usepackage{arydshln}
\usepackage{subfig}
\usepackage{color}
\usepackage{multirow}
\usepackage{rotating} %

\usepackage[colorlinks]{hyperref}

\usepackage{lipsum}
\usepackage{amsfonts}
\usepackage{graphicx}
\usepackage{epstopdf}
\usepackage{algorithmic}
\usepackage{array}
\usepackage{booktabs}
\usepackage{multirow}
\usepackage{amsbsy}
\usepackage{amstext}

\definecolor{celadon}{rgb}{0.67, 0.88, 0.69}
\definecolor{hellgelb}{rgb}{1,1,0.85} 
\definecolor{colKeys}{rgb}{0,0,1} 
\definecolor{colIdentifier}{rgb}{0,0,0} 
\definecolor{colComments}{rgb}{0,0.5,0} 
\definecolor{colString}{rgb}{0.81,0.12,0.95}
\DeclareFixedFont{\ttb}{T1}{txtt}{bx}{n}{12} %
\DeclareFixedFont{\ttm}{T1}{txtt}{m}{n}{12}  %
\definecolor{deepblue}{rgb}{0,0,1}
\definecolor{deepred}{rgb}{0.6,0,0}
\definecolor{deepgreen}{rgb}{0,0.5,0}

\definecolor{myteal}{RGB}{0,129,169}

\newcommand{\tikzbox}[2][black,fill=red]{\tikz[baseline=0.0ex, line width=0.2mm]\draw[#1] [#1] (0,0) rectangle (#2,#2);}%
\newcommand{\tikzcircle}[2][red,fill=red]{\tikz[baseline=-0.6ex]\draw[#1,radius=#2] (0,0) circle ;}%
\AtBeginDocument{%
\hypersetup{
	colorlinks,
	citecolor=myteal,
	linkcolor=myteal,
	urlcolor=myteal}}

\definecolor{MINVO_color}{RGB}{207,242,207}
\definecolor{Bernstein_color}{RGB}{206,206,232}
\definecolor{BSpline_color}{RGB}{246,246,206}
\definecolor{problem1_color}{RGB}{255,255,255} 
\definecolor{problem2_color}{RGB}{255,255,255}
\definecolor{problem3_color}{RGB}{255,255,255}
\definecolor{problem4_color}{RGB}{255,255,255} 
\newcommand{\nthdegree}{$n^{\text{th}}$-degree}

\begin{document}
	\begin{frontmatter}
		\title{ \textbf{MINVO Basis: Finding Simplexes with Minimum Volume Enclosing Polynomial Curves}}
		\author{Jesus Tordesillas}
		\author{Jonathan P. How}
		\ead{\{jtorde,jhow\}@mit.edu}
		\address{Aerospace Controls Laboratory, Massachusetts Institute of Technology, Cambridge, MA, USA.\vspace{-0.1cm}}
		\begin{abstract}
			This paper studies 
			\addcad{the polynomial basis that generates} the smallest $n$-simplex enclosing a given $n^{\text{th}}$-degree polynomial curve \addcad{in $\mathbb{R}^n$}. 
			Although the Bernstein and B-Spline polynomial bases provide feasible solutions to this problem,  the simplexes obtained by these bases are not the smallest possible, which leads to \addcad{overly} conservative results in many \addcad{CAD (computer-aided design) applications}. 
			We first prove that the polynomial basis that solves this problem (MINVO basis) also solves for the $n^\text{th}$\add{-degree} polynomial curve with largest convex hull enclosed in a given $n$-simplex. Then, we present a formulation that is \emph{independent} of the $n$-simplex or $n^{\text{th}}$-degree polynomial curve given. By using Sum-Of-Squares (SOS) programming, branch and bound, and moment relaxations, we obtain high-quality feasible solutions for any $n\in\mathbb{N}$\addcad{, and prove (numerical) global optimality for $n=1,2,3$ and (numerical) local optimality for $n=4$}.  
			The results obtained for $n=3$ show that, for \textit{any} given $3^{\text{rd}}$\add{-degree} polynomial curve \addcad{in $\mathbb{R}^3$}, the MINVO basis is able to obtain an enclosing simplex whose volume is $2.36$ and $254.9$ times smaller than the ones obtained by the Bernstein and B-Spline bases, respectively. When $n=7$, these ratios increase to $902.7$ and $2.997\cdot10^{21}$, respectively.
		\end{abstract}
	\end{frontmatter}
     
    \urlstyle{same}
	\noindent %
	\textbf{Video}: \textcolor{blue}{\url{https://youtu.be/TXR8mXCaMNg}}\\%\hspace{0.1cm}
		\textbf{Code}: \textcolor{blue}{\url{https://github.com/mit-acl/minvo}}

	\vspace{-0.4cm}
	\section{Introduction\label{sec:Introduction}}

    \addcad{    Polyhedral enclosures of a given polynomial curve 
    have a crucial role  in a large number of CAD algorithms  
    to compute curve intersections, perform ray tracing, or obtain minimum distances between convex shapes~\cite{efremov2005robust,sederberg1990curve,schulz2009bezier,gilbert1988fast,cichella2017optimal}. These polyhedral enclosures are also used in
    rasterization~\cite{hersch1994font,nguyen2007gpu}, mesh generation~\cite{cardoze2004bezier}, path planning for numerical control machines~\cite{chuang1997tool,chuang1999one}, and trajectory optimization for robots~\cite{tordesillas2019faster,tordesillas2020mader,tordesillas2021panther,preiss2017trajectory,sahingoz2014generation}. Many of these works leverage the convex hull property of the Bernstein basis (polynomial basis used by B\'{e}zier curves) to obtain these  polyhedral enclosures, although some works use the \mbox{B-Spline} basis instead~\cite{ding2018trajectory,zhou2019robust}.}

	\addcad{Although both the Bernstein basis and B-Spline basis have many useful properties, they are not designed to generate the smallest $n$-simplex that encloses a given \nthdegree{} polynomial curve in $\mathbb{R}^n$.
	This directly translates into undesirably conservative results in	many of the aforementioned applications. Polyhedral enclosures with more than $n+1$ \addrevision{vertices} (i.e., not simplexes) can provide tighter volume approximations, but at the expense of a larger number of \addrevision{vertices}, which can eventually increase the computation time in real-time applications. The main focus of this paper is therefore on simplex enclosures.}

	\addcad{Moreover, and rather than designing an iterative algorithm that needs to be run for each different curve to obtain the smallest simplex enclosure, this paper studies a novel polynomial basis that is designed by construction to minimize the volume of this simplex. Compared to an iterative algorithm, this basis benefits from the properties of \emph{linearity} (the simplex is a linear transformation of the coefficient matrix of the curve, avoiding therefore the need of expensive iterative algorithms) and  \emph{independence with respect to the curve} (the matrix that defines this linear transformation is the same one for all the curves of the same degree). These two properties are highly desirable in real-time computing and/or when used in an optimization problem.}

	\addcad{In summary, this paper studies the polynomial basis that generates the $n$-simplex with minimum volume enclosing a given $n^{\text{th}}$-degree polynomial curve. Additionally, we also investigate the related problem of obtaining the $n^{\text{th}}$-degree polynomial curve with largest convex hull enclosed in a given $n$-simplex. The contributions of this paper are summarized as follows:}
	\begin{itemize}
		\item Formulation of \add{the optimization} problem 
		\add{whose minimizer is the polynomial basis that generates} the smallest 
		$n$-simplex that encloses \textbf{any} given $n^{\text{th}}$\add{-degree} polynomial curve. We show that this \add{basis also obtains} 
		the  $n^{\text{th}}$\add{-degree} polynomial curve 
		with largest convex hull 
		enclosed in \textbf{any} given $n$-simplex. \add{Another formulation that imposes} a specific structure on the polynomials of the basis is
		also presented. %
		\item We \add{derive high-quality feasible solutions for any $n\in\mathbb{N}$, }%
		obtaining simplexes
		that, for $n=3$, are $2.36$ and $254.9$ times smaller than the
		ones obtained using the Bernstein and B-Spline bases respectively. For $n=7$, these values increase to 902.7 and $2.997\cdot10^{21}$\addcad{, respectively}.\vspace{0.0cm}
		\item \add{Numerical global} optimality \add{(with respect to the volume)} is proven for $n=1,2,3$ 
		using SOS, branch and bound, and moment relaxations. \add{Numerical local optimality is proven for $n=4$, and feasibility is guaranteed for $n\ge5$.}
		\item \add{Extension to polynomial curves embedded in subspaces of higher dimensions}, and to some rational curves. 
	\end{itemize}

	\section{Related work}\label{sec:related_work}
	
	\add{Herron \cite{herron1989polynomial} attempted to find (for $n=2$ and $n=3$) the smallest $n$-simplex enclosing an \add{$n^{\text{th}}$-degree} polynomial curve. The approach of \cite{herron1989polynomial} imposed a specific
		structure on the polynomials of the basis and then solved}
	the associated nonconvex optimization problem over the roots of those
	polynomials. For this specific structure of the polynomials, a global
	minimizer was found for $n=2$, and a local minimizer was found for
	$n=3$. However, global optimality over all possible polynomials was not proven, and only the cases $n=2$ and $n=3$ were studied. Similarly, in the results of~\cite{kuti2014minimal}, Kuti
	et al.~use the algorithm proposed in~\cite{li2008minimum} to obtain
	a minimal 2-simplex that encloses a specific $2^{\text{nd}}$\add{-degree} polynomial
	curve. However, this approach requires running the algorithm for \textit{each} different curve, no global optimality is shown, and only one case with $n=2$ was analyzed. 
	\addrevision{ There are also works that have derived bounds on the distance between the control polygon and the curve~\cite{nairn1999sharp, karavelas2004bounding, ma2006efficient}, while others propose the SLEFE (subdividable linear efficient function enclosure) to enclose spline curves via subdivision~\cite{peters2004sleves, mccann2004exploring, lutterkort1999tight, lutterkort2001tight}. However, the SLEFE depends \textit{pseudo-linearly} on the coefficients of the polynomial curve (i.e., linearly except for a min/max operation)~\cite{peters2004sleves}, which is disadvantageous when the curve is a decision variable in a time-critical optimization problem. This paper focuses instead on enclosures that depend \textit{linearly} on the coefficients of the curve. 
	}
	
	\iffalse %
	Our work presented in this paper goes further, and using SOS programming, it
	first proposes the most general formulation that does not impose any
	specific structure in the form of the polynomials (apart from the
	necessary condition of being SOS), proving global and local optimality
	for some $n$. Then, by imposing a structure in the polynomials, we
	are able to able to find global optima for $n=1,2,3$ and local optima
	$n=4,5,6,7$.
	\fi %

	\add{Other} works have focused on the properties of the smallest $n$-simplex
	that \add{encloses a given generic convex body. For example, \cite{kanazawa2014minimal,galicer2019minimal} derived some bounds for the volume of this simplex, while Klee \cite{klee1986facet} showed}
	that any locally optimal simplex (with respect to its volume) that
	encloses a convex body must be a centroidal simplex. In other words, the centroid
	of its facets must belong to the convex body. Applied to a curve~$P$,
	this \addcad{means} that the centroid of its facets must belong to $\text{conv}\left(P\right)$. Although this is a necessary condition, it is \textit{not} sufficient for local optimality. 
	
	When the convex body is a polyhedron (or equivalently the convex hull
	of a finite set of points), \cite{vegter1993finding} classifies
	the possible minimal circumscribing simplexes, and this classification
	is later used by \cite{zhou2002algorithms} to derive a
	$O(k^{4})$ algorithm that computes the smallest simplex enclosing
	a finite set of $k$ points. This problem is also crucial for the hyperspectral unmixing in remote sensing, to be able to find 
	the proportions or abundances of each macroscopic material (endmember)
	contained in a given image pixel \cite{uezato2019hyperspectral,hendrix2013minimum}, and many different iterative algorithms have been proposed towards this end \cite{rogge2006iterative,iordache2009unmixing,velez2003iterative}. All these works focus \add{on obtaining the enclosing simplex for a generic discrete set of points. Our work focuses instead on polynomial curves and, by leveraging their structure, we can avoid the need to iterate and/or discretize the curve. } 
	
	The convex hull of curves has also been studied in the literature. \add{For instance,}
	\cite{ranestad2009convex,sedykh1986structure,derry1956convex} studied
	the boundaries of these convex hulls, while \cite{ciripoi2018computing}
	focused on the patches of the convex hull of trajectories of polynomial
	dynamical systems. For a \addrevision{moment curve $\arraycolsep=1.4pt \left[\begin{array}{cccc}
		t & t^{2} & \cdots & t^{n}\end{array}\right]^{T}$ (where $t$ is in some interval~$[a,b]$)}, \cite{mazur2017convex} found
	that the number of points needed to represent every point in the convex
	hull of this curve is $\frac{n+1}{2}$, giving therefore a tighter
	bound than the $n+1$ points found using the Carathéodory's Theorem
	\cite{caratheodory1907variabilitatsbereich,steinitz1913bedingt}.
	This particular curve and the volume of its convex hull were also analyzed by \cite{karlin1953geometry}
	in the context of moment spaces and orthogonal polynomials. Although many useful properties of the convex hull of a curve are shown in all these previous works, none of them addresses the problem of finding the polynomial curve with largest convex hull enclosed in a given simplex. 
	
	\section{Preliminaries\label{sec:Preliminaries}}
	
	\subsection{Notation and Definitions}\label{sec:Notation-and-Definitions}
	
	The notation used throughout
	the paper is summarized in Table \ref{tab:Notation-used-in}. Unless otherwise noted, all the indexes in this paper start at
	$0$. For instance, $\boldsymbol{M}_{0,3}$ is the fourth element of the first
	row of the matrix $\boldsymbol{M}$. Let
	us also introduce the two following common definitions and their respective
	notations:
	\vspace{0.2cm} 
	
	\noindent\fbox{\begin{minipage}[t]{1\columnwidth - 2\fboxsep - 2\fboxrule}%
			\textbf{Polynomial curve $P$} \textbf{of degree} $n$ \textbf{and dimension}
			$k$: 
			\vskip -0.2cm 
			$$P:=\left\{ \boldsymbol{p}(t)\;|\;t\in[a,b],a\in\mathbb{R},b\in\mathbb{R},b>a\right\} $$%
			where  
			$\arraycolsep=1.9pt\boldsymbol{p}(t):=\add{\left[\begin{array}{ccc}
					p_{0}(t) & \cdots & p_{k-1}(t)\end{array}\right]^T}:=\boldsymbol{P}\boldsymbol{t}\in\mathbb{R}^{k}$\add{, $p_{i}(t)$ is a polynomial \addmorerevision{in} $\mathbb{R}[t]$, and $n\in\mathbb{N}$ is the highest degree of all the $p_{i}(t)$}. The $k\times(n+1)$ matrix $\boldsymbol{P}$
			is the coefficient matrix.
			Without loss of generality we will use the interval $t\in[-1,1]$ (i.e., $a=-1$ and $b=1$), and assume that the parametrization of $\boldsymbol{p}(t)$ has minimal degree (i.e., no other polynomial parametrization with lower degree produces the same spatial curve $P$). The subspace
			containing $P$ and that has the smallest dimension will be denoted
			as $\mathcal{M}\subseteq\mathbb{R}^{k}$, and its dimension will be $m$. We will
			work with the case $n=m=k$, and refer to such curves simply as $n^\text{th}$\textbf{\add{-degree} polynomial curves}. \addcad{Note that we will use the term \emph{polynomial curve} to refer to a curve with only one segment, and not to a curve with several polynomial segments.} The set of all possible $n^\text{th}$\add{-degree} polynomial curves will be denoted as $\mathcal{P}^n$. \add{Section~}\ref{sec:other_curves}
			will then extend the results to curves with arbitrary $n$, $m$ and $k$. 
	\end{minipage}}
	
	\begin{table}
		\setlength\extrarowheight{0.88pt}
		\begin{centering}
			\caption{Notation used in this paper. \label{tab:Notation-used-in}}
			\begin{tabular}{|>{\centering}m{0.17\columnwidth}|>{\raggedright}m{0.76\columnwidth}|}
				\hline 
				\textbf{Symbol} & \textbf{Meaning}\tabularnewline
				\hline 
				\hline 
				$a$, $\boldsymbol{a}$, $\boldsymbol{A}$ & Scalar, column vector, matrix\tabularnewline
				\hline 
				$|\boldsymbol{A}|$, $\text{tr}(\boldsymbol{A})$ & Determinant of $\boldsymbol{A}$, trace of $\boldsymbol{A}$\tabularnewline
				\hline 
				$\boldsymbol{t}$ & $\arraycolsep=1.4pt\left[\begin{array}{ccc}
					t^{r} & t^{r-1}\cdots & 1\end{array}\right]^{T}$ ($r$ given by the context)\tabularnewline
				\hline 
				$\hat{\boldsymbol{t}}$ & $\arraycolsep=1.4pt\left[\begin{array}{cccc}
					1 & \cdots & t^{r-1} & t^{r}\end{array}\right]^{T}$ ($r$ given by the context)\tabularnewline
				\hline
				\add{$\mathbb{R}[t]$}&\add{Set of univariate polynomials in $t$ with coefficients \addmorerevision{in} $\mathbb{R}$}\tabularnewline
				\hline 
				$\boldsymbol{p}(t)$ & Column vector whose coordinates \addmorerevision{are polynomials in} \add{$\mathbb{R}[t]$}\tabularnewline
				\hline 
				$P$ & Polynomial curve $P:=\left\{ \boldsymbol{p}(t)\;|\;t\in[-1,1]\right\} $\tabularnewline
				\hline 
				$\boldsymbol{P}$ & Coefficient matrix of $P$. $\boldsymbol{p}(t)=\boldsymbol{P}\boldsymbol{t}$\tabularnewline
				\hline 
				$\mathcal{P}^{n}$ & Set of all possible \add{$n^{\text{th}}$-degree} polynomial curves\tabularnewline
				\hline 
				$\text{conv}(P)$ & Convex hull of $P$\tabularnewline
				\hline 
				$n$ & Maximum degree of the entries of $\boldsymbol{p}(t)$\tabularnewline
				\hline 
				$k$ & Number of rows of $\boldsymbol{p}(t)$\tabularnewline
				\hline 
				$\mathcal{M}$ & Subspace with the smallest dimension that contains $P$. $\mathcal{M}\subseteq\mathbb{R}^{k}$\tabularnewline
				\hline 
				$m$ & Dimension of $\mathcal{M}$\tabularnewline
				\hline 
				$S$ & Simplex\tabularnewline
				\hline 
				$\mathcal{S}^{n}$ & Set of all possible $n$-simplexes\tabularnewline
				\hline 
				$\boldsymbol{V}$ & Matrix whose columns are the \addrevision{vertices} of a simplex. \addrevision{This definition will be generalized in Section~\ref{sec:other_curves_nmk}}\tabularnewline
				\hline
				\addcad{$\boldsymbol{0}$, $\boldsymbol{1}$} & \addcad{Column vectors of zeros and ones}\tabularnewline
				\hline
				\addcad{$\boldsymbol{a}\ge\boldsymbol{b}$} & \addcad{Element-wise	inequality}\tabularnewline
				\hline
				\addmorerevision{$s$} & \addmorerevision{Number of intervals the polynomial curve is subdivided into ($s=1$ means no subdivision)}  \tabularnewline
				\hline
				\addmorerevision{SL\textsubscript{$h$}} & \addmorerevision{SLEFE of a polynomial curve using $h$ breakpoints in each interval of subdivision}\tabularnewline
				\hline
				\addcad{$\propto$} & \addcad{Proportional to}\tabularnewline
				\hline
				\addcad{$\partial\;\cdot$} & \addcad{Frontier of a set}\tabularnewline
				\hline
				\addcad{$\left\lfloor  \cdot \right\rfloor $} & \addcad{Floor function}\tabularnewline
				\hline
				\addcad{$\text{abs}(\cdot)$} & \addcad{Absolute value}\tabularnewline
				\hline
				\addcad{$\ensuremath{\cdot_{a\times b}}$} & \addcad{Size of a matrix ($a$ rows $\times$ $b$ columns)}\tabularnewline
				\hline 
				$\boldsymbol{e}$ & $\arraycolsep=1.4pt\left[\begin{array}{ccccc}
					0 & 0 & \cdots & 0 & 1\end{array}\right]^{T}$ (size given by the context)\tabularnewline
				\hline 
				$\boldsymbol{I}_{n}$ & Identity matrix of size $n\times n$\tabularnewline
				\hline 
				$\boldsymbol{M}_{:,c:d}$ & Matrix formed by columns $c,c+1,\ldots,d$ of $\boldsymbol{M}$\tabularnewline
				\hline 
				$\mathbb{S}_{+}^{a}$ & Positive semidefinite cone (set of all symmetric positive semidefinite
				$a\times a$ matrices)\tabularnewline
				\hline 
				$\text{vol}(\cdot)$ & Volume (Lebesgue measure)\tabularnewline
				\hline 
				$\boldsymbol{\pi}$ & Hyperplane\tabularnewline
				\hline 
				$\text{dist}(\boldsymbol{a},\boldsymbol{\pi})$ & Distance between the point $\boldsymbol{a}$ and the hyperplane $\boldsymbol{\pi}$ \tabularnewline
				\hline 
				\add{$\text{odd}(a,b)$} & \add{1 if both $a$ and $b$ are odd, 0 otherwise} \tabularnewline
				\hline 
				\add{NLO, NGO} & \add{Numerical Local Optimality, Numerical Global Optimality}\tabularnewline%
				\hline 
				$\text{MV},\text{Be},\text{BS}$ & MINVO, Bernstein, and B-Spline\tabularnewline
				\hline 
				\vspace{4pt}
				\tikzbox[black,fill=MINVO_color]{0.3cm} \tikzbox[black,fill=Bernstein_color]{0.3cm} \tikzbox[black,fill=BSpline_color]{0.3cm}  & Color notation for the \add{MV, Be, and BS} bases respectively\tabularnewline
				\hline 
			\end{tabular}
			\par\end{centering}
		\vspace{-0.5cm}
	\end{table}
	
	\vspace{0.3cm}
	
	\noindent\fbox{\begin{minipage}[t]{1\columnwidth - 2\fboxsep - 2\fboxrule}%
			\textbf{$n$-simplex}: Convex hull of $n+1$ \addrevision{affinely independent} points $\boldsymbol{v}_{0},\ldots,\boldsymbol{v}_{n}\in\mathbb{R}^{n}$.
			These points are \addrevision{the \textbf{vertices}} of the simplex, and will be stacked
			in the \textbf{matrix of \addrevision{vertices}} $\arraycolsep=1.9pt\boldsymbol{V}:=\left[\begin{array}{ccc}
				\boldsymbol{v}_{0} & \cdots & \boldsymbol{v}_{n}\end{array}\right]$. The letter $S$ will denote a particular simplex, while $\mathcal{S}^{n}$ will denote the set of all possible $n$-simplexes. A simplex with $\arraycolsep=1.9pt\boldsymbol{V}=\left[\protect\begin{array}{cc}
				\boldsymbol{0} & \boldsymbol{I}_{n} \protect\end{array}\right]$ will be called the \textbf{standard $n$-simplex}.  %
	\end{minipage}}
	\vspace{0.2cm}

	We will use the basis matrix of a segment of a non-clamped uniform B-Spline for \add{comparison \cite{schoenberg1973cardinal, de1978practical, qin2000general}},
	and refer to this basis simply as the \textbf{B-Spline basis}.
	
	\add{Moreover, throughout this paper we will use the term \emph{numerical local optimality} (NLO) to classify a solution for which the first-order optimality measure and the maximum constraint violation are smaller than a predefined small tolerance~\cite{matlabOptToolbox}. Similarly, we will use the term  \emph{numerical global optimality} (NGO) to classify a feasible solution for which the difference between its objective value and a lower bound on the global minimum (\addcad{typically} obtained via a relaxation of the problem) is less than a specific small tolerance. All the tolerances and parameters of the numerical solvers used are available in the attached code.}
	
	\add{\addcad{Finally, and for purposes of clarity}, we will use the term \emph{MINVO basis} to refer to both the global minimizers of the problem (proved for $n=1,2,3$) and the proposed locally-optimal/feasible solutions (obtained for $n\ge4$).}

	\subsection{Volume of the Convex Hull of a Polynomial Curve}\label{sec:Volume-of-the-convex-hull}
	
	At several points throughout the paper, we will make use of the following
	theorem, that we prove \addcad{in~\ref{sec:AppVolume}}:
	
	\vspace{0.1cm}
	\noindent\fcolorbox{black}{white}{\begin{minipage}[t]{1\columnwidth - 2\fboxsep - 2\fboxrule}%
			
			\textbf{Theorem 1:} The volume of the convex hull of $P\in\mathcal{P}^n$ ($t\in[-1,1]$), is given
			by:
			\[
			\text{vol}\left(\text{conv}\left(P\right)\right)=\frac{\text{abs}\left(\left|\boldsymbol{P}_{:,0:n-1}\right|\right)}{n!}2^{\frac{n(n+1)}{2}}\prod_{0\le i<j\le n}\left(\frac{j-i}{j+i}\right)
			\]

			\textbf{Proof}: \addcad{See~\ref{sec:AppVolume}}. \hfill $\square$
			
	\end{minipage}}
	
	\vspace{0.2cm}
	
	Note that, as the curve $P$ satisfies $n=m=k$ (see \addcad{Section~}\ref{sec:Notation-and-Definitions}), the volume of its convex hull is nonzero and therefore $\left|\boldsymbol{P}_{:,0:n-1}\right|\neq0$.
	
	\section{Problems definition}\label{sec:problems-definition}
	
	As explained in \addcad{Section~}\ref{sec:Introduction}, the goal of this paper is to find the smallest simplex $S\in \mathcal{S}^{n}$ enclosing a given polynomial curve $P\in\mathcal{P}^n$, and to find the polynomial curve $P\in\mathcal{P}^n$ with largest convex hull enclosed in a given simplex $S\in \mathcal{S}^n$. 
	
	\subsection{Given $P\in\mathcal{P}^n$, find $S\in \mathcal{S}^{n}$}\label{sec:GivenPfindS}
	
	\vspace{0.1cm}
	
	\noindent\fcolorbox{black}{problem1_color}{
		\begin{minipage}[t]{0.99\columnwidth - 2\fboxsep - 2\fboxrule}%
			
			\textbf{Problem 1}: Given the polynomial curve $P\in\mathcal{P}^n$,
			find the simplex $S\in \mathcal{S}^{n}$ with minimum volume that contains $P$. In other words:
			\[
			\arraycolsep=1.4pt\min_{S\in \mathcal{S}^{n}}\quad \addrevision{\text{vol}(S)}
			\]
			\[
			\begin{array}{cc}
				\text{s.t.} &  P\subset S\end{array}
			\]
			
	\end{minipage}}
	
	\vspace{0.2cm}
	
	For $n=2$, Problem 1 tries to find the triangle with the smallest
	area that contains \add{a planar $2^\text{nd}$-degree polynomial curve}.
	For $n=3$, it tries to find the tetrahedron with the smallest volume
	that \add{contains a $3^\text{rd}$-degree polynomial curve in 3D.}
	Similar geometric interpretations apply for higher $n$.

	\addrevision{Letting $f_1$ denote the objective function of this problem, we have that  $\arraycolsep=1.4pt f_{1}:=\text{vol}(S)\propto\text{abs}\left(\left|\left[\begin{array}{cc} \boldsymbol{V}^{T} & \boldsymbol{1}\end{array}\right]\right|\right)$.} Note that, as the volume of the convex hull of $P$ is nonzero (see \addcad{Section~}\ref{sec:Volume-of-the-convex-hull})\addrevision{, then it is guaranteed that $\arraycolsep=1.4pt\left|\left[\begin{array}{cc}	\boldsymbol{V}^{T} & \boldsymbol{1}\end{array}\right]\right|\neq0$.}
	
	\subsection{Given $S\in \mathcal{S}^{n}$, find $P\in\mathcal{P}^n$}
	
	\vspace{0.1cm}
	
	\noindent\fcolorbox{black}{problem2_color}{\begin{minipage}[t]{1\columnwidth - 2\fboxsep - 2\fboxrule}%
			
			\textbf{Problem 2:} Given a simplex $S\in \mathcal{S}^n$, find the polynomial curve $P\in\mathcal{P}^n$
			contained in $S$, whose convex hull has maximum volume:
			\[
			\min_{P\in\mathcal{P}^n}\quad \addrevision{-\text{vol}(\text{conv}(P))}
			\]
			\[
			\begin{array}{cc}
				\text{s.t.} & P\subset S\end{array}
			\]
			
	\end{minipage}}

	\vspace{0.2cm}

	\addrevision{By the definition of a simplex (see Section~\ref{sec:Notation-and-Definitions}), its vertices are affinely independent and therefore
	the matrix of vertices of the given simplex} $S$ satisfies $\arraycolsep=1.4pt\left|\left[\begin{array}{cc} \boldsymbol{V}^{T} & \boldsymbol{1}\end{array}\right]\right|\neq0$. 
	
	\addrevision{Letting $f_2$ denote the objective function of this problem, we have that  $\arraycolsep=1.4pt f_{2}:=-\text{vol}(\text{conv}(P))\propto-\text{abs}\left(\left|\boldsymbol{P}_{:,0:n-1}\right|\right)$.} Now note that the optimal solution	for this problem is guaranteed to satisfy $\left|\boldsymbol{P}_{:,0:n-1}\right|\neq0$, which can be easily proven by noting that we are maximizing the absolute
	value of $\left|\boldsymbol{P}_{:,0:n-1}\right|$, and that there exists
	at least one feasible solution (for example the B\'{e}zier curve whose control
	points are the \addrevision{vertices} of $S$) with
	$\left|\boldsymbol{P}_{:,0:n-1}\right|\neq0$.
	
	\section{Equivalent Formulation}
	
	Let us now study the constraints and the objective functions of Problems
	1 and 2.
	
	\subsection{Constraints of Problems 1 and 2\label{subsec:Constraints-of-Problem}}
	
	Both problems share the same constraint $P\subset S$ (i.e., $\boldsymbol{p}(t)\in S\;\;\forall t\in[-1,1]$),
	which is equivalent to $\boldsymbol{p}(t)$ being a convex combination
	of the \addrevision{vertices} $\boldsymbol{v}_{i}$ of the simplex for $t\in[-1,1]$:
	\begin{equation} \label{eq:PinS}
		P\subset S\equiv\left\{ \begin{aligned}
			&\boldsymbol{p}(t)=\textstyle\sum_{i=0}^{n}\lambda_{i}(t)\boldsymbol{v}_{i}\\
			&\textstyle\sum_{i=0}^{n}\lambda_{i}(t)=1\quad\forall t\\
			&\lambda_{i}(t)\ge0\quad\forall i=0,\dots,n\;\;\forall t\in[-1,1]
		\end{aligned}\right. 
	\end{equation}
	
	The variables $\lambda_{i}(t)$ are usually called barycentric \addrevision{coordinates \cite{berger2009geometry, hormann2017generalized, ericson2004real}, and} their geometric interpretation is as follows.
	Let us first define $\boldsymbol{\pi}_{i}$ as the hyperplane that passes
	through the points $\{\boldsymbol{v}_{0},\boldsymbol{v}_{1},\ldots,\boldsymbol{v}_{n}\}\backslash\addrevision{\{\boldsymbol{v}_{i}\}}$,
	and $\boldsymbol{n}_{i}$ as its normal vector that 
	\add{points towards the interior of the simplex}.
	Choosing now $q\in\{0,\ldots,n\}\backslash \addrevision{\{i\}}$, and using the fact that $\sum_{j=0}^{n}\lambda_{j}(t)=1$, we have that \add{$\boldsymbol{p}(t)=\boldsymbol{v}_{q}+\sum_{
			j=0
		}^{n}\lambda_{j}(t)\left(\boldsymbol{v}_{j}-\boldsymbol{v}_{q}\right)$}. 
	\begin{figure}
		\begin{centering}
			\includegraphics[width=1.0\columnwidth]{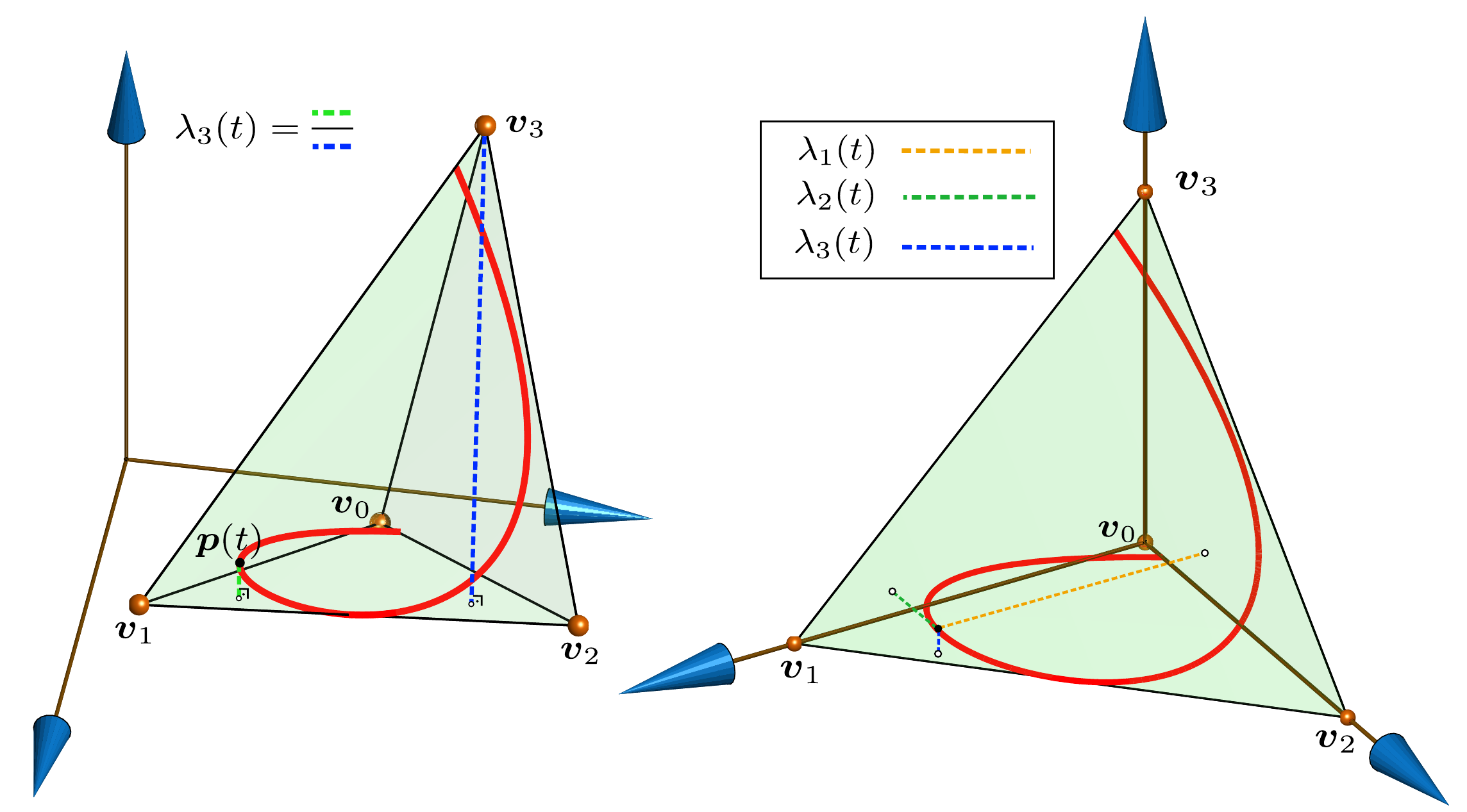}
			\par\end{centering}
		\caption{Geometric interpretation of $\lambda_{i}(t)$. Each $\lambda_{i}(t)$
			represents the distance between the curve $\boldsymbol{p}(t)$
			and the hyperplane formed by the \addrevision{vertices} $\{\boldsymbol{v}_{0},\boldsymbol{v}_{1},\ldots,\boldsymbol{v}_{n}\}\backslash\addrevision{\{\boldsymbol{v}_{i}\}}$,
			divided by the distance from the vertex $\boldsymbol{v}_{i}$
			to that hyperplane (left). For the standard $3$-simplex in 3D (i.e., $\arraycolsep=1.9pt\boldsymbol{V}=\left[\protect\begin{array}{cc}
				\boldsymbol{0} & \boldsymbol{I}_{3} \protect\end{array}\right]$), the curve in red has $\arraycolsep=1.4pt\boldsymbol{p}(t)=\left[\protect\begin{array}{ccc}
				\lambda_{1}(t) & \lambda_{2}(t) & \lambda_{3}(t)\protect\end{array}\right]^{T}$ (right).\label{fig:Geometric-interpretation-of}}
		\vskip -3ex
	\end{figure}
	Therefore:
	\addcad{$$	%
			\text{dist}\left(\boldsymbol{p}(t),\boldsymbol{\pi}_{i}\right)  :=\left(\boldsymbol{p}(t)-\boldsymbol{v}_{q}\right)^{T}\boldsymbol{n}_{i}=\displaystyle\sum_{j=0}^{n}\lambda_{j}(t)\left(\boldsymbol{v}_{j}-\boldsymbol{v}_{q}\right)^{T}\boldsymbol{n}_{i}=
		$$
		\vskip -0.3cm
		$$
		=\lambda_{i}(t)\left(\boldsymbol{v}_{i}-\boldsymbol{v}_{q}\right)^{T}\boldsymbol{n}_{i}=\lambda_{i}(t)\;\text{dist}\left(\boldsymbol{v}_{i},\boldsymbol{\pi}_{i}\right)	\addrevision{\, ,}
		$$
	}
	which implies that
	\begin{equation}
		\lambda_{i}(t)=\frac{\text{dist}\left(\boldsymbol{p}(t),\boldsymbol{\pi}_{i}\right)}{\text{dist}\left(\boldsymbol{v}_{i},\boldsymbol{\pi}_{i}\right)} \addrevision{\, .} \label{eq:normalized_distance}
	\end{equation}
	
	Hence, $\lambda_{i}(t)$ represents the ratio between the distance
	from the point $\boldsymbol{p}(t)$ of the curve to the hyperplane $\boldsymbol{\pi}_i$ and the distance from $\boldsymbol{v}_{i}$
	to that hyperplane $\boldsymbol{\pi}_i$ (see \add{Figure}~\ref{fig:Geometric-interpretation-of}
	for the case $n=3$)\footnote{Note that multiplying numerator and denominator of \add{Eq.~}\ref{eq:normalized_distance} by the area of the facet that lies on the plane \add{$\boldsymbol{\pi}_i$}, each $\lambda_{i}(t)$ can also be defined as a ratio of volumes, as in \cite{ericson2004real}.}. From
	\add{Eq.~}\ref{eq:normalized_distance} it is clear that each $\lambda_{i}(t)$
	is an $n^\text{th}$-degree polynomial, that we will write as $\lambda_{i}(t):=\boldsymbol{\lambda}_{i}^{T}\boldsymbol{t}$, where $\boldsymbol{\lambda}_{i}$ is its vector of coefficients.
	Matching now the coefficients of $\boldsymbol{p}(t)$ with the ones of $\sum_{i=0}^{n}\lambda_{i}(t)\boldsymbol{v}_{i}$, the first constraint of \add{Eq.~}\ref{eq:PinS} can be rewritten as $\addrevision{\boldsymbol{P}=\boldsymbol{V} \boldsymbol{A}}$, where
	\addrevision{
		$$\arraycolsep=1.4pt
	\boldsymbol{A}:=\left[\begin{array}{c}
		\boldsymbol{\lambda}_{0}^{T} \\
		\boldsymbol{\lambda}_{1}^{T} \\
		\vdots \\
		\boldsymbol{\lambda}_{n}^{T}
	\end{array}\right]=\left[\begin{array}{llll}
		\boldsymbol{\lambda}_{0} & \boldsymbol{\lambda}_{1} & \cdots & \boldsymbol{\lambda}_{n}
	\end{array}\right]^{T} \addrevision{\, .}
	$$
}
	
	\addrevision{Note that $\boldsymbol{A}$ is a $(n+1)\times(n+1)$ matrix whose} $i^\text{th}$ row contains the coefficients of the polynomial $\lambda_{i}(t)$ 	in decreasing order. The second and third constraints of \add{Eq.~}\ref{eq:PinS} can be rewritten as $\boldsymbol{A}^{T}\boldsymbol{1}=\boldsymbol{e}$ and $\boldsymbol{A}\boldsymbol{t}\ge0\;\forall t\in[-1,1]$ respectively. We conclude therefore that 
	\begin{equation}\label{eq:PinS2}
		P\subset S\equiv\left\{ \begin{aligned}
			&\boldsymbol{P}=\boldsymbol{V}\boldsymbol{A}\\
			&\boldsymbol{A}^{T}\boldsymbol{1}=\boldsymbol{e}\\
			&\boldsymbol{A}\boldsymbol{t}\ge0\;\;\forall t\in[-1,1]
		\end{aligned}\right.  
	\end{equation}
	
	\subsection{Objective Function of Problem 1}
	
	Using the constraints in \add{Eq.~}\ref{eq:PinS2}, and noting that the matrix $\boldsymbol{A}$ is invertible (as proven \addcad{in~\ref{sec:AppInvertibility}}), we can write
	\addcad{
	$$
		\arraycolsep=1.5pt f_{1}\propto\text{abs}\left(\left|\left[\begin{array}{cc}
		\boldsymbol{V}^{T} & \boldsymbol{1}\end{array}\right]\right|\right)=\text{abs}\left(\left|\boldsymbol{A}^{-T}\left[\begin{array}{cc}
		\boldsymbol{P}^{T} & \boldsymbol{e}\end{array}\right]\right|\right)\propto
	$$
	\vskip -0.3cm
	$$
\propto\text{abs}\left(\left|\boldsymbol{A}^{-1}\right|\right)=\frac{1}{\text{abs}\left(\left|\boldsymbol{A}\right|\right)} \addrevision{\, ,}
	$$
}where we have used the fact that everything inside $\arraycolsep=1.5pt \left[\begin{array}{cc}
		\boldsymbol{P}^{T} & \boldsymbol{e}\end{array}\right]$ is given (i.e., not a decision variable of the optimization problem),
	and the fact that $\left|\boldsymbol{A}\right|=|\boldsymbol{A}^{T}|$. \add{We can therefore  
		minimize $-\text{abs}\left(\left|\boldsymbol{A}\right|\right)$}. Note that now the objective function $f_1$ is \textbf{independent} of the \addrevision{given} curve $P$. 
	
	\subsection{Objective Function of Problem 2}
	
	Similar to the previous subsection, and noting that $\boldsymbol{V}$ is given in Problem 2, we have that
	\addcad{
		$$\arraycolsep=1.6pt
		f_{2}\propto-\text{abs}\left(\left|\boldsymbol{P}_{:,0:n-1}\right|\right)=-\text{abs}\left(\left|\left[\begin{array}{c}
			\boldsymbol{P}\\
			\boldsymbol{e}^{T}
		\end{array}\right]\right|\right)=
		$$
		\vskip -0.3cm
		$$\arraycolsep=1.6pt
		=-\text{abs}\left(\left|\left[\begin{array}{c}
			\boldsymbol{V}\\
			\boldsymbol{1}^{T}
		\end{array}\right]\boldsymbol{A}\right|\right)\propto-\text{abs}\left(\left|\boldsymbol{A}\right|\right) \addrevision{\, ,}
		$$
	}%
	and therefore the objective function $f_2$ is now \textbf{independent} of the \addrevision{given simplex} $S$. 
	
	\subsection{Equivalent Formulation for Problems 1 and 2}\label{subsec:Equiv_formulation}
	
	Note that now the dependence on the \addrevision{given} polynomial curve (for Problem 1) or on the \addrevision{given} simplex (for Problem 2) appears only in the constraint $\boldsymbol{P}=\boldsymbol{V}\boldsymbol{A}$. As $\boldsymbol{A}$ is invertible \add{(\addcad{see~\ref{sec:AppInvertibility}})}, we can safely remove this constraint from the optimization, leaving $\boldsymbol{A}$ as the only decision variable, and then use $\boldsymbol{P}=\boldsymbol{V}\boldsymbol{A}$ to obtain $\boldsymbol{V}$ (for Problem 1) or $\boldsymbol{P}$ (for Problem 2). We end up therefore with the following optimization problem\footnote{Note that in the objective function of Problem 3 the $\text{abs}(\cdot)$ is not necessary,
		since any permutation of the rows of $\boldsymbol{A}$ will change
		the sign of $\left|\boldsymbol{A}\right|$. We keep it simply for
		consistency purposes, since later in the solutions we will show a
		specific order of the rows of $\boldsymbol{A}$ for which (for some
		$n$) $\left|\boldsymbol{A}\right|<0$, but that allows us to highlight
		the similarities and differences between this matrix and the ones the
		Bernstein and B-Spline bases use. }:
	\vspace{0.2cm}
	
	\noindent\fcolorbox{black}{problem3_color}{\begin{minipage}[t]{1\columnwidth - 2\fboxsep - 2\fboxrule}%
			
			\textbf{Problem 3: }
			\[
			\min_{\boldsymbol{A}\in\mathbb{R}^{(n+1)\times(n+1)}}\quad \addrevision{-\text{abs}\left(\left|\boldsymbol{A}\right|\right)}
			\]
			\vspace{-0.5cm}
			\add{\begin{alignat*}{1} 
					\qquad\qquad\text{s.t.} \quad& \boldsymbol{A}^{T}\boldsymbol{1}=\boldsymbol{e}\\
					&\boldsymbol{A}\boldsymbol{t}\ge\boldsymbol{0}\quad\forall t\in[-1,1]
			\end{alignat*}}
			\vspace{-0.5cm}
	\end{minipage}}
	
	\vspace{0.2cm}
	
	\emph{Remark: }As detailed above, Problem 3 does not depend on the specific \addrevision{given} $n^\text{th}$\add{-degree} polynomial curve (for Problem 1) or on the specific \addrevision{given} $n$-simplex (for Problem 2). Hence, its optimal solution $\boldsymbol{A}^*$ for a specific $n$ can be applied to obtain the optimal solution of Problem 1 for \textbf{any} given polynomial curve $P\in\mathcal{P}^n$ 
	(by using $\boldsymbol{V}^*=\boldsymbol{P}\left(\boldsymbol{A}^*\right)^{-1}$) and to obtain the optimal solution of Problem 2 for \textbf{any} given simplex $S\in\mathcal{S}^n$
	(by using $\boldsymbol{P}^*=\boldsymbol{V}\boldsymbol{A}^*$).
	
	As the objective function of Problem 3 is the determinant of the nonsymmetric
	matrix $\boldsymbol{A}$, it is clearly a nonconvex problem. We can rewrite the constraint $\boldsymbol{A}\boldsymbol{t}\ge\boldsymbol{0}\quad\forall t\in[-1,1]$ of Problem~3 using Sum-Of-Squares programming \add{\cite{blekherman2012semidefinite}}: %
	\begin{itemize}
		\item If $n$ is odd, $\lambda_{i}(t)\ge0\;\;\forall t\in[-1,1]$ \addcad{if and only if}
		$$\add{\left\{ 
			\begin{aligned}
				&\lambda_{i}(t)=\hat{\boldsymbol{t}}^{T}\left((t+1)\boldsymbol{W}_{i}+(1-t)\boldsymbol{V}_{i}\right)\hat{\boldsymbol{t}}\\
				&\boldsymbol{W}_{i}\in\mathbb{S}_{+}^{\frac{n+1}{2}},\boldsymbol{V}_{i}\in\mathbb{S}_{+}^{\frac{n+1}{2}}
			\end{aligned}
			\right.}
		$$
		
		\item If $n$ is even, $\lambda_{i}(t)\ge0\;\;\forall t\in[-1,1]$ \addcad{if and only if}
		$$\add{\left\{ 
			\begin{aligned}
				&\lambda_{i}(t)=\hat{\boldsymbol{t}}^{T}\boldsymbol{W}_{i}\hat{\boldsymbol{t}}+\hat{\boldsymbol{t}}^{T}(t+1)(1-t)\boldsymbol{V}_{i}\hat{\boldsymbol{t}}\\
				&\boldsymbol{W}_{i}\in\mathbb{S}_{+}^{\frac{n}{2}+1},\boldsymbol{V}_{i}\in\mathbb{S}_{+}^{\frac{n}{2}}
			\end{aligned}
			\right.}
		$$
	\end{itemize}
	Note that the \emph{if and only if} condition applies because $\lambda_{i}(t)$ is
	a univariate polynomial~\add{\cite{blekherman2012semidefinite}}. The decisions variables would be the positive semidefinite matrices $\boldsymbol{W}_{i}$
	and $\boldsymbol{V}_{i}$, $i=0,\ldots,n$. \addrevision{Another} option is to use the Markov–Luk\'{a}cs \addrevision{Theorem} (\cite[Theorem 1.21.1]{szeg1939orthogonal},\cite[Theorem 2.2]{kreuin1977markov},\cite{roh2006discrete}): 
	
	\begin{itemize}
		\item If $n$ is odd, $\lambda_{i}(t)\ge0\;\;\forall t\in[-1,1]$ \addcad{if and only if}\add{ $$\lambda_i(t)=(t+1)g_i^2(t)+(1-t)h_i^2(t)} \addrevision{\, .}$$
		\item If $n$ is even, $\lambda_{i}(t)\ge0\;\;\forall t\in[-1,1]$ \addcad{if and only if}\add{ $$\lambda_i(t)=g_i^2(t)+(t+1)(1-t)h_i^2(t)} \addrevision{\, .}$$ 
	\end{itemize}
	where $g_i(t)$ and $h_i(t)$ are polynomials of degrees $\operatorname{deg}(g_i(t)) \leq\lfloor n / 2\rfloor$ and $\operatorname{deg}(h_i(t)) \leq\lfloor(n-1)/2\rfloor$. The decision variables would be the coefficients of the polynomials $g_i(t)$ and $h_i(t)$, $i=0,\ldots,n$. \addcad{In~\ref{sec:AppKKT}} we derive the Karush–Kuhn–Tucker (KKT) conditions of Problem 3 using this theorem.

	Regardless of the choice of the representation of the constraint $\boldsymbol{A}\boldsymbol{t}\ge\boldsymbol{0}\;\;\forall t\in[-1,1]$ (SOS or the Markov–Luk\'{a}cs \addrevision{Theorem}), no generality has been lost so far. However, these formulations easily
	become intractable for large $n$ due to the high number of decision variables. We
	can reduce the number of decision variables of Problem 3 by
	imposing a structure in $\lambda_{i}(t)$. As Problem 1 is trying
	to minimize the volume of the simplex, we can impose that the facets
	of the $n$-simplex be tangent to several internal points $\boldsymbol{p}(t)$ of the curve (with $t\in(-1,1)$), and in contact with the first and last points of the curve ($\boldsymbol{p}(-1)$ and $\boldsymbol{p}(1)$)  \cite{zhou2002algorithms,klee1986facet}. Using the geometric
	interpretation of the $\lambda_{i}(t)$ given in \addcad{Section~}\ref{subsec:Constraints-of-Problem},
	this means that each $\lambda_{i}(t)$ should have either real double
	roots in $t\in(-1,1)$ (where the curve is tangent to a facet), and/or roots at \addrevision{$t\in\{-1,1\}$}. These conditions, together with an additional symmetry condition between different $\lambda_{i}(t)$, translate into the formulation shown in Problem 4, in which the decisions variables are the roots of $\lambda_{i}(t)$ and the coefficients $b_i$.
	
	\vspace{0.15cm}
	\noindent\fcolorbox{black}{problem4_color}{
	\resizebox{0.95\columnwidth}{!}{
		\begin{minipage}[t]{1.0\columnwidth}
			\textbf{Problem 4:}
			\[
			\min_{\boldsymbol{A}\in\mathbb{R}^{(n+1)\times(n+1)}}\quad \addrevision{-\text{abs}\left(\left|\boldsymbol{A}\right|\right)\quad} \text{subject to:}
			\]
			\textbf{\uline{If \mbox{$n$} is odd:}} \vskip-4ex
			\addtolength{\jot}{-0.3em}
			\add{\begin{alignat*}{2} 
					&\lambda_{i}(t)=-b_{i}(t-1)\textstyle\prod_{j=1}^{\frac{n-1}{2}}(t-t_{ij})^{2}\hspace{0.2cm} &&i=0,2,\ldots,n-1\\
					&\lambda_{i}(t)=\lambda_{n-i}(-t)  \hspace{3cm} &&i=1,3,\ldots,n\\
					&b_{i}\ge0\quad &&i=0,2,\ldots,n-1\\
					&\boldsymbol{A}^{T}\boldsymbol{1}=\boldsymbol{e} &&
			\end{alignat*}}
			\textbf{\uline{If \mbox{$n$} is even:}}  \vskip-4ex
			
			\begin{alignat*}{2} 
					&\lambda_{i}(t)=-b_{i}(t+1)(t-1)\textstyle\prod_{j=1}^{\frac{n-2}{2}}(t-t_{ij})^{2} \quad&&\addcad{i\text{ odd integer}\in \left[0,n/2-1\right]}\\
					&\lambda_{i}(t)=b_{i}\textstyle\prod_{j=1}^{\frac{n}{2}}(t-t_{ij})^{2} &&\addcad{i\text{ even integer}\in \left[0,n/2-1\right]}\\
					&\lambda_{i}(t)=\lambda_{n-i}(-t) &&i=n/2+1,\ldots,n\\
					&b_{i}\ge0\quad &&i=0,1,\dots,n/2\\
					&\boldsymbol{A}^{T}\boldsymbol{1}=\boldsymbol{e} &&
			\end{alignat*} \vskip-7ex
			\begin{alignat*}{3} 
			&\text{}&&\addcad{\lambda_{i}(t)=-b_{i}(t+1)(t-1)\textstyle\prod_{j=1}^{\frac{n-2}{4}}(t-t_{ij})^{2}(t+t_{ij})^{2}}\quad&&i=n/2,\; i\text{ odd}\\
			&\text{}&&\addcad{\lambda_{i}(t)=b_{i}\textstyle\prod_{j=1}^{\frac{n}{4}}(t-t_{ij})^{2}(t+t_{ij})^{2}} &&i=n/2,\; i\text{ even}
		    \end{alignat*}
	\end{minipage}
}
}
	\vspace{0.9cm}
	
\begin{figure}
		\begin{centering}
			\includegraphics[width=1.0\columnwidth]{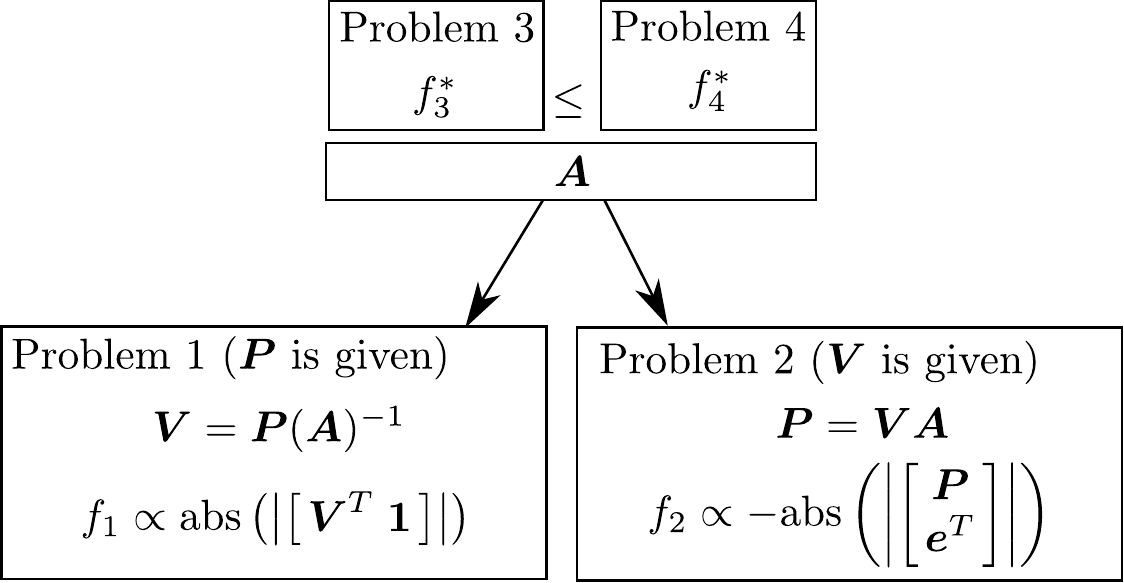}
			\par\end{centering}
		\caption{Relationship between Problems 1, 2, 3, and 4: Problem~3 and~4 have the same objective function, but the feasible set of Problem~4 is contained in the feasible set of Problem~3, and therefore $f_{3}^{*}\le f_{4}^{*}$.
			Both Problem~3 and 4 generate a solution
			$\boldsymbol{A}$, which can be applied to \textbf{any} polynomial curve $P\in\mathcal{P}^n$
			to find the simplex $S\in\mathcal{S}^n$ in Problem 1, or to \textbf{any} simplex $S\in\mathcal{S}^n$ to find the polynomial
			curve $P\in\mathcal{P}^n$ in Problem 2. \label{fig:relationships_problems}}
	\end{figure}

	\addrevision{Letting $f_i$ denote the objective function of Problem $i$}, the relationship between Problems 1, 2, 3 and 4 is given in \add{Figure}~\ref{fig:relationships_problems}.
	First note that the constraints and structure imposed on $\lambda_{i}(t)$ in Problem
	4 guarantee that they are nonnegative for $t\in[-1,1]$ and that they sum up to 1. Hence the
	feasible set of Problem 4 is contained in the feasible set of Problem
	3, and therefore, $f_{3}^{*}\le f_{4}^{*}$ holds. The matrix $\boldsymbol{A}$
	found in Problem 3 or 4 can be used to find the \addrevision{vertices} of the simplex
	in Problem 1 (by simply using $\boldsymbol{V}=\boldsymbol{P}\left(\boldsymbol{A}\right)^{-1}$,
	where $\boldsymbol{P}$ is the coefficient matrix of the polynomial curve
	given), or to find the coefficient matrix of the polynomial curve in Problem
	2 (by using $\boldsymbol{P}=\boldsymbol{V}\boldsymbol{A}$,
	where $\boldsymbol{V}$ contains the \addrevision{vertices} of the \addrevision{given} simplex).
	
	\section{Results}
	\subsection{Results for $n=1,2,\dots,7$}\label{sec:Resultsn17}
	\begin{table}%
	\caption{Roots of each $\lambda_{i}(t)$ of the MINVO basis. $\boldsymbol{r}(\lambda_{i}(t))$
		is the column vector that contains the roots of $\lambda_{i}(t)$.
		All the roots~\addmorerevision{lying in $(-1,1)$} are double roots, while the ones~\addmorerevision{in}~$\{-1,1\}$
		are single roots. Each $\lambda_{i}(t)$ has $n$ real roots in total. These roots are plotted in \add{Figure}~\ref{fig:roots_distribution_all}.
		\label{tab:roots}}
	\centering
	\noindent\resizebox{0.8\columnwidth}{!}{%
		\centering
		\begin{tabular}{|c|l|}
			\hline 
			\textbf{\hspace{0.1cm}$\boldsymbol{n}$\hspace{0.1cm}} & \qquad{}Roots of $\lambda_{i}(t),\;t\in[-1,1]$\tabularnewline
			\hline 
			\hline 
			$1$ & {\scriptsize{}$\left[\begin{array}{c}
					\boldsymbol{r}(\lambda_{0})^{T}\\
					\boldsymbol{r}(\lambda_{1})^{T}
				\end{array}\right]=\left[\begin{array}{c}
					1.0\\
					-1.0
				\end{array}\right]$}\tabularnewline
			\hline 
			$2$ & {\scriptsize{}$\left[\begin{array}{c}
					\boldsymbol{r}(\lambda_{0})^{T}\\
					\boldsymbol{r}(\lambda_{1})^{T}\\
					\boldsymbol{r}(\lambda_{2})^{T}
				\end{array}\right]=\left[\begin{array}{cc}
					\frac{1}{\sqrt{3}} & \mathrm{}\\
					-1.0 & 1.0\\
					-\frac{1}{\sqrt{3}} & \mathrm{}
				\end{array}\right]$}\tabularnewline
			\hline 
			$3$ & {\scriptsize{}$\left[\begin{array}{c}
					\boldsymbol{r}(\lambda_{0})^{T}\\
					\boldsymbol{r}(\lambda_{1})^{T}\\
					\boldsymbol{r}(\lambda_{2})^{T}\\
					\boldsymbol{r}(\lambda_{3})^{T}
				\end{array}\right]\approx\left[\begin{array}{cc}
					0.03088 & 1.0\\
					-1.0 & 0.7735\\
					-0.7735 & 1.0\\
					-1.0 & -0.03088
				\end{array}\right]$}\tabularnewline
			\hline 
			$4$ & {\scriptsize{}$\left[\begin{array}{c}
					\boldsymbol{r}(\lambda_{0})^{T}\\
					\boldsymbol{r}(\lambda_{1})^{T}\\
					\boldsymbol{r}(\lambda_{2})^{T}\\
					\boldsymbol{r}(\lambda_{3})^{T}\\
					\boldsymbol{r}(\lambda_{4})^{T}
				\end{array}\right]\approx\left[\begin{array}{ccc}
					-0.2872 & 0.835 & \mathrm{}\\
					-1.0 & 0.3657 & 1.0\\
					-0.8618 & 0.8618 & \mathrm{}\\
					-1.0 & -0.3657 & 1.0\\
					-0.835 & 0.2872
				\end{array}\right]$}\tabularnewline
			\hline 
			$5$ & {\scriptsize{}$\left[\begin{array}{c}
					\boldsymbol{r}(\lambda_{0})^{T}\\
					\boldsymbol{r}(\lambda_{1})^{T}\\
					\boldsymbol{r}(\lambda_{2})^{T}\\
					\boldsymbol{r}(\lambda_{3})^{T}\\
					\boldsymbol{r}(\lambda_{4})^{T}\\
					\boldsymbol{r}(\lambda_{5})^{T}
				\end{array}\right]\approx\left[\begin{array}{ccc}
					-0.4866 & 0.5121 & 1.0\\
					-1.0 & 0.04934 & 0.8895\\
					-0.9057 & 0.5606 & 1.0\\
					-1.0 & -0.5606 & 0.9057\\
					-0.8895 & -0.04934 & 1.0\\
					-1.0 & -0.5121 & 0.4866
				\end{array}\right]$}\tabularnewline
			\hline 
			$6$ & {\scriptsize{}$\left[\begin{array}{c}
					\boldsymbol{r}(\lambda_{0})^{T}\\
					\boldsymbol{r}(\lambda_{1})^{T}\\
					\boldsymbol{r}(\lambda_{2})^{T}\\
					\boldsymbol{r}(\lambda_{3})^{T}\\
					\boldsymbol{r}(\lambda_{4})^{T}\\
					\boldsymbol{r}(\lambda_{5})^{T}\\
					\boldsymbol{r}(\lambda_{6})^{T}
				\end{array}\right]\approx\left[\begin{array}{cccc}
					-0.6135 & 0.2348 & 0.9137 & \mathrm{}\\
					-1.0 & -0.1835 & 0.6449 & 1.0\\
					-0.9317 & 0.2822 & 0.9214 & \mathrm{}\\
					-1.0 & -0.6768 & 0.6768 & 1.0\\
					-0.9214 & -0.2822 & 0.9317 & \mathrm{}\\
					-1.0 & -0.6449 & 0.1835 & 1.0\\
					-0.9137 & -0.2348 & 0.6135 & \mathrm{}
				\end{array}\right]$}\tabularnewline
			\hline 
			$7$ & {\scriptsize{}$\left[\begin{array}{c}
					\boldsymbol{r}(\lambda_{0})^{T}\\
					\boldsymbol{r}(\lambda_{1})^{T}\\
					\boldsymbol{r}(\lambda_{2})^{T}\\
					\boldsymbol{r}(\lambda_{3})^{T}\\
					\boldsymbol{r}(\lambda_{4})^{T}\\
					\boldsymbol{r}(\lambda_{5})^{T}\\
					\boldsymbol{r}(\lambda_{6})^{T}\\
					\boldsymbol{r}(\lambda_{7})^{T}
				\end{array}\right]\approx\left[\begin{array}{cccc}
					-0.7 & 0.008364 & 0.7132 & 1.0\\
					-1.0 & -0.3509 & 0.4068 & 0.9355\\
					-0.9481 & 0.05239 & 0.7315 & 1.0\\
					-1.0 & -0.753 & 0.4399 & 0.9408\\
					-0.9408 & -0.4399 & 0.753 & 1.0\\
					-1.0 & -0.7315 & -0.05239 & 0.9481\\
					-0.9355 & -0.4068 & 0.3509 & 1.0\\
					-1.0 & -0.7132 & -0.008364 & 0.7
				\end{array}\right]$}\tabularnewline
			\hline 
		\end{tabular}
	}
\end{table}
	
\begin{figure*}
		\begin{centering}
			\includegraphics[width=0.95\textwidth]{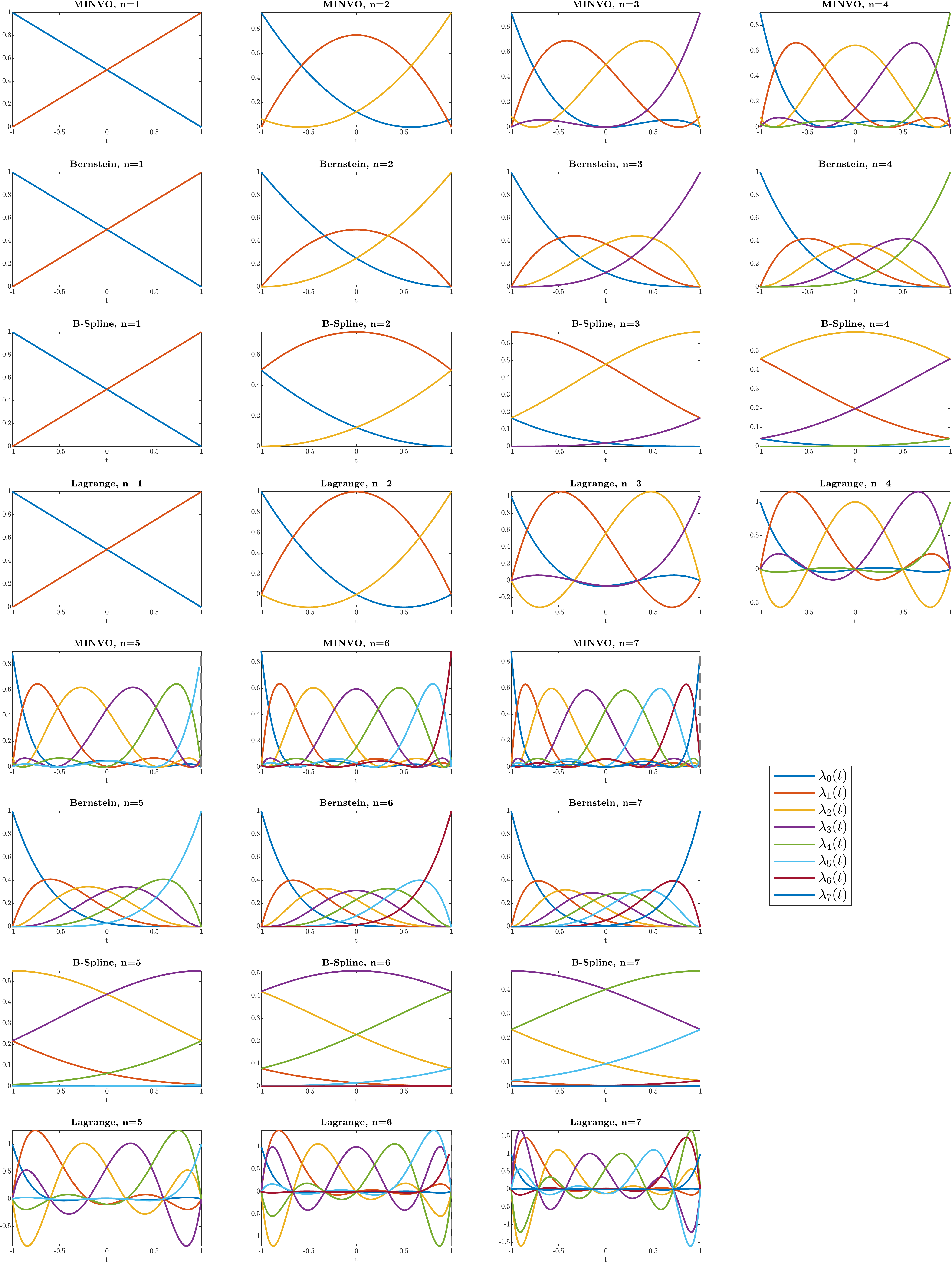}
			\par\end{centering}
		\caption{Comparison between the MINVO, Bernstein, B-Spline, and Lagrange bases
			for $n=1,2,\dots,7$. All these bases satisfy $\sum_{i=0}^{n}\lambda_{i}(t)=1$,
			and the MINVO, Bernstein, and B-Spline bases also satisfy $\lambda_{i}(t)\ge0\;\;\forall t\in[-1,1]$.
			\label{fig:minvo_and_bernstein}}
		\vskip-3ex
	\end{figure*}

\begin{figure*}[t]
		\begin{centering}
			\includegraphics[width=0.92\textwidth]{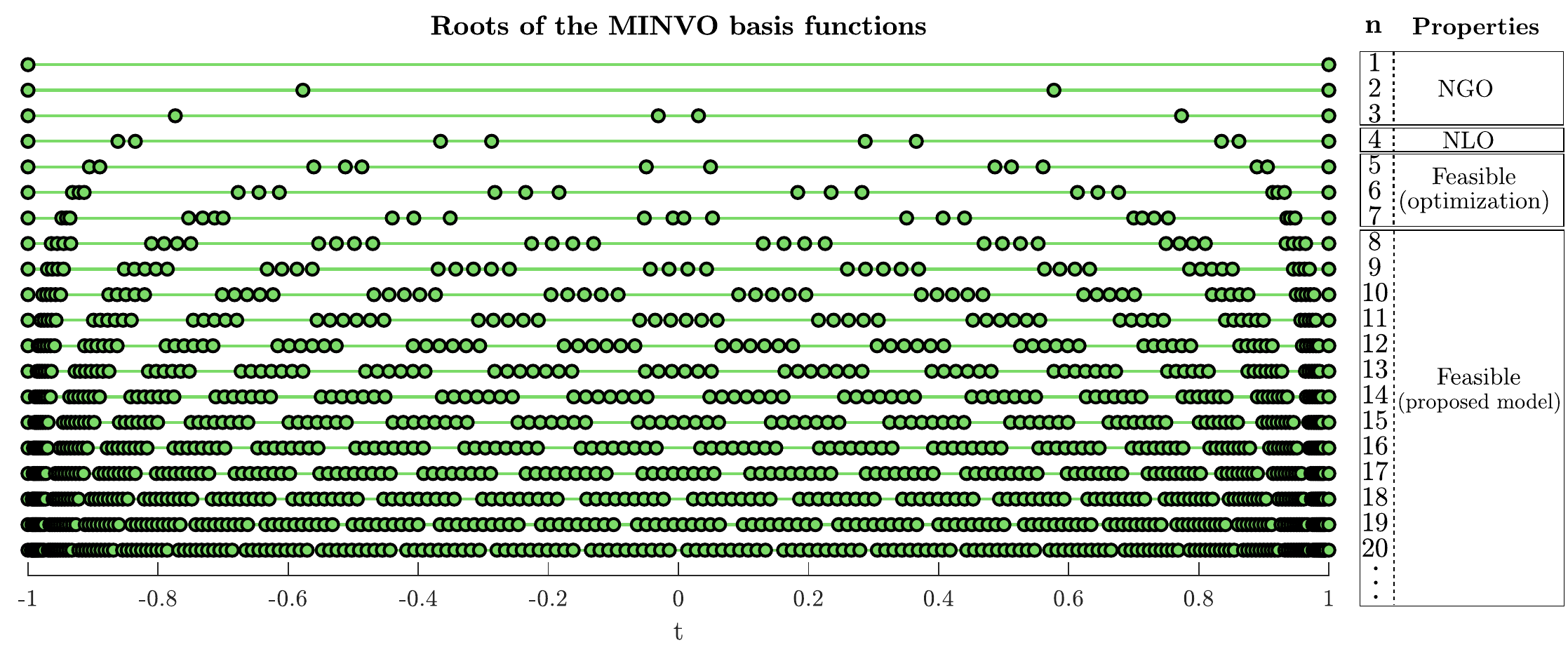}
			\par\end{centering}
		\vskip -1ex
		\caption{\add{Distribution of the roots of the MINVO basis functions for different $n$. The results for $n\le7$ were obtained by solving the optimization problems (\addcad{Section~}\ref{sec:Resultsn17}, see also Table~\ref{tab:roots}), while the results for $n\ge8$ were obtained using the model proposed in \addcad{Section~}\ref{sec:resultsnGT7}.}\label{fig:roots_distribution_all}}
	\end{figure*}

	Using the nonconvex solvers \emph{fmincon} \cite{matlabOptToolbox} and \emph{SNOPT} \cite{GilMS05, snopt77} (with the \emph{YALMIP} interface \cite{Lofberg2004,Lofberg2009}), 
	we were able to find \add{NLO solutions} for Problem 4 for $n=1,2,\dots,7$,
	and the same \add{NLO solutions} were found in Problem 3 for $n=1,2,3,4$. Problem 3 and 4 become intractable for $n\ge5$ and $n\ge8$ respectively. The optimal matrices
	$\boldsymbol{A}$ found are shown in Table~\ref{tab:table_matrices},
	and are denoted as $\boldsymbol{A}_{\text{MV}}$\footnote{\addrevision{Note that any permutation
		in the rows of $\boldsymbol{A}_{\text{MV}}$ will not change the objective
		value, since only the sign of the determinant is affected. Despite this multiplicity of solutions, we will refer to any matrix shown in Table~\ref{tab:table_matrices} as, e.g., \textit{the} optimal solution $\boldsymbol{A}_\text{MV}$, \textit{the} feasible solution $\boldsymbol{A}_\text{MV}$, \addmorerevision{etc.}}}. \add{Their determinants}
	$\left|\boldsymbol{A}_{\text{MV}}\right|$ \add{are} also compared with the
	one of the Bernstein and B-Spline matrices (denoted as $\boldsymbol{A}_{\text{Be}}$
	and $\boldsymbol{A}_{\text{BS}}$ respectively). The corresponding
	plots of the MINVO basis functions are shown in \add{Figure}~\ref{fig:minvo_and_bernstein},
	together with the Bernstein, B-Spline and Lagrange bases for comparison. \add{All of these bases} satisfy $\sum_{i=0}^{n}\lambda_{i}(t)=1$, and the MINVO, Bernstein, and B-Spline bases also satisfy $\lambda_{i}(t)\ge0\;\;\forall t\in[-1,1]$. 
	The roots of each of the MINVO basis functions $\lambda_{i}(t)$ \add{for $n=1,\hdots,7$ are}
	shown in Table~\ref{tab:roots} and plotted in \add{Figure}~\ref{fig:roots_distribution_all}.
	
	{\setlength\extrarowheight{1pt}
		\setlength\tabcolsep{0pt}
		
		\begin{table*}[t]
			\caption{Results for the MINVO basis. $\boldsymbol{A}_{\text{MV}}$, $\boldsymbol{A}_{\text{Be}}$
				and $\boldsymbol{A}_{\text{BS}}$ denote the coefficient matrix of
				the MINVO, Bernstein, and B-Spline bases respectively ($t\in[-1,1]$).
				The greater the absolute value of the determinant, the smaller the
				associated simplex (for Problem 1) and the larger the convex hull
				of the curve (for Problem 2). The matrices $\boldsymbol{A}_{\text{MV}}$
				found are \textbf{independent} of the \addrevision{given} polynomial curve (in Problem
				1), or of the \addrevision{given} simplex (in Problem 2). \add{NGO and NLO denote numerical Global/Local Optimality.} \label{tab:table_matrices}}
			\noindent\resizebox{\textwidth}{!}{%
				\begin{centering}
					\begin{tabular}{|c|c|c|c|c|>{\columncolor{problem3_color}\centering}c|>{\columncolor{problem4_color}\centering}c|}
						\hline 
						\textbf{\hspace{0.1cm}$\boldsymbol{n}$\hspace{0.1cm}} & $\boldsymbol{A}_{\text{MV}}$ & $\text{abs}\left(\left|\boldsymbol{A}_{\text{MV}}\right|\right)$ & $\;\frac{\text{abs}\left(|\boldsymbol{A}_{\text{MV}}|\right)}{\text{abs}\left(|\boldsymbol{A}_{\text{Be}}|\right)}\;$ & $\frac{\text{abs}\left(|\boldsymbol{A}_{\text{MV}}|\right)}{\text{abs}\left(|\boldsymbol{A}_{\text{BS}}|\right)}$ & \textbf{$\;$Problem 3$\;$} & \textbf{$\;$Problem 4$\;$}\tabularnewline
						\hline 
						\hline 
						$1$ & {\scriptsize{}$\frac{1}{2}\left[\begin{array}{cc}
								-1 & 1\\
								1 & 1
							\end{array}\right]$} & $0.5$ & $1.0$ & $1.0$ & \add{NGO} & \add{NGO}\tabularnewline
						\hline 
						$2$ & {\scriptsize{}$\frac{1}{8}\left[\begin{array}{ccc}
								3 & -2\sqrt{3} & 1\\
								-6 & 0 & 6\\
								3 & 2\sqrt{3} & 1
							\end{array}\right]$} & $0.3248$ & $1.299$ & $5.196$ & \add{NGO} & \add{NGO}\tabularnewline
						\hline 
						$3$ & {\scriptsize{}$\left[\begin{array}{cccc}
								-0.4302 & 0.4568 & -0.02698 & 0.0004103\\
								0.8349 & -0.4568 & -0.7921 & 0.4996\\
								-0.8349 & -0.4568 & 0.7921 & 0.4996\\
								0.4302 & 0.4568 & 0.02698 & 0.0004103
							\end{array}\right]$} & $0.3319$ & $2.360$ & $254.9$ & \add{NGO} & \add{NGO}\tabularnewline
						\hline 
						$4$ & {\scriptsize{}$\left[\begin{array}{ccccc}
								0.5255 & -0.5758 & -0.09435 & 0.1381 & 0.03023\\
								-1.108 & 0.8108 & 0.9602 & -0.8108 & 0.1483\\
								1.166 & 0 & -1.732 & 0 & 0.643\\
								-1.108 & -0.8108 & 0.9602 & 0.8108 & 0.1483\\
								0.5255 & 0.5758 & -0.09435 & -0.1381 & 0.03023
							\end{array}\right]$} & $0.5678$ & $6.057$ & $1.675\cdot10^{5}$ & $\begin{array}{c}
							\text{\add{NLO}}\\
							\text{(at least)}
						\end{array}$ & $\begin{array}{c}
							\text{\add{NLO}}\\
							\text{(at least)}
						\end{array}$\tabularnewline
						\hline 
						$5$ & {\scriptsize{}$\left[\begin{array}{cccccc}
								-0.7392 & 0.7769 & 0.3302 & -0.3773 & -0.0365 & 0.04589\\
								1.503 & -1.319 & -1.366 & 1.333 & -0.121 & 0.002895\\
								-1.75 & 0.5424 & 2.777 & -0.9557 & -1.064 & 0.4512\\
								1.75 & 0.5424 & -2.777 & -0.9557 & 1.064 & 0.4512\\
								-1.503 & -1.319 & 1.366 & 1.333 & 0.121 & 0.002895\\
								0.7392 & 0.7769 & -0.3302 & -0.3773 & 0.0365 & 0.04589
							\end{array}\right]$} & $1.6987$ & $22.27$ & $1.924\cdot10^{9}$ & $\begin{array}{c}
							\text{Feasible}\\
							\text{(at least)}
						\end{array}$ & $\begin{array}{c}
							\text{\add{NLO}}\\
							\text{(at least)}
						\end{array}$\tabularnewline
						\hline 
						$6$ & {\scriptsize{}$\left[\begin{array}{ccccccc}
								1.06 & -1.134 & -0.7357 & 0.8348 & 0.1053 & -0.1368 & 0.01836\\
								-2.227 & 2.055 & 2.281 & -2.299 & -0.08426 & 0.2433 & 0.0312\\
								2.59 & -1.408 & -4.27 & 2.468 & 1.58 & -1.081 & 0.152\\
								-2.844 & 0 & 5.45 & 0 & -3.203 & 0 & 0.5969\\
								2.59 & 1.408 & -4.27 & -2.468 & 1.58 & 1.081 & 0.152\\
								-2.227 & -2.055 & 2.281 & 2.299 & -0.08426 & -0.2433 & 0.0312\\
								1.06 & 1.134 & -0.7357 & -0.8348 & 0.1053 & 0.1368 & 0.01836
							\end{array}\right]$} & $9.1027$ & $117.8$ & $4.750\cdot10^{14}$ & $\begin{array}{c}
							\text{Feasible}\\
							\text{(at least)}
						\end{array}$ & $\begin{array}{c}
							\text{\add{NLO}}\\
							\text{(at least)}
						\end{array}$\tabularnewline
						\hline 
						$7$ & {\scriptsize{}$\left[\begin{array}{cccccccc}
								-1.637 & 1.707 & 1.563 & -1.682 & -0.3586 & 0.4143 & -0.006851 & 2.854\cdot10^{-5}\\
								3.343 & -3.285 & -3.947 & 4.173 & 0.6343 & -0.9385 & -0.02111 & 0.05961\\
								-4.053 & 2.722 & 6.935 & -4.96 & -2.706 & 2.269 & -0.2129 & 0.00535\\
								4.478 & -1.144 & -9.462 & 2.469 & 6.311 & -1.745 & -1.312 & 0.435\\
								-4.478 & -1.144 & 9.462 & 2.469 & -6.311 & -1.745 & 1.312 & 0.435\\
								4.053 & 2.722 & -6.935 & -4.96 & 2.706 & 2.269 & 0.2129 & 0.00535\\
								-3.343 & -3.285 & 3.947 & 4.173 & -0.6343 & -0.9385 & 0.02111 & 0.05961\\
								1.637 & 1.707 & -1.563 & -1.682 & 0.3586 & 0.4143 & 0.006851 & 2.854\cdot10^{-5}
							\end{array}\right]$} & $89.0191$ & $902.7$ & $\;2.997\cdot10^{21}$ & $\begin{array}{c}
							\text{Feasible}\\
							\text{(at least)}
						\end{array}$ & $\begin{array}{c}
							\text{\add{NLO}}\\
							\text{(at least)}
						\end{array}$\tabularnewline
						\hline 
					\end{tabular}
					\par\end{centering}
			}
			
			\vskip 0.1cm	
		\end{table*}
		
	}
	
	\definecolor{orange_cone}{RGB}{232,174,126} %
	\definecolor{blue_parallelepiped}{RGB}{217,254,234} %
	
	\begin{figure}[hbt!]
		\begin{centering}
			\includegraphics[width=0.8\columnwidth]{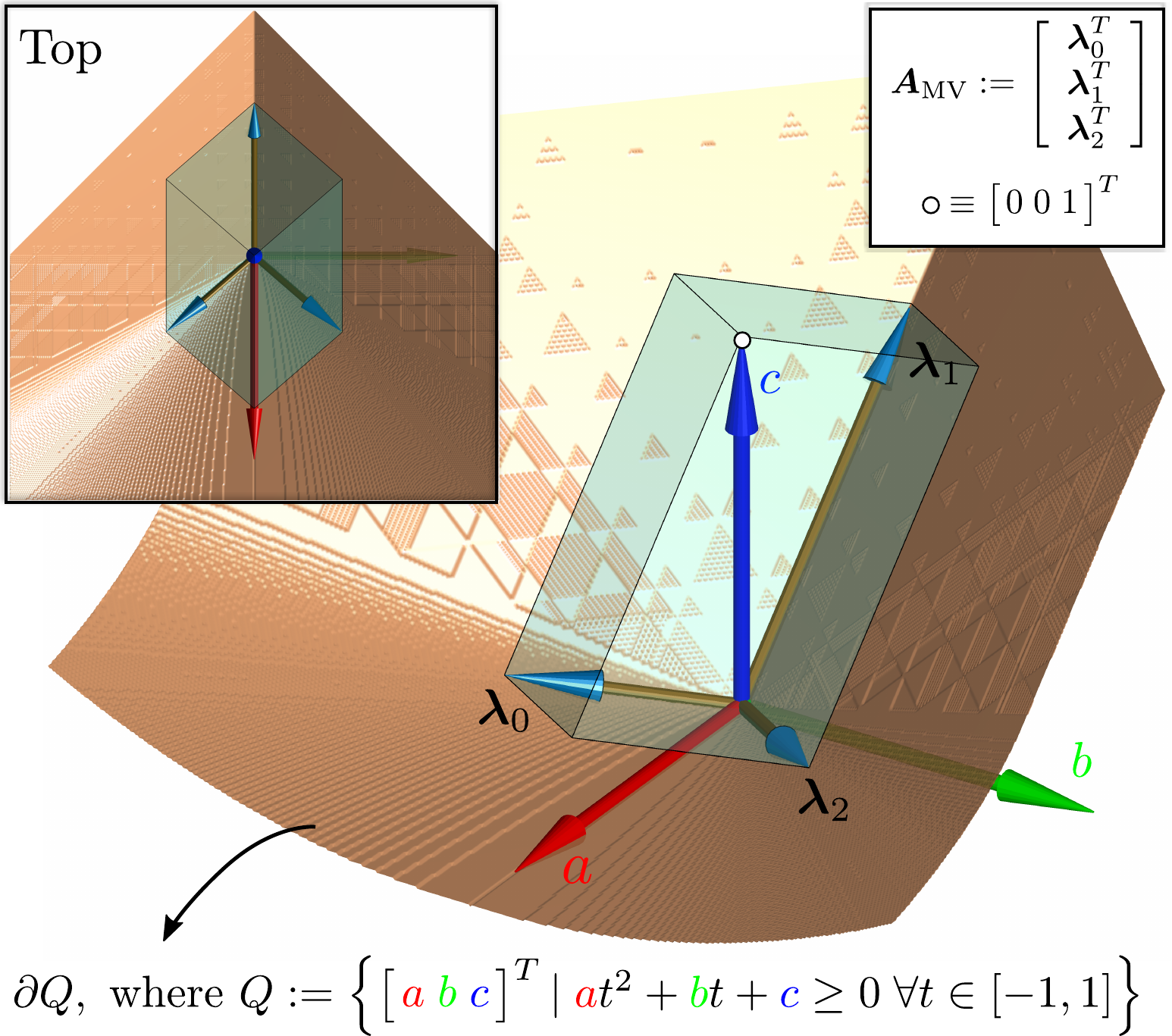}
			\par\end{centering}
		\caption{For $n=2$, the MINVO basis has $\arraycolsep=1.4pt\boldsymbol{A}_{\text{MV}}=\left[\protect\begin{array}{ccc}
				\boldsymbol{\lambda}_{0} & \boldsymbol{\lambda}_{1} & \boldsymbol{\lambda}_{2}\protect\end{array}\right]^{T}$, where $\boldsymbol{\lambda}_{0}, \boldsymbol{\lambda}_{1}$, and $\boldsymbol{\lambda}_{2}$ are vectors that span the parallelepiped \tikzbox[black,fill=blue_parallelepiped]{0.2cm}, and whose sum is  $\arraycolsep=1.0pt\left[\protect\begin{array}{ccc}
				0 &0 & 1\protect\end{array}\right]^{T}$. \addcad{Here, $\partial Q$ (\tikzbox[black,fill=orange_cone]{0.2cm}) is the}   frontier of the cone \addcad{$Q$} formed by the coefficients of the polynomials that are nonnegative \addmorerevision{for all} $ t\in[-1,1]$. Note how the globally-optimal vectors $\boldsymbol{\lambda}_0,\boldsymbol{\lambda}_1$, and $\boldsymbol{\lambda}_2$  belong to \addcad{$\partial Q$}.}
		\vskip 0.2cm
		\label{fig:cone_A2}
	\end{figure}	
	\begin{figure}[hbt!]
		\begin{centering}
			\includegraphics[width=1.0\columnwidth]{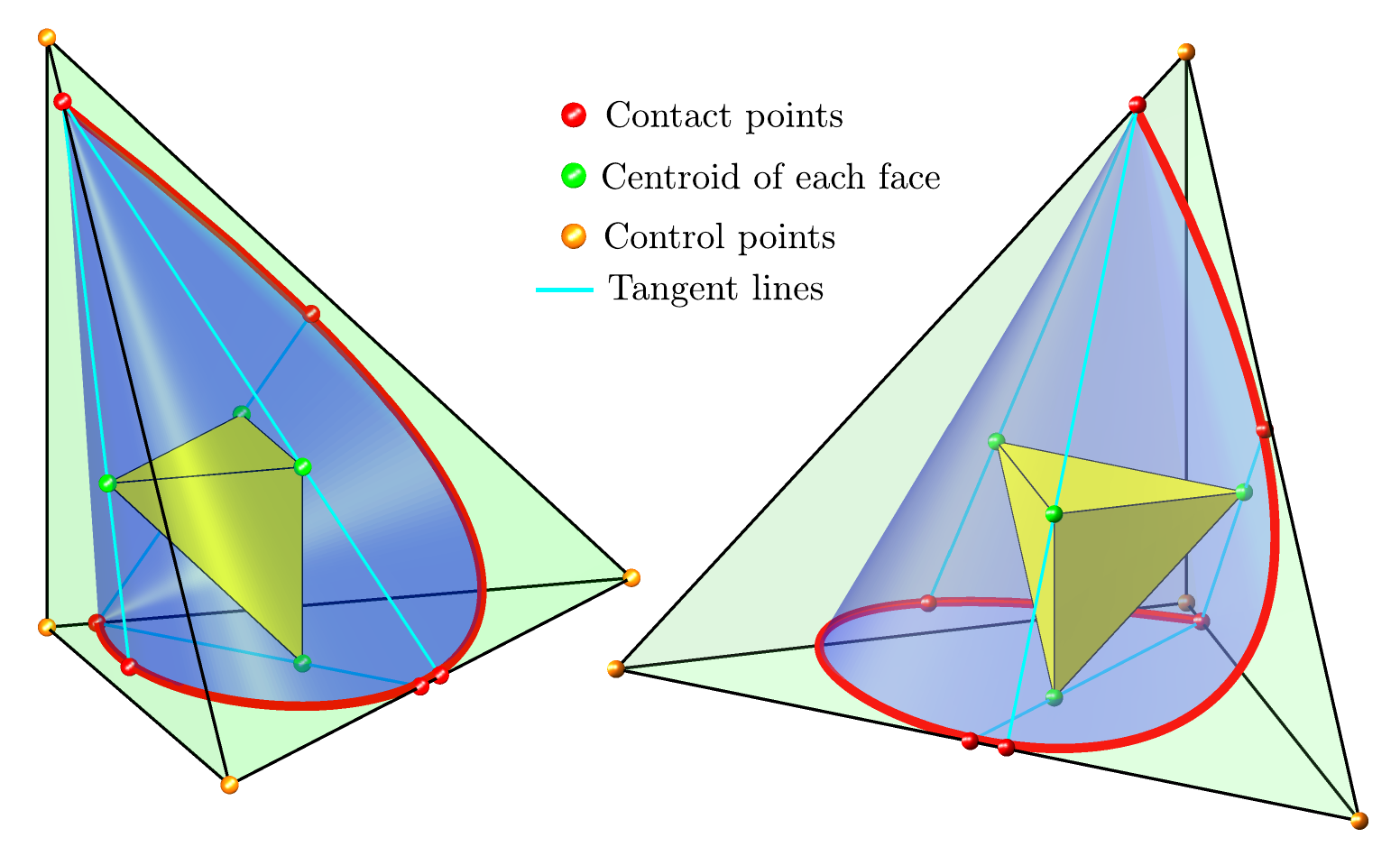}
			\par\end{centering}
		\caption{Solution obtained by the MINVO basis for $n=3$. Note how the centroids of \add{each of the facets} (\tikzcircle[green,fill=green]{2pt}, \addrevision{vertices} of the yellow tetrahedron) belong to $\text{conv}\left(P\right)$, which is a necessary condition for \add{an extremal simplex} \cite{klee1986facet}. Moreover, note that $\text{conv}\left(P\right)$ is tangent to the simplex along the blue lines. The red points \tikzcircle[red,fill=red]{2pt} denote the contact points between the curve and the simplex, which happen at the roots of the MINVO basis functions. 
			\label{fig:centroid_facets}}
		\vskip 1.0cm
	\end{figure}
	
	\renewcommand{\floatpagefraction}{1.0}%
	\begin{figure*}[hbt!]
		\centering
		\subfloat[For any given curve $P\in\mathcal{P}^2$, the MINVO basis
		finds an enclosing $2$-simplex that is $1.3$ and $5.2$
		times smaller than the one found by the Bernstein and B-Spline bases
		respectively. \label{fig:Comparison2d_poly_given.pdf}]{ \includegraphics[width=0.8\columnwidth]{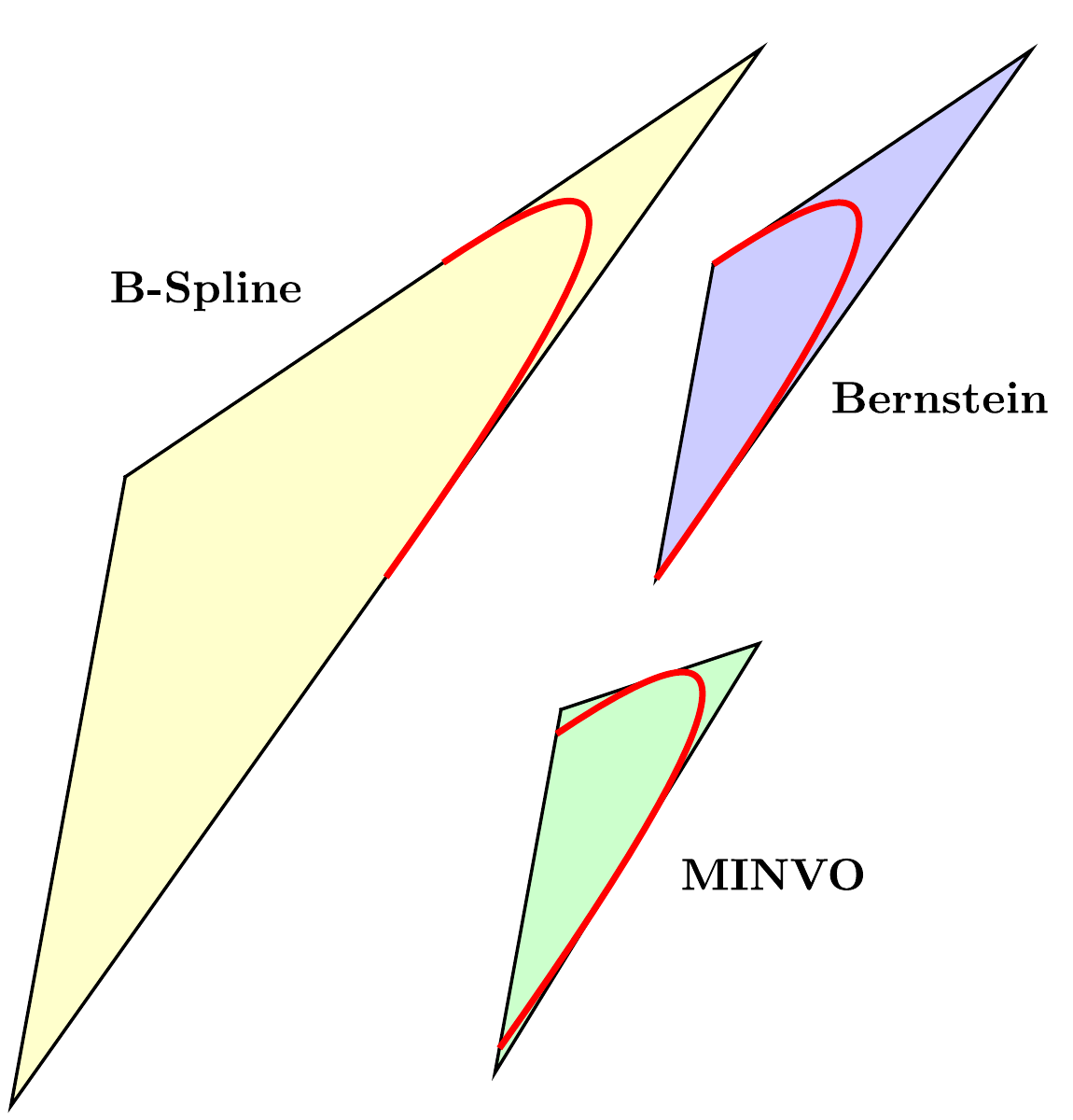}		}
		\qquad
		\subfloat[For any given $2$-simplex, the MINVO basis finds a curve $P\in\mathcal{P}^2$ inscribed in the simplex, and whose convex hull is $1.3$ and
		$5.2$ times larger than the one found by the Bernstein and
		B-Spline bases respectively. \label{fig:Comparison2d_simplex_given.pdf}]{ \includegraphics[width=1.0\columnwidth]{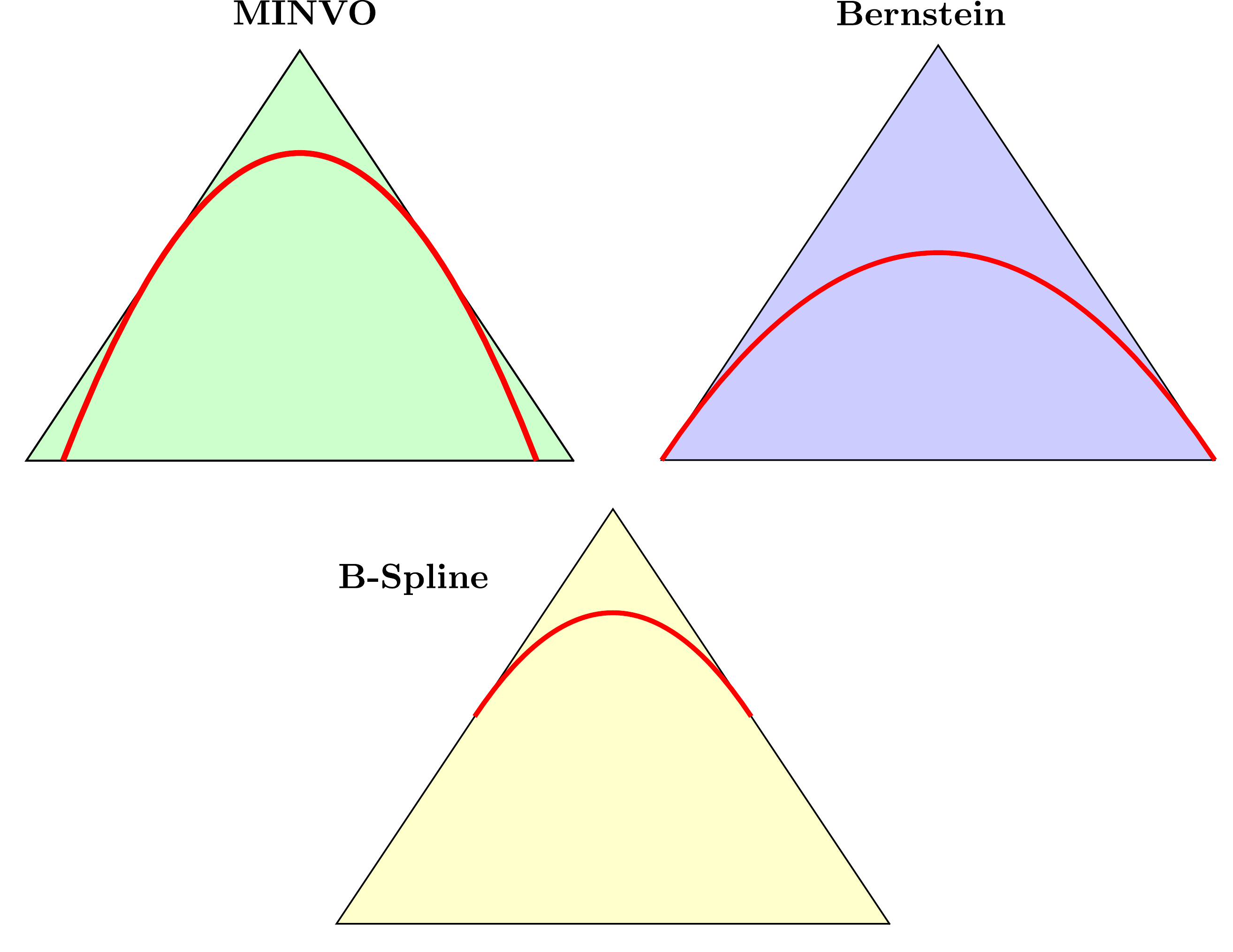}		}\caption{Comparison between the MINVO, Bernstein, and B-Spline bases for $n=2$. The MINVO basis obtains \add{numerically} globally optimal results for $n=1,2,3$.
			\label{fig:Comparison2d}}
		\centering
		\subfloat[For \textbf{any} given $3^{\text{rd}}$\add{-degree} polynomial curve, the MINVO basis finds an enclosing $3$-simplex that is 2.36 and 254.9 times smaller than the one found by the Bernstein and B-Spline bases respectively. \label{fig:Comparison3d_poly_given}]{\includegraphics[width=0.45\textwidth]{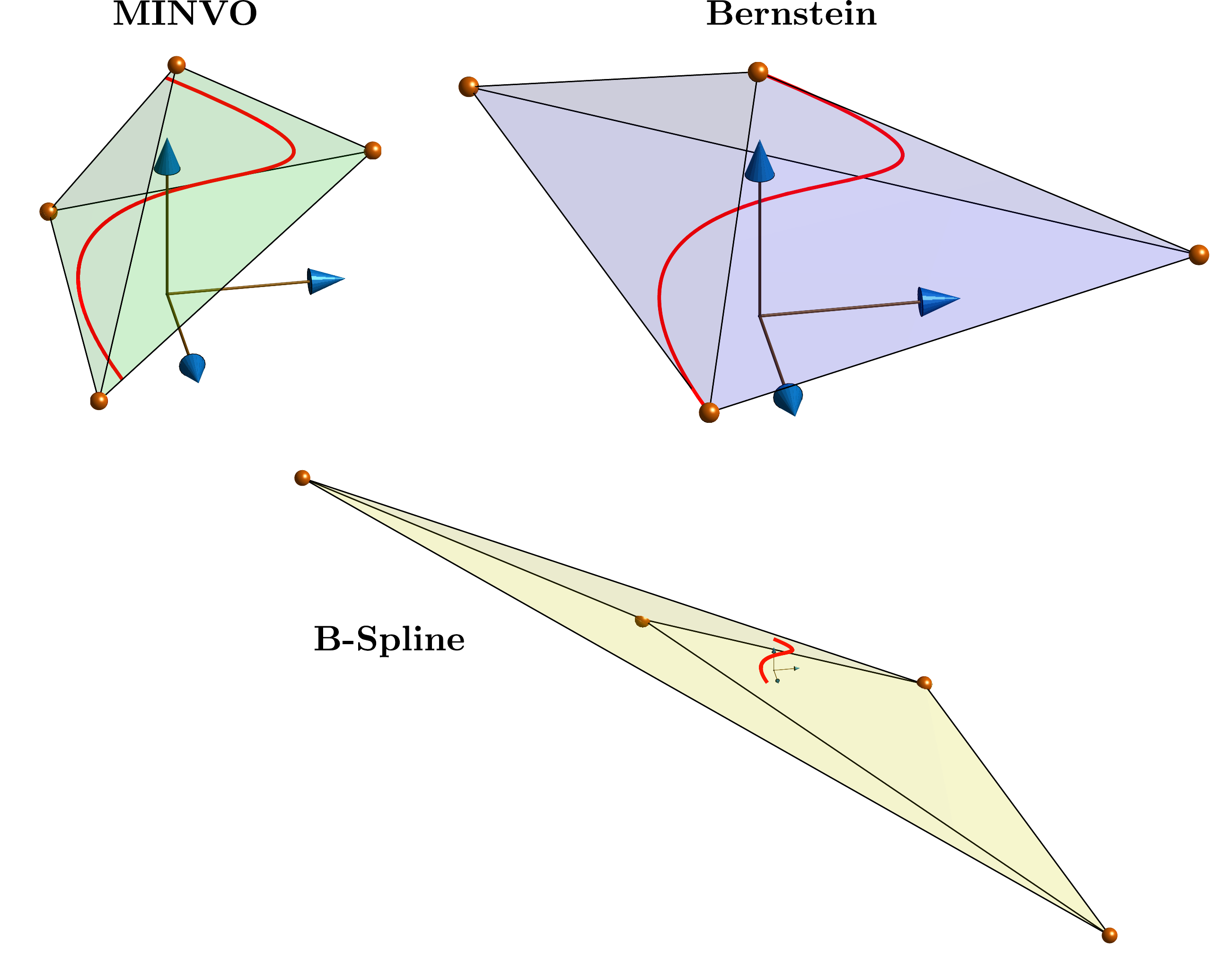}}
		\qquad
		\subfloat[For \textbf{any} given $3$-simplex, the MINVO basis finds a  $3^{\text{rd}}$\add{-degree} polynomial curve
		inscribed in the simplex, and whose convex hull is 2.36 and 254.9
		times larger than the one found by the Bernstein and B-Spline bases
		respectively. \label{fig:Comparison3d_simplex_given}]{
			\includegraphics[width=0.45\textwidth]{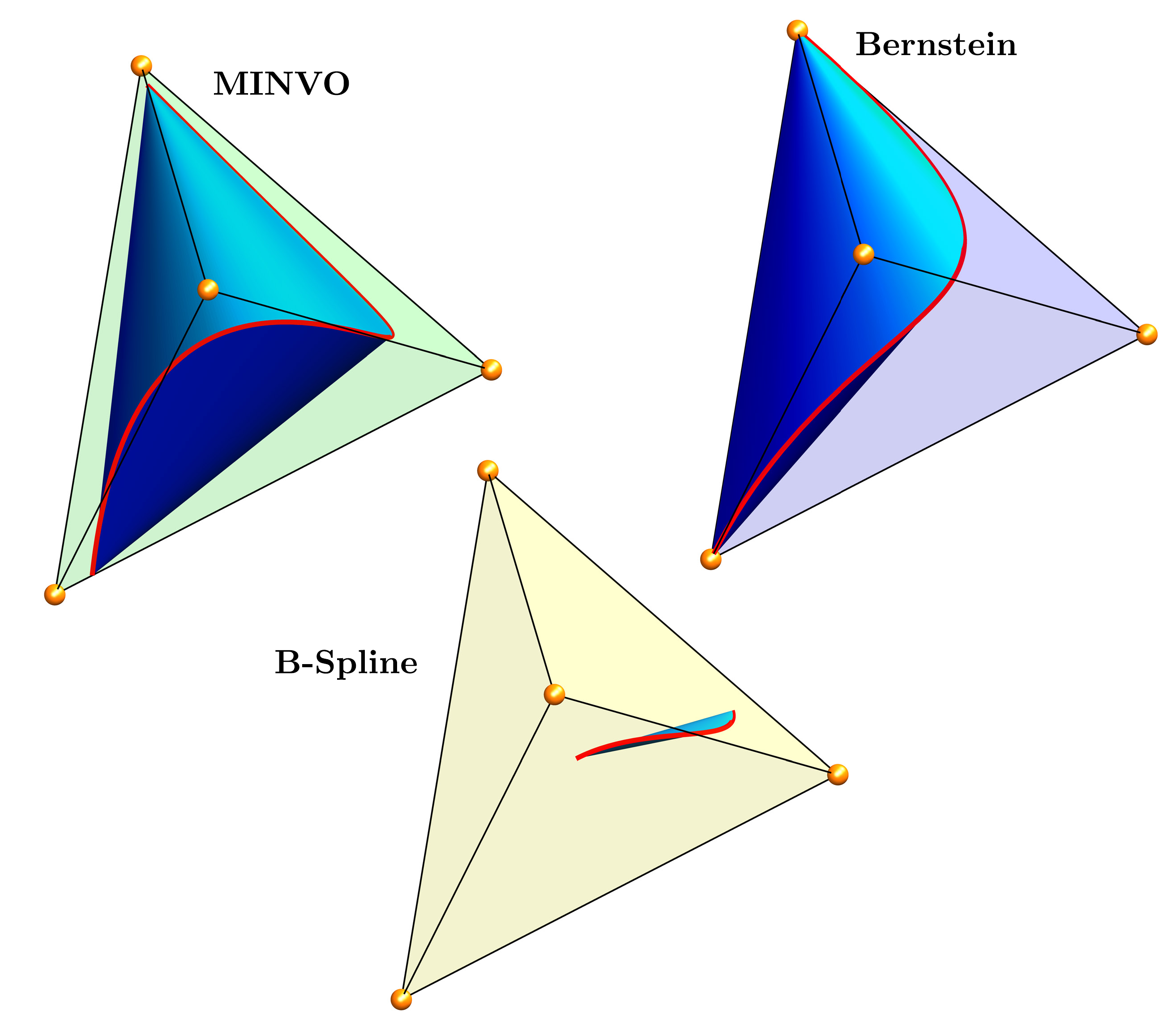}}
		\caption{Comparison between the MINVO, Bernstein, and B-Spline bases for $n=3$.
			The MINVO basis obtains \add{numerically} globally optimal results for $n=1,2,3$. }
		\label{fig:Comparison3d}
	\end{figure*}
	
		One natural question to ask is whether the basis found constitutes
	a global minimizer for either Problem 3 or Problem 4. To answer this,
	first note that both Problem 3 and Problem 4 are polynomial optimization
	problems. Therefore, we can make use of Lasserre's moment method \cite{lasserre2001},
	and increase the order of the moment relaxation to find tighter lower
	bounds of the original nonconvex polynomial optimization problem. Using this technique, we were able to obtain, \add{for $n=1,2,3$ and for Problem 4}, the same objective value \add{as the NLO solutions found before, proving therefore numerical global optimality for these cases}. For Problem 3, the moment relaxation
	technique becomes intractable due to the high number of variables. Hence, to prove \add{numerical} global optimality in Problem 3 we  instead use the branch-and-bound algorithm, which proves global optimality by reducing to zero the gap
	between the upper bounds found by a nonconvex solver and the lower
	bounds found using convex relaxations~\cite{bmibnb2020}. This technique proved to be tractable for cases $n=1,2,3$ in Problem 3, and zero optimality gap was obtained.
	
		\begin{figure*}
		\centering
		\subfloat[Simplexes found by the MINVO basis for four different \addrevision{given} $3^{\text{rd}}$\add{-degree}
		polynomial curves (Problem 1). \label{fig:ManyComparisons3d_poly_given}]{\centering{}\includegraphics[width=0.45\textwidth]{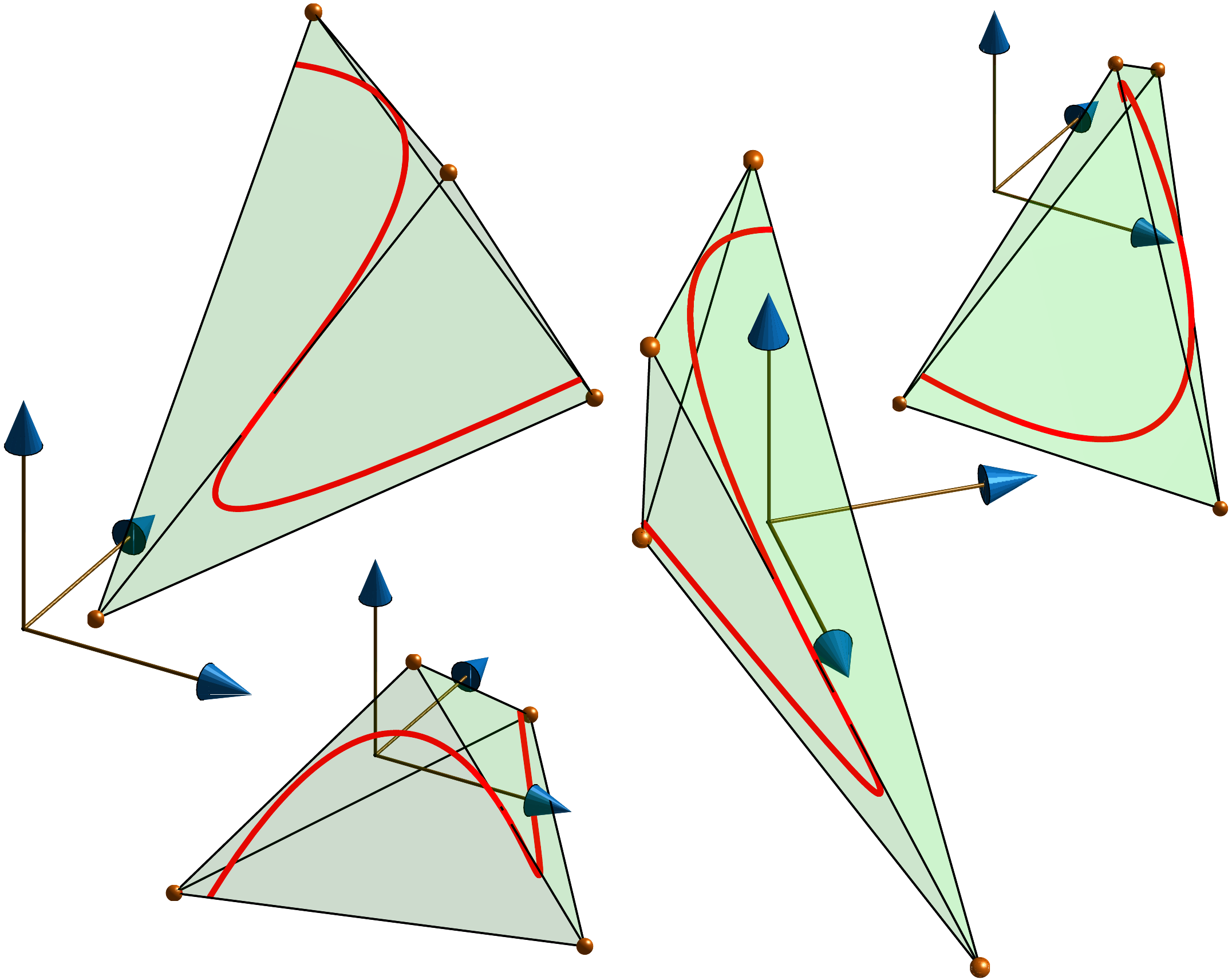}}
		\qquad
		\subfloat[Polynomial curves (and their convex hulls in blue) obtained using the MINVO basis for four different \addrevision{given}	$3$-simplexes (Problem 2).\label{fig:ManyComparisons3d_simplex_given-1}]{
			\includegraphics[width=0.45\textwidth]{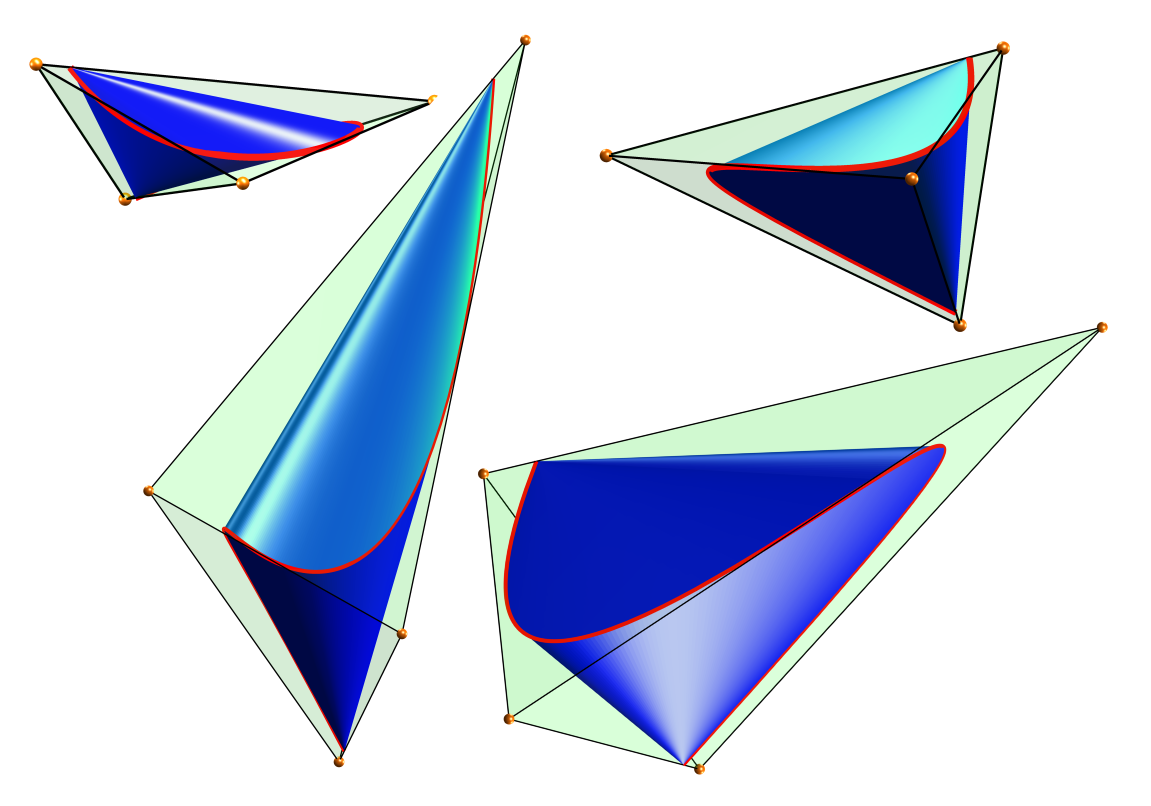}}
		\caption{MINVO results for $n=3$, where \add{numerical} global optimality is guaranteed\label{fig:ManyComparisons3d}.}
	\end{figure*}

		All these results lead us to the following conclusions, which are
	also summarized in Table \ref{tab:table_matrices}:
	\begin{itemize}
		\item The matrices $\boldsymbol{A}_{\text{MV}}$ found for $n=1,2,3$ are \addcad{(numerical) global optima}
		of both Problem 3 and Problem 4.
		\item The matrix $\boldsymbol{A}_{\text{MV}}$ found for $n=4$ is at least a \addcad{(numerical) local optimum} of both Problem 3 and Problem 4.
		\item The matrices $\boldsymbol{A}_{\text{MV}}$ found for $n=5,6,7$ are at least \addcad{(numerical) local optima} for Problem 4, and are at least feasible solutions for
		Problem~3.
	\end{itemize}

	The geometric interpretation of Problem 3 (for $n=2$) is shown in \add{Figure}~\ref{fig:cone_A2}. The rows of $\boldsymbol{A}$ are vectors that lie in the cone of the polynomials that are nonnegative in $t\in[-1,1]$ (and whose frontier is shown in orange in the figure). As Problem~3 is maximizing the volume of the parallelepiped spanned by these vectors, the optimal minimizer is obtained in the frontier of the cone, while guaranteeing that the sum of these vectors is $\arraycolsep=1.4pt\left[\begin{array}{ccc} 0&0 & 1\end{array}\right]^{T}$.

	In \add{Figure}~\ref{fig:centroid_facets} we check that the centroids of \add{each of the facets} of the simplex belongs to  $\text{conv}\left(P\right)$, which is a necessary condition for that simplex to be minimal \cite{klee1986facet}. Note also that $\text{conv}\left(P\right)$ is tangent to the simplex along four lines (in blue in the figure), and that the contact points of the curve with the simplex happen at the roots of the MINVO basis functions.

	When the polynomial curve is given (i.e., Problem 1), the ratio between
	the volume of the simplex $S_{\alpha}$ obtained by a basis $\alpha$
	and the volume of the simplex $S_{\beta}$ obtained by a basis $\beta$
	($\alpha,\beta\in\{\text{MV},\text{Be},\text{BS}\}$) is given by
	\[
	\frac{\text{vol}\left(S_{\alpha}\right)}{\text{vol}(S_{\beta})}=\frac{\text{abs}(|\boldsymbol{A}_{\beta}|)}{\text{abs}\left(|\boldsymbol{A}_{\alpha}|\right)} \addrevision{\, .}
	\]
	Similarly, when the simplex is given (i.e., Problem 2), the ratio between
	the volume of the convex hull of the polynomial curve $P_{\alpha}$
	found by a basis $\alpha$ and the volume of the convex hull of the
	polynomial curve $P_{\beta}$ found by a basis $\beta$
	($\alpha,\beta\in\{\text{MV},\text{Be},\text{BS}\}$) is given by
	\[
	\frac{\text{vol}\left(\text{conv}\left(P_{\alpha}\right)\right)}{\text{vol}(\text{conv}(P_{\beta}))}=\frac{\text{abs}\left(|\boldsymbol{A}_{\alpha}|\right)}{\text{abs}(|\boldsymbol{A}_{\beta}|)} \addrevision{\, .}
	\]
	These ratios are shown in Table \ref{tab:table_matrices}, and they mean the following for Problem 1 (Problem~2 respectively):
	\begin{itemize}
		\item For $n=3$, the MINVO basis finds a simplex that has a volume (a polynomial
		curve whose convex hull has a volume) $\approx2.36$ and $\approx254.9$
		times smaller (larger) than the one the Bernstein and B-Spline bases
		find respectively. 
		\item For $n=7$, the MINVO basis finds a simplex that has a volume (a polynomial
		curve whose convex hull has a volume) $\approx902.7$ and $\approx2.997\cdot10^{21}$
		times smaller (larger) than the one the Bernstein and B-Spline bases
		find respectively. 
	\end{itemize}
	An analogous reasoning applies to the volume ratios of other $n$. These comparisons are shown in \add{Figure}~\ref{fig:Comparison2d} (for $n=2$), and in \add{Figure}~\ref{fig:Comparison3d}
	(for $n=3$). More examples of the MINVO bases applied to different polynomial curves and simplexes are shown in \add{Figure}~\ref{fig:ManyComparisons3d}. 

	\begin{figure*}
		\begin{centering}
			\includegraphics[width=1.0\textwidth]{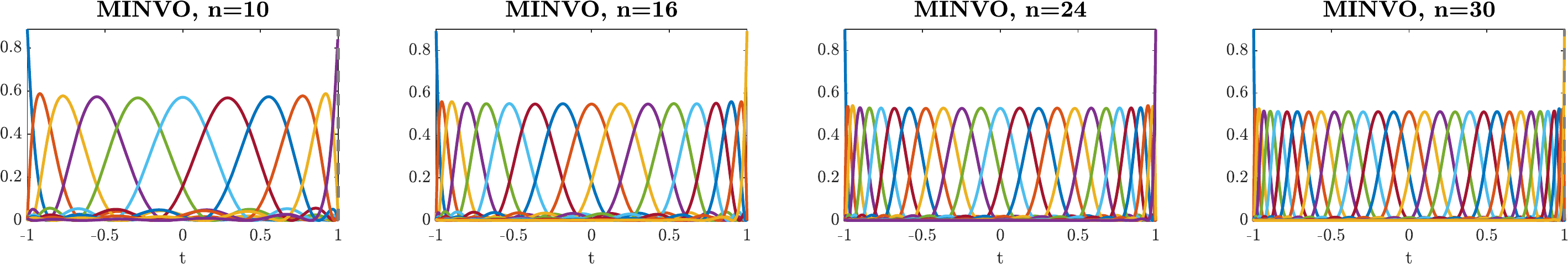}
			\par\end{centering}
		\vskip-1ex
		\caption{\add{MINVO basis functions obtained for $n=10,16,24,30$ using the model proposed in \addcad{Section~}\ref{sec:resultsnGT7}. }\label{fig:plots_basis_high_degree}}
		\vskip 0ex
	\end{figure*}

	\begin{table*}[hbt!]
	\setlength\extrarowheight{0.6pt}
	\caption{\add{Comparison of the results obtained from the optimization (available only for $n=1,\hdots,7$) with the results obtained using the model proposed in \addcad{Section~}\ref{sec:resultsnGT7} (available \addmorerevision{for all} $n\in \mathbb{N}$). The results obtained using the proposed model are also compared with the Bernstein and B-Spline bases.\label{tab:comparison_fitted_model_opt}}  }
	\vskip-2ex
	\noindent\resizebox{\textwidth}{!}{%
		\begin{centering}
			\begin{tabular}{>{\centering}p{0.04\textwidth}>{\centering}p{0.16\textwidth}>{\centering}m{0.03\textwidth}>{\centering}m{0.06\textwidth}>{\centering}m{0.06\textwidth}>{\centering}m{0.06\textwidth}>{\centering}m{0.05\textwidth}>{\centering}m{0.05\textwidth}>{\centering}m{0.05\textwidth}>{\centering}m{0.05\textwidth}>{\centering}m{0.05\textwidth}>{\centering}m{0.05\textwidth}>{\centering}m{0.05\textwidth}}
				\cmidrule{3-13} \cmidrule{4-13} \cmidrule{5-13} \cmidrule{6-13} \cmidrule{7-13} \cmidrule{8-13} \cmidrule{9-13} \cmidrule{10-13} \cmidrule{11-13} \cmidrule{12-13} \cmidrule{13-13} 
				&  & \multicolumn{11}{c}{\textbf{Degree n}}\tabularnewline
				\cmidrule{3-13} \cmidrule{4-13} \cmidrule{5-13} \cmidrule{6-13} \cmidrule{7-13} \cmidrule{8-13} \cmidrule{9-13} \cmidrule{10-13} \cmidrule{11-13} \cmidrule{12-13} \cmidrule{13-13} 
				&  & \textbf{1} & \textbf{2} & \textbf{3} & \textbf{4} & \textbf{5} & \textbf{6} & \textbf{7} & \textbf{8} & \textbf{10} & \textbf{14} & \textbf{18}\tabularnewline
				\midrule
				\midrule 
				\textbf{Opt.} & $\text{abs}\left(\left|\boldsymbol{A}_{\text{MV}}\right|\right)$ & 0.5 & 0.325 & 0.332 & 0.568 &  1.7 &  9.1 & 89.0 & - & - & - & -\tabularnewline
				\midrule 
				\multirow{3}{0.04\textwidth}{\rotatebox[origin=c]{90}{\textbf{Proposed\hspace{0.2cm}}}\rotatebox[origin=c]{90}{\textbf{model\hspace{0.3cm}}}} & $\text{abs}\left(\left|\boldsymbol{A}_{\text{MV}}\right|\right)$ & 0.5 & 0.325 & 0.332 & 0.567 & 1.69 & 9.08 & 88.2 & {\footnotesize{}1590.0} & {\footnotesize{}3.3e6} & {\footnotesize{}2.7e16} & {\footnotesize{}5.7e30}\tabularnewline
				\cmidrule{2-13} \cmidrule{3-13} \cmidrule{4-13} \cmidrule{5-13} \cmidrule{6-13} \cmidrule{7-13} \cmidrule{8-13} \cmidrule{9-13} \cmidrule{10-13} \cmidrule{11-13} \cmidrule{12-13} \cmidrule{13-13} 
				& {\scriptsize{}$\log_{10}\text{abs}\left(\frac{\left|\boldsymbol{A}_{\text{MV}}\right|}{\left|\boldsymbol{A}_{\text{Be}}\right|}\right)$} &   0 & 0.114 & 0.373 & 0.782 & 1.35 & 2.07 & 2.95 &    4.0 & 6.57 &     13.6 &     23.2\tabularnewline
				\cmidrule{2-13} \cmidrule{3-13} \cmidrule{4-13} \cmidrule{5-13} \cmidrule{6-13} \cmidrule{7-13} \cmidrule{8-13} \cmidrule{9-13} \cmidrule{10-13} \cmidrule{11-13} \cmidrule{12-13} \cmidrule{13-13} 
				& {\scriptsize{}$\log_{10}\text{abs}\left(\frac{\left|\boldsymbol{A}_{\text{MV}}\right|}{\left|\boldsymbol{A}_{\text{BS}}\right|}\right)$} &   0 & 0.716 &  2.41 &  5.22 & 9.28 & 14.7 & 21.5 &   29.7 &   50.9 & 113.0 & 203.0\tabularnewline
				\bottomrule
			\end{tabular}
			\par\end{centering}
	}
	\vskip-2ex
\end{table*}
	
			\definecolor{diag_color}{RGB}{255,149,149}
	\newcommand{\one}{\cellcolor{MINVO_color}}
	\definecolor{n_smaller_than_k_color}{RGB}{191,255,242}
	\newcommand{\two}{\cellcolor{white}}
	\definecolor{n_greater_than_k_color}{RGB}{255,149,10}
	\newcommand{\three}{\cellcolor{white}}
	\begin{table}[h!]
		\captionof{table}{All the possible cases of polynomial curves of degree $n$, dimension $k$, and embedded in a subspace $\mathcal{M}$ of dimension $m$. Note that $m\le\text{min}(k,n)$ always holds.\label{tab:diff_k_and_n}}
		\vskip-2ex
		\begin{centering}
			\begin{tabular}{|c|>{\centering}m{0.08\columnwidth}|>{\centering}p{0.112\columnwidth}|>{\centering}p{0.112\columnwidth}|>{\centering}p{0.112\columnwidth}|>{\centering}p{0.112\columnwidth}|>{\centering}p{0.112\columnwidth}|}
				\cline{3-7} \cline{4-7} \cline{5-7} \cline{6-7} \cline{7-7} 
				\multicolumn{1}{c}{} &  & \multicolumn{5}{c|}{\textbf{\footnotesize{}Degree}{\footnotesize{} $n$}}\tabularnewline
				\cline{3-7} \cline{4-7} \cline{5-7} \cline{6-7} \cline{7-7} 
				\multicolumn{1}{c}{} &  & 1 & 2 & 3 & 4 & $\ge5$\tabularnewline
				\hline 
				\multirow{5}{*}{\begin{turn}{90}
						\textbf{\footnotesize{}Dimension}{\footnotesize{} $k$}
				\end{turn}} & 1 & {\small{}\one{}NGO} & {\small{}\three{}F} & {\small{}\three{}F} & {\small{}\three{}F} & {\small{}\three{}F}\tabularnewline
				\cline{2-7} \cline{3-7} \cline{4-7} \cline{5-7} \cline{6-7} \cline{7-7} 
				& 2 & {\small{}\two{}NGO} & {\small{}\one{}NGO} & {\small{}\three{}F} & {\small{}\three{}F} & {\small{}\three{}F}\tabularnewline
				\cline{2-7} \cline{3-7} \cline{4-7} \cline{5-7} \cline{6-7} \cline{7-7} 
				& 3 & {\small{}\two{}NGO} & {\small{}\two{}NGO} & {\small{}\one{}NGO} & {\small{}\three{}F} & {\small{}\three{}F}\tabularnewline
				\cline{2-7} \cline{3-7} \cline{4-7} \cline{5-7} \cline{6-7} \cline{7-7} 
				& 4 & {\small{}\two{}NGO} & {\small{}\two{}NGO} & {\small{}\two{}NGO} & {\small{}\one{}NLO} & {\small{}\three{}F}\tabularnewline
				\cline{2-7} \cline{3-7} \cline{4-7} \cline{5-7} \cline{6-7} \cline{7-7} 
				& $\ge5$ & {\small{}\two{}NGO} & {\small{}\two{}NGO} & {\small{}\two{}NGO} & {\small{}\two{}NLO} & {\small{}\one{}F}\tabularnewline
				\hline 
				\multicolumn{7}{|>{\raggedright}m{0.955\columnwidth}|}{\textbf{\footnotesize{}In all the cases}{\footnotesize{} (and for
						any $m$): there are $n+1$ control points, and conv(control points)
						is a polyhedron $\subset\mathcal{M}\subseteq\mathbb{R}^{k}$ that
						encloses the curve and that has at most $n+1$ \addrevision{vertices}. \\}\textbf{\footnotesize{}In
						and below the diagonal}{\footnotesize{}: When $n=m$, the polyhedron
						is \addrevision{an $n$-simplex embedded in $\mathbb{R}^k$} that is at least \add{numerically globally optimal (NGO)}}\textbf{\footnotesize{},
					}{\footnotesize{}\add{numerically locally optimal (NLO), or feasible (F)}.
						\vspace{-0.4cm} \\}}\tabularnewline
				\hline 
			\end{tabular}
			\par\end{centering}
		\vskip-0ex
	\end{table}
	
		\begin{figure}
		\begin{centering}
			\includegraphics[width=1.0\columnwidth]{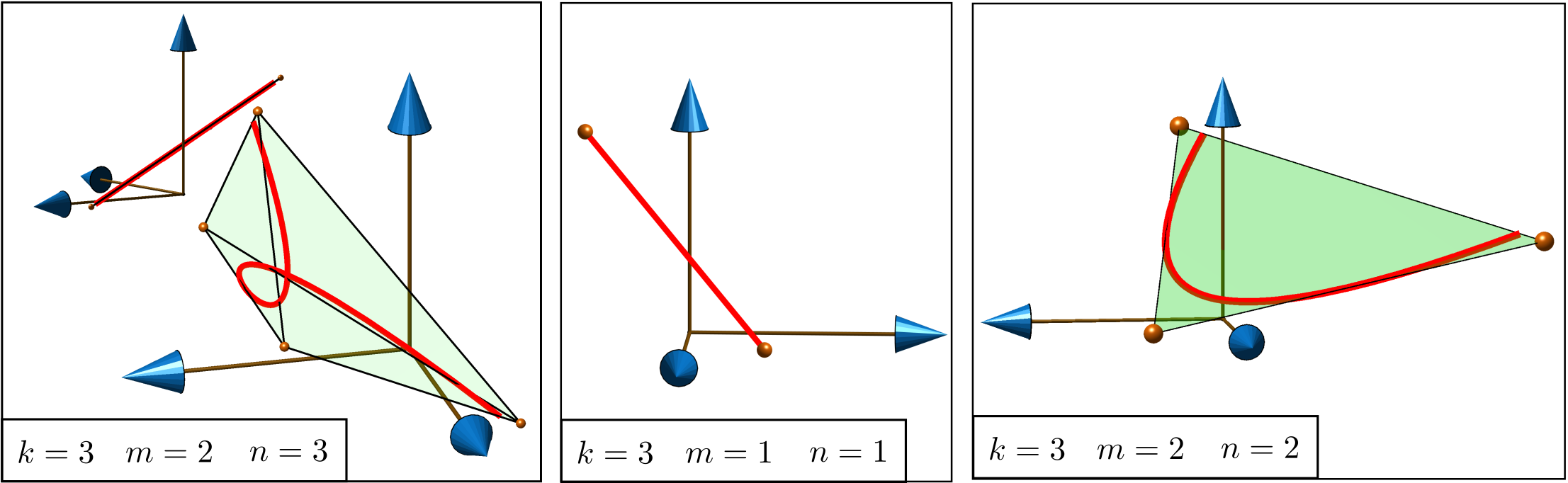}
			\par\end{centering}
		\caption{The MINVO basis applied for different curves with $k=3$ and different values of $m$ and $n$: A cubic curve embedded in a two-dimensional subspace \add{(left)}, a segment embedded in a one-dimensional subspace \add{(middle)} and a quadratic curve embedded in a two-dimensional subspace \add{(right)}\label{fig:embedded_curves}.}
		\vskip-3ex
	\end{figure}	
	
			\begin{figure*}[hbt!]
		\begin{centering}
			\includegraphics[width=1.0\textwidth]{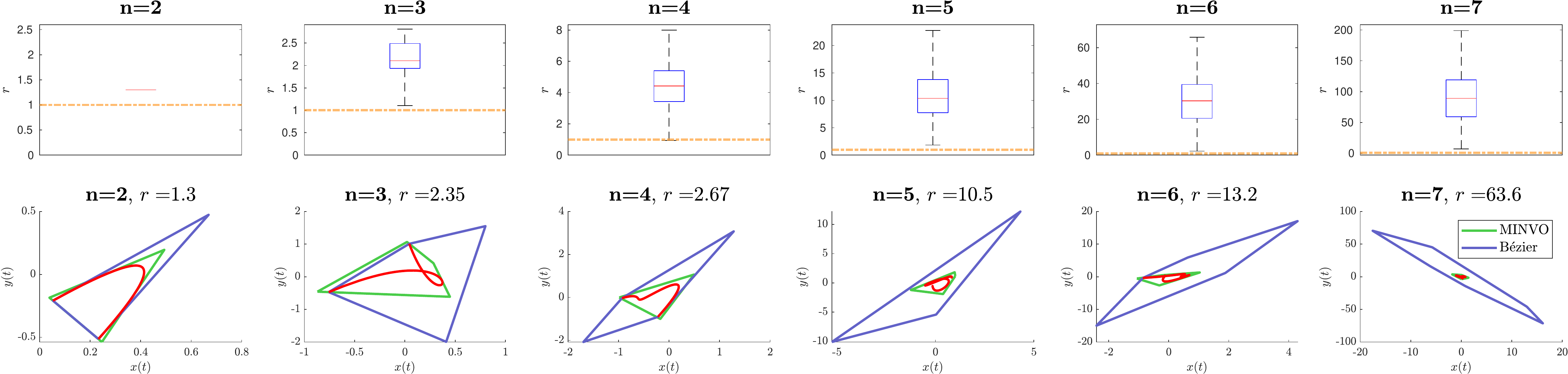}
			\par\end{centering}
		\caption{Comparison of the convex hull of the MINVO and B\'{e}zier control points for $k=m=2$ and different $n$. Here $r$ denotes the ratio of the areas \addrevision{$\frac{\text{Area}_\text{Be}}{\text{Area}_\text{MV}}$}. The boxplots (top) have been obtained from $10^4$ \add{polynomials} passing through $n+1$ random points in the square $[-1,1]^2$. The yellow dashed line highlights the value \addrevision{$r=\frac{\text{Area}_\text{Be}}{\text{Area}_\text{MV}}=1$}. Some of these random curves and the associated convex hulls are shown at the bottom.
			\label{fig:diff_degree2D}}
		\vskip-0ex
		\vskip 2ex
		\begin{centering}
			\includegraphics[width=1.0\textwidth]{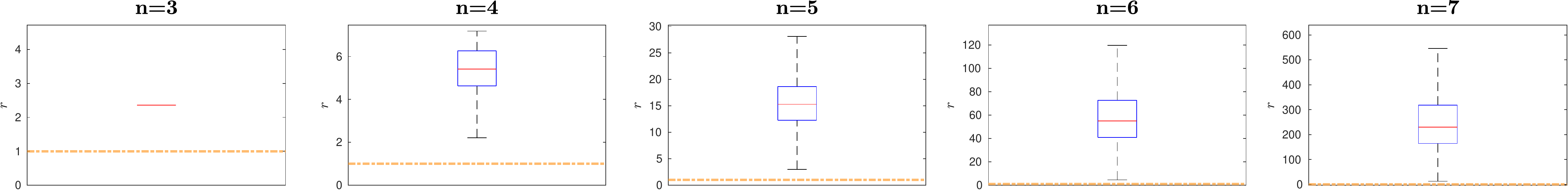}
			\includegraphics[width=1.0\textwidth]{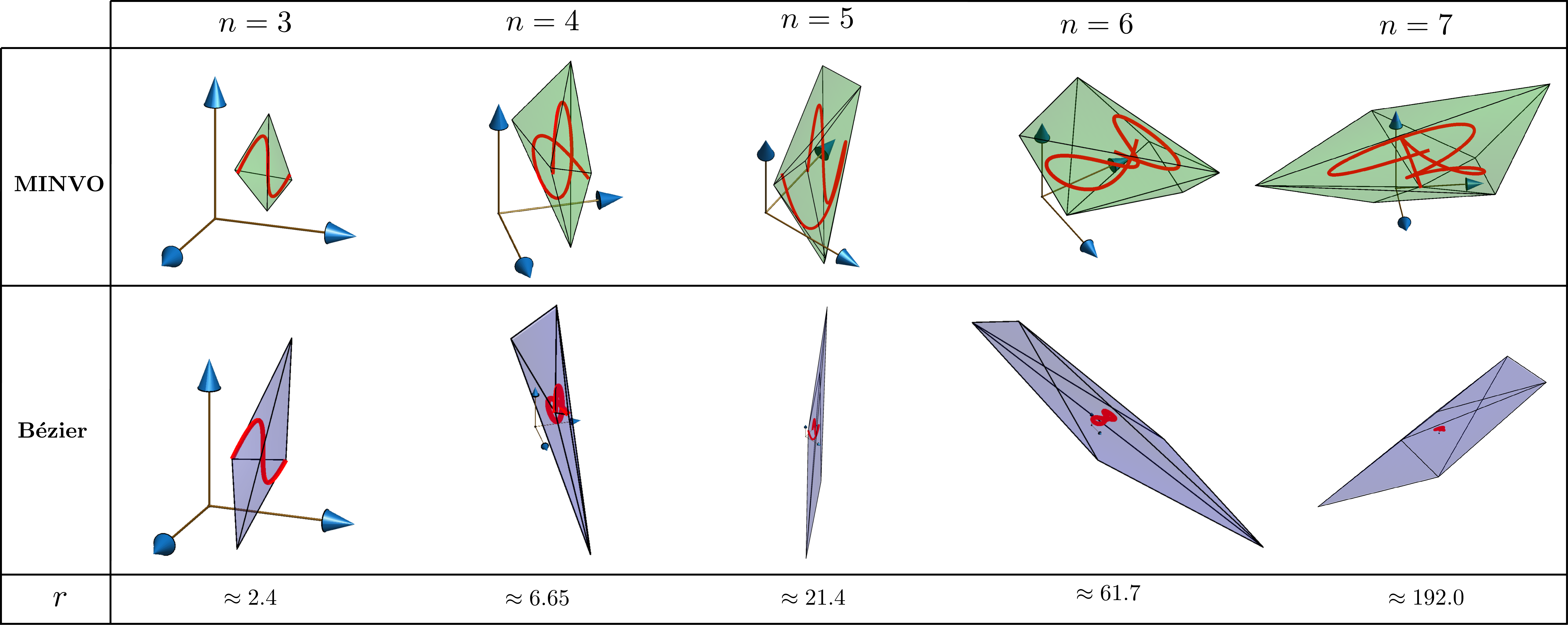}
			\par\end{centering}
		\caption{Comparison of the convex hull of the MINVO and B\'{e}zier control points for $k=m=3$ and different $n$. Here $r$ denotes the ratio of the volumes \addrevision{$\frac{\text{Vol}_\text{Be}}{\text{Vol}_\text{MV}}$}. The boxplots (top) have been obtained from $10^4$ polynomial \additionalrevision{curves} passing through $n+1$ random points in the cube $[-1,1]^3$ were used. The yellow dashed line highlights the value \addrevision{$r=\frac{\text{Vol}_\text{Be}}{\text{Vol}_\text{MV}}=1$}. Some of these random curves and the associated convex hulls are shown at the bottom.
			\label{fig:diff_degree3D}}
		\vskip-0ex
	\end{figure*}	
	
	\subsection{Results for $n>7$}\label{sec:resultsnGT7}
	\add{In	\addcad{Section~}\ref{sec:Resultsn17}, we obtained the results for $n=1,\hdots,7$ by using the optimization problems. However, solving these problems becomes intractable when $n>7$. To address this problem, we present a model that finds high-quality feasible solutions by  extrapolating for  $n>7$ the pattern found for the roots of the MINVO basis (see the cases  $n=1,\hdots,7$ in %
		\addrevision{Figure}~\ref{fig:roots_distribution_all}). 
		Specifically, and noting that the double roots for a degree $n$ tend to be distributed in $n$ clusters,  
		we found that  the MINVO double roots in the interval $(-1,1)$ for the degree $n$ can be approximated by 
		\begin{equation} \label{eq:extrapolation}
		\sin\left(\frac{c_{0}\left(k-\frac{s_{j,n} -1}{2}\right)+c_{1}\left(j-\frac{n-1}{2}\right)}{n+c_{2}}\right) \addrevision{\, ,}
		\end{equation}
		where $s_{j,n}:=\left\lfloor \frac{n+\text{odd}\left(j,n\right)}{2}\right\rfloor$ models the number of roots per cluster \addcad{$j\in \{0,\hdots,(n-1)\}$}, and \addcad{$k\in \{0,\hdots,(s_{j,n} -1)\}$} is the index of the root inside a specific cluster\footnote{\addrevision{The intuition behind the design of Eq.~\ref{eq:extrapolation} is as follows: The $\sin\left(  \frac{\cdot}{n+\cdot}\right)$ forces every root to be in $[-1,1]$, and makes the centers of the clusters more densely distributed near the extremes $t\in\{-1,1\}$, and less around $t=0$. The numerator inside the $\sin(\cdot)$ is a weighted sum of the index of the root inside the cluster (centered around $\frac{s_{j,n}-1}{2}$), and the index of the cluster (centered around $\frac{n-1}{2}$). Finally, note that, by construction, this formula enforces symmetry with respect to $t=0$.}}. 
		
		Here, $c_0\approx0.2735$, $c_1\approx3.0385$, and $c_2\approx0.4779$ were found by optimizing the associated nonlinear least-squares problem. This proposed model, with only three parameters, is able to obtain a least-square residual of $5.02\cdot10^{-3}$ with respect to the MINVO roots \addrevision{lying in} $(-1,1)$ found for $n=2,\hdots,7$. The distribution of roots generated by this proposed model ($n\ge8$) is shown in Figure~\ref{fig:roots_distribution_all}. Each root can then be assigned to a polynomial $i$ of the basis by simply following the same assignment pattern found for $n=1,...,7$. 
		Then, and by solving a linear system, the polynomials can be scaled to enforce $\sum_{i=0}^{n}\lambda_{i}(t)=1\;\forall t$
		(or equivalently, $\boldsymbol{A}^T\boldsymbol{1}=\boldsymbol{e}$). Note that this proposed model, although not guaranteed to be optimal, is guaranteed to be feasible  by construction. Some examples of the MINVO basis functions are shown in Figure~\ref{fig:plots_basis_high_degree}. The comparison of $\left|\boldsymbol{A}_{\text{MV}}\right|$ between the proposed model and the optimization results of \addcad{Section~}\ref{sec:Resultsn17} is shown in Table~\ref{tab:comparison_fitted_model_opt}. %
		For $n=1,\hdots,7$, the relative error between the objective value obtained using the optimization and the one obtained using the proposed model is always $<9.2\cdot10^{-3}$. The proposed model also produces much smaller simplexes than the Bernstein and B-Spline bases. }

	\begin{figure*}
		\begin{centering}
			\includegraphics[width=0.465\textwidth]{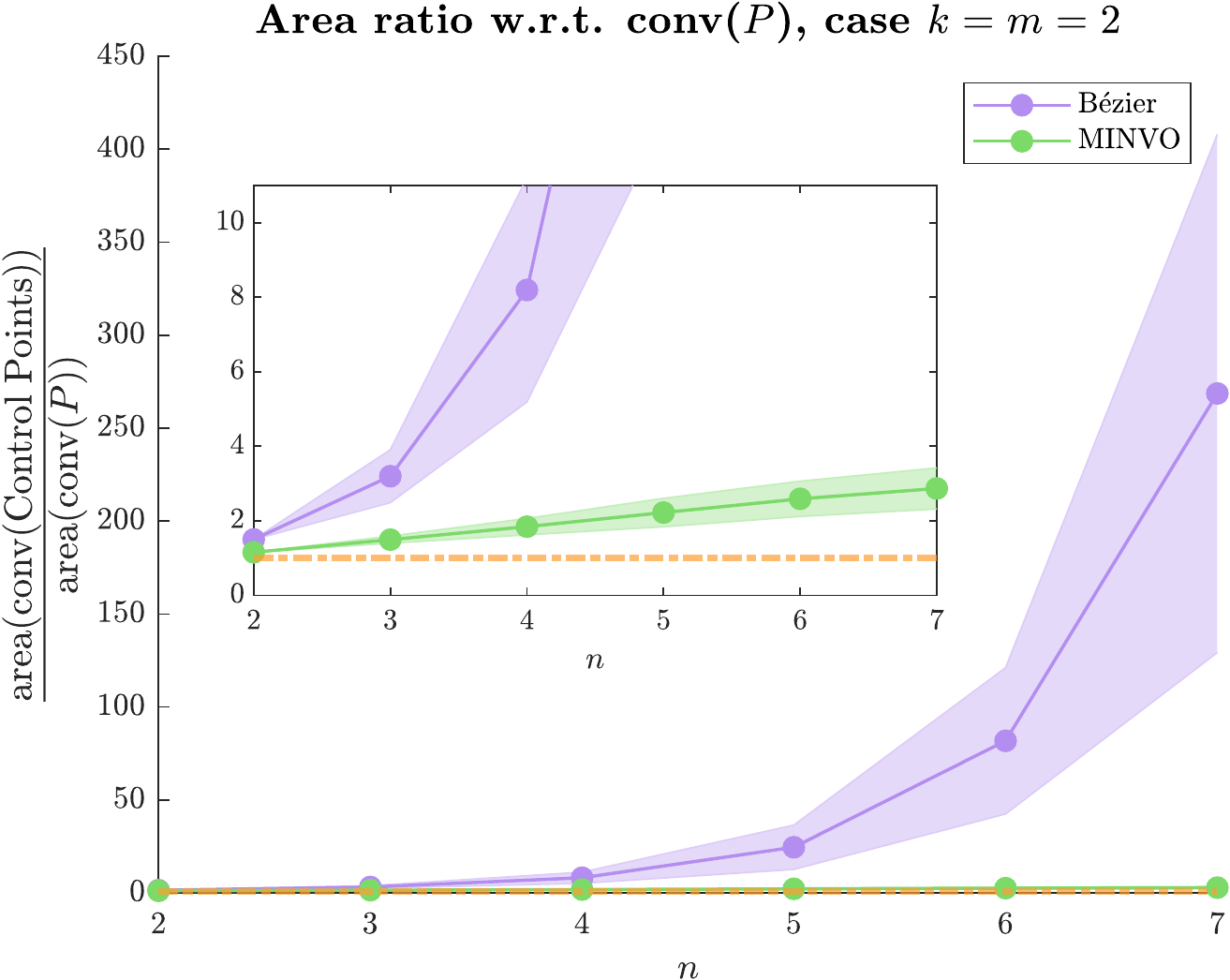}
			\includegraphics[width=0.49\textwidth]{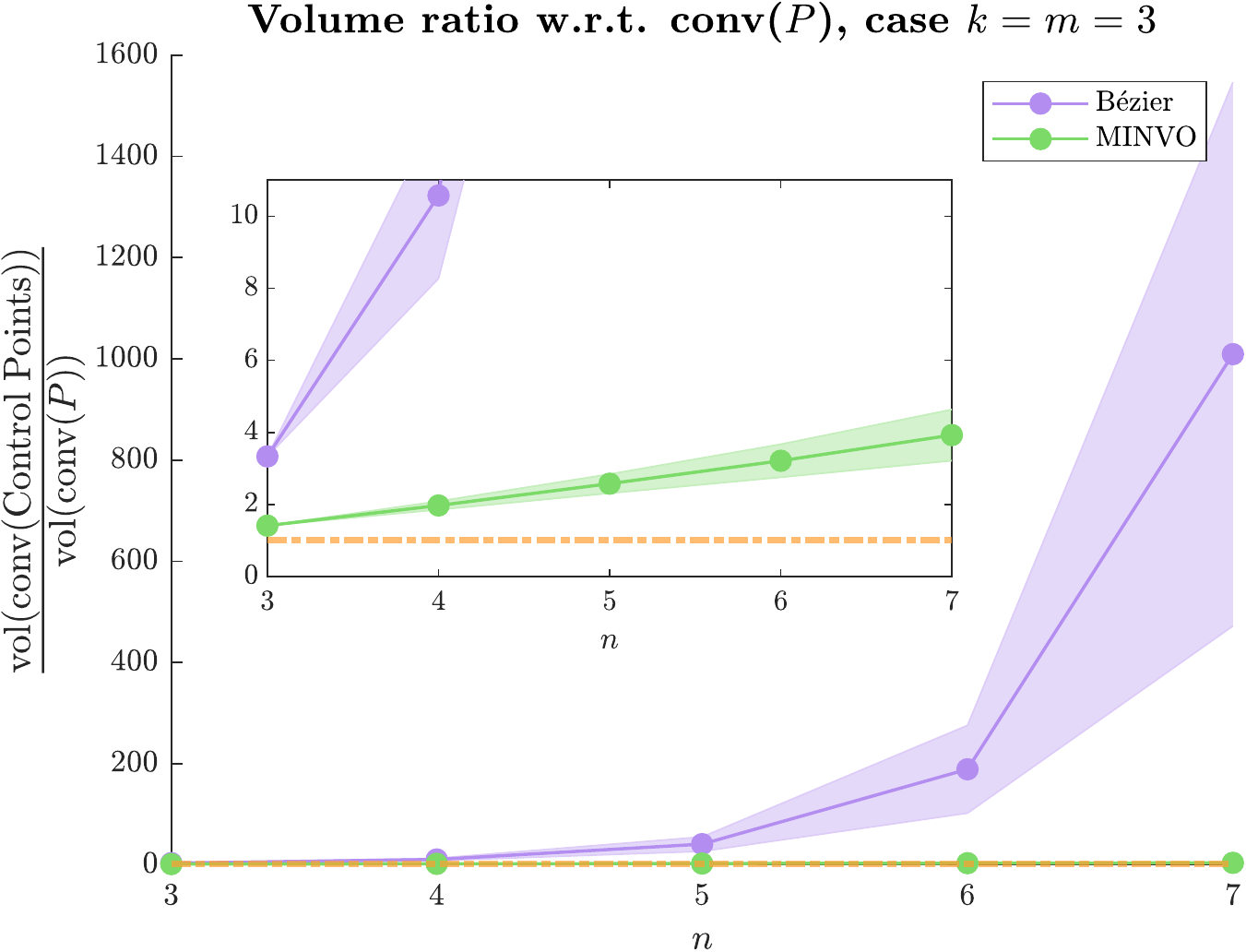}
			\par\end{centering}
		\caption{Comparison of the convex hull of the MINVO and B\'{e}zier control points with respect to $\text{conv}(P)$ for the cases $k=m=2$ (left) and $k=m=3$ (right).  For each $n$, a total of $10^3$ polynomial \additionalrevision{curves} passing through $n+1$ random points in the cube $[-1,1]^k$ were used. The shaded area is the 1$\sigma$ interval, where $\sigma$ is the standard deviation. The yellow dashed line marks a ratio of 1. Note how the growth of the ratio for the MINVO basis is approximately linear with $n$, while for the B\'{e}zier basis it is approximately exponential. \label{fig:area_volume_wrt_convP}}
		\vskip-0ex
	\end{figure*}
	
	\section{MINVO basis applied to other curves} \label{sec:other_curves}
	
	\subsection{Polynomial curves of degree $n$, dimension $k$, and embedded in a subspace $\mathcal{M}$ of dimension $m$}\label{sec:other_curves_nmk}
	
	So far we have studied the case of $n=m=k$ (i.e., a polynomial curve of degree $n$ and dimension $k=n$ and for which $n$ is also the dimension of $\mathcal{M}$, see \add{Section}~\ref{sec:Notation-and-Definitions}). The most general case would be any $k,n$ and $m$, as shown in Table~\ref{tab:diff_k_and_n}. In all these cases, and using the $(n+1)\times(n+1)$ matrix $\boldsymbol{A}_{\text{MV}}$, we can still apply the equation
	$
	\boldsymbol{V}_{k\times(n+1)}=\boldsymbol{P}_{k\times(n+1)}\boldsymbol{A}_{\text{MV}}^{-1}
	$
	to obtain all the $n+1$ MINVO control points \addmorerevision{in} $\mathbb{R}^{k}$ of the given curve (columns of the matrix $\boldsymbol{V}$). The convex hull of the control points is \addrevision{a polyhedron that is} guaranteed to contain the curve because the curve is a convex combination of the control points. Note also that, when $n=m$, all the cases below the diagonal of Table~\ref{tab:diff_k_and_n} have the same optimality properties (NGO/NLO/Feasible) as the diagonal element that has the same $n$.

	\addcad{	For $k=3$, \add{Figure}~\ref{fig:embedded_curves} shows a cubic curve embedded in a two-dimensional subspace ($m=2$, $n=3$), a segment embedded in a one-dimensional subspace ($m=n=1$) and a quadratic curve embedded in a two-dimensional subspace ($m=n=2$).}
	
	For $k=m=2$, the comparison between the area of the convex hull of the MINVO control points (\addrevision{$\text{Area}_{\text{MV}}$}) and the area of the convex hull of the B\'{e}zier control points (\addrevision{$\text{Area}_{\text{Be}}$}) is shown in \add{Figure}~\ref{fig:diff_degree2D}. Note that this ratio is constant for any polynomial curve for the case $n=2$, but depends on the \addrevision{given curve} for the cases $n>2$. To generate the boxplots of \add{Figure}~\ref{fig:diff_degree2D}, we used a total of $10^4$ polynomial \additionalrevision{curves} passing through $n+1$ points randomly sampled from the square $[-1,1]^2$. Although it is not guaranteed that \addrevision{$\text{Area}_{\text{MV}}<\text{Area}_{\text{Be}}$} for any polynomial with $n>2$, the Monte Carlo analysis performed using these random polynomial \additionalrevision{curves} shows that \addrevision{$\text{Area}_{\text{MV}}<\text{Area}_{\text{Be}}$} holds for the great majority of them, with improvements up to $\approx 200$ times for the case $n=7$. 
	
	Similarly, the results for $k=m=3$ are shown in \add{Figure}~\ref{fig:diff_degree3D}, where we used a total of $10^4$ polynomial \additionalrevision{curves} passing through $n+1$ points randomly sampled from the cube $[-1,1]^3$. Again, it is not guaranteed that \addrevision{$\text{Vol}_{\text{MV}}<\text{Vol}_{\text{Be}}$} for any polynomial with $n>3$, but the Monte Carlo results obtained show that this is true in most of these random polynomials. For the case $n=7$, the MINVO basis obtains convex hulls up to $\approx 550$ times smaller than the B\'{e}zier basis. 
	
	\addcad{Qualitatively, and for the comparisons shown above, the MINVO enclosures are much smaller than the B\'{e}zier enclosures when used in ``tangled'' curves. In these curves, the B\'{e}zier control points tend to be spread out and  far from the curve, leading therefore to large and conservative B\'{e}zier enclosures.
	}  
		\begin{figure}
		\begin{centering}
			\includegraphics[width=1.0\columnwidth]{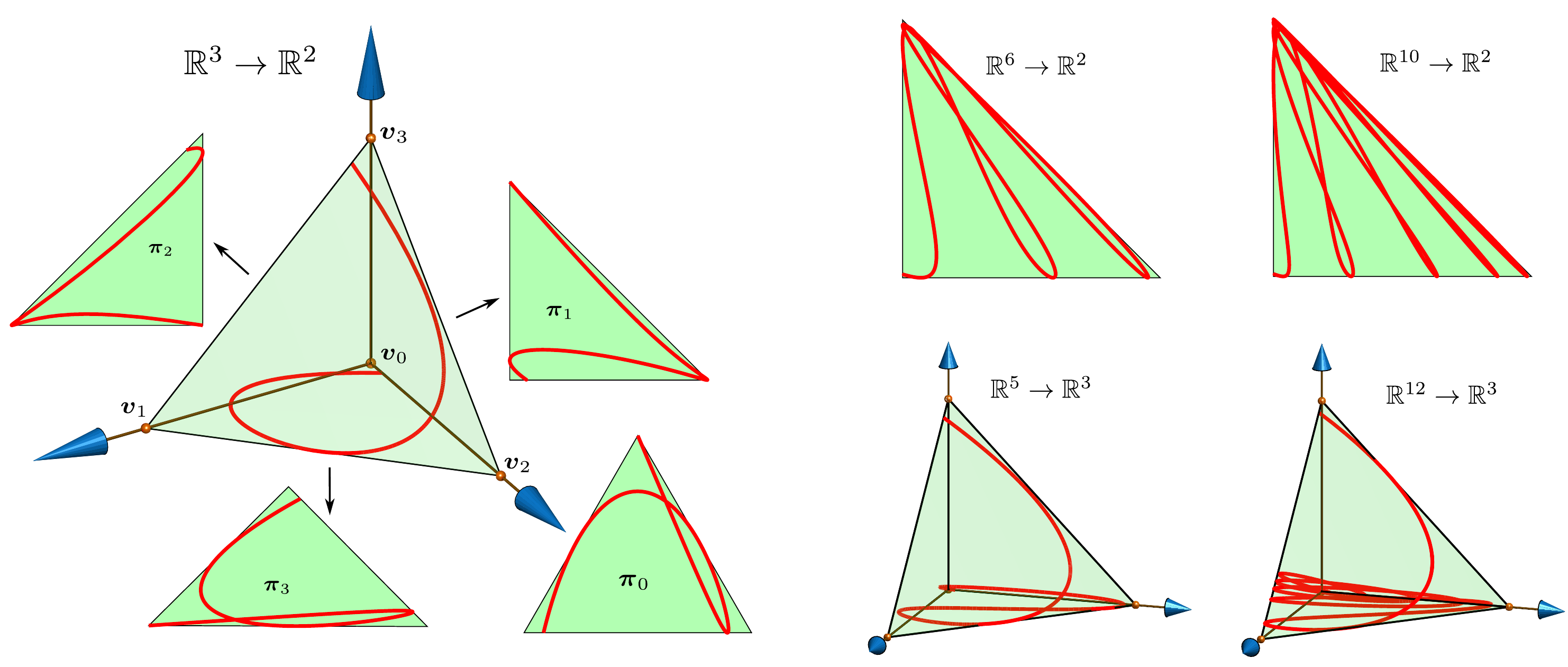}
			\par\end{centering}
		\caption{\add{The MINVO basis also obtains simplexes that tightly enclose some rational curves (curves whose coordinates are the quotient of two polynomials). On the left, the standard simplex is the smallest 3-simplex containing the 3D curve in red. This means that each facet $i$ (contained in the plane $\boldsymbol{\pi}_i$) is also the smallest 2-simplex enclosing the projection of the curve onto that facet using the opposite vertex $\boldsymbol{v}_i$ as viewpoint. On the right, different successive projections to $\mathbb{R}^2$ and $\mathbb{R}^3$ are shown. \label{fig:projections_with_higher_degree}}}
		\vskip-0ex
	\end{figure}
	Finally, we compare in \add{Figure}~\ref{fig:area_volume_wrt_convP} how these polyhedral convex hulls, obtained by either the MINVO or B\'{e}zier control points, approximate  $\text{conv}(P)$, which is the convex hull of the curve $P$. Similar to the previous cases, here we used a total of $10^3$ polynomial \additionalrevision{curves} passing through $n+1$ points randomly sampled from the cube $[-1,1]^k$. The error in the MINVO outer polyhedral approximation is approximately linear as $n$ increases, but it is exponential for the B\'{e}zier basis. For instance, when $n=7$ and $k=3$, the B\'{e}zier control points generate a polyhedral outer approximation that is $\approx 1010$ times \add{larger} than the volume of $\text{conv}(P)$, while the polyhedral outer approximation obtained by the MINVO control points is only  $\approx 3.9$ times \add{larger}.

	\newcommand{\notationSLh}{SL\textsubscript{$h$} denotes the SLEFE computed using $h$ breakpoints per subinterval (i.e., $h-1$ linear segments per subinterval)}

\newcommand{\conclusionRawPoints}{The MINVO enclosures have fewer raw points than all~SL\textsubscript{$h$} ($h\in[2,8]$) for $n\in[2,6]$. Compared to B\'{e}zier, the MINVO enclosure has $s-1$ additional raw points, but achieves areas that are up to \addextrarevision{$30$} times smaller (see Figure~\ref{fig:comparison_slefes_colored_matrices_smooth})}

		\begin{figure*}
	\begin{centering}
		\addmorerevisiongraphics{\includegraphics[width=1.0\textwidth]{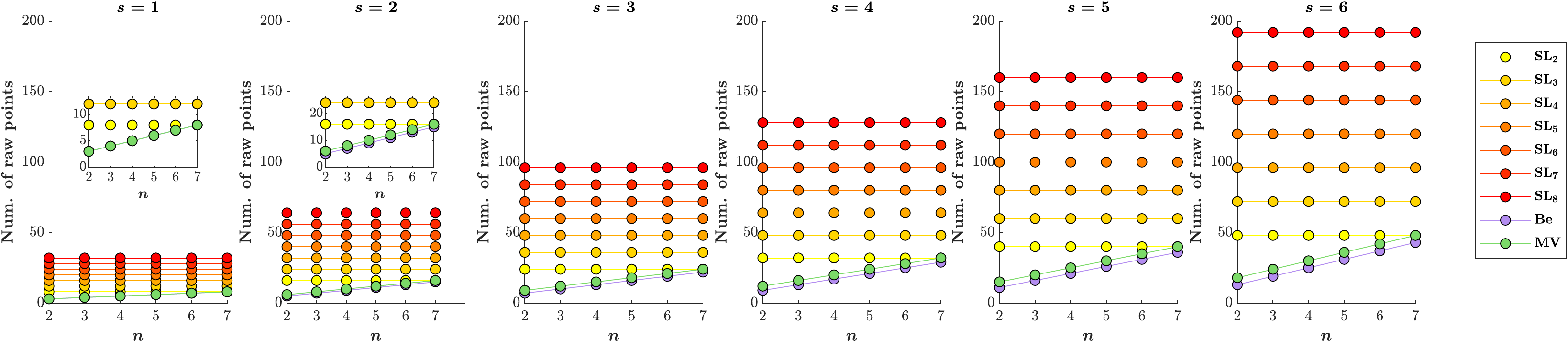}}
		\par\end{centering}
	\vskip-1ex
	\caption{\addmorerevision{Comparison of the number of \textit{raw} points produced by the MINVO (MV) enclosure, B\'{e}zier (Be) enclosure, and SLEFE (SL) for \nthdegree{} 2D polynomial curves. Here, $s$ is the number of subintervals the curve is divided into and \notationSLh{}. \conclusionRawPoints{}. } \label{fig:comparison_raw_points}}
	\vskip 0ex
\end{figure*}

	\subsection{Rational Curves} \label{subsec:rational curves}
	Via projections, the MINVO basis is also able to obtain the smallest simplex that encloses some rational curves, which are curves whose coordinates are the quotients of two polynomials. \add{For instance, given the $n$-simplex obtained by the MINVO basis for a given $n^{\text{th}}$-degree polynomial curve $P$, we can project every point $\boldsymbol{p}(t)$ of the curve via a perspective projection, using a vertex as the viewpoint, and the opposite facet as the projection plane. Note that this perspective projection of the polynomial curve will be a rational curve. If the  $n$-simplex is the smallest one enclosing the polynomial curve, then each facet is also the smallest $(n-1)$-simplex that encloses the projection. This can be easily proven by contradiction, since if the facet were not a minimal $(n-1)$-simplex for the projected curve, then the $n$-simplex would not be minimal for the original 3D curve (see \cite{sayrafiezadeh1992inductive} for instance).}
	\add{Let us define $\arraycolsep=1.4pt\left[\protect\begin{array}{ccc}
			\boldsymbol{v}_{0} & \hdots & \boldsymbol{v}_{n} \protect\end{array}\right]:=\left[\protect\begin{array}{cc}
			\boldsymbol{0} & \boldsymbol{I}_{n} \protect\end{array}\right]$, and let $\boldsymbol{\pi}_i$ denote the plane that passes though the \addrevision{vertices} $\{\boldsymbol{v}_{0},\boldsymbol{v}_{1},\ldots,\boldsymbol{v}_{n}\}\backslash\addrevision{\{\boldsymbol{v}_{i}\}}$. Then, for a standard $n$-simplex, the perspective projection of $\arraycolsep=1.4pt\boldsymbol{p}(t):=\left[\begin{array}{ccc}
			\boldsymbol{\lambda}_{1} & \cdots & \boldsymbol{\lambda}_{n}\end{array}\right]^{T}\boldsymbol{t}$ onto the plane $\boldsymbol{\pi}_i$, using $\boldsymbol{v}_i$ as the viewpoint, is given by:}
	\addcad{
	\begin{equation*}
				\arraycolsep=1.4pt\begin{cases}
			\frac{1}{1-\boldsymbol{\lambda}_{0}^{T}\boldsymbol{t}}\left[\begin{array}{ccc}
				\boldsymbol{\lambda}_{1} & \cdots & \boldsymbol{\lambda}_{n}\end{array}\right]^{T}\boldsymbol{t} & \text{Projection onto \ensuremath{\boldsymbol{\pi}_{0}}}\\
			\frac{1}{1-\boldsymbol{\lambda}_{i}^{T}\boldsymbol{t}}\left[\begin{array}{cccccc}
				\boldsymbol{\lambda}_{1}\cdots & \boldsymbol{\lambda}_{i-1} & 0 & \boldsymbol{\lambda}_{i+1} & \cdots & \boldsymbol{\lambda}_{n}\end{array}\right]^{T}\boldsymbol{t} & \text{\text{Projection onto }}\boldsymbol{\pi}_{i}, i>0
		\end{cases}
\end{equation*}}

	\add{This projection can also be applied successively from $\mathbb{R}^i$ to $\mathbb{R}^j$ ($i>j\ge1$). Figure~\ref{fig:projections_with_higher_degree} shows the case $\mathbb{R}^3\rightarrow\mathbb{R}^2$ (for all the four possible projections), and some projections of the cases $\mathbb{R}^6\rightarrow\mathbb{R}^2$, $\mathbb{R}^{10}\rightarrow\mathbb{R}^2$, $\mathbb{R}^5\rightarrow\mathbb{R}^3$, and $\mathbb{R}^{12}\rightarrow\mathbb{R}^3$.}

	\section{\addrevision{Comparison with SLEFEs\label{sec:SLEFEs}}}
	
	\addmorerevision{		
		In this section, we compare the enclosures for \nthdegree{} 2D polynomial curves obtained using these three techniques with and without subdivision:
	\begin{itemize}
		\item \textbf{MINVO}: The curve is divided into $s$ subintervals, and then the MINVO enclosure for each of the subintervals is computed.
		\item \textbf{B\'{e}zier}: The curve is divided into $s$ subintervals, and then the B\'{e}zier enclosure for each of the subintervals is computed.
		\item \textbf{SLEFE}: The curve is divided into $s$ subintervals, and the SLEFE (subdividable linear efficient function enclosure~\cite{lutterkort2001optimized, lutterkort1999tight,peters2004sleves}) is obtained using  $h$ breakpoints\footnote{\additionalrevision{As an example, if the time subinterval is $[0.6, 1]$, then three uniformly-distributed breakpoints would be $\left\{0.6,0.8,1\right\}$, and the SLEFE for that subinterval will consist of a convex enclosure for the part of the curve in $t\in[0.6,0.8]$, and another convex enclosure for the part of the curve in $t\in[0.8,1]$.
		}} per subinterval (i.e., $h-1$ linear segments per subinterval). A SLEFE obtained with $h$ breakpoints per subinterval will be denoted as SL\textsubscript{$h$}.
	\end{itemize}
	}
	
	\addmorerevision{Note that $s=1$ corresponds to the case where no subdivision is performed. When $s>1$, the subintervals of the curve are obtained by evenly splitting the time interval. Moreover, SL\textsubscript{$2$} (i.e., $h=2$) corresponds to a SLEFE with only one linear segment (i.e., two breakpoints) per subinterval.} 

   We compare \addextrarevision{the width}, the union, and the convex hull 	\addextrarevision{(defined in~\ref{sec:appendix_definitions})}  for the different enclosures. \addextrarevision{The comparison of the width of the enclosures produced is available in~\ref{sec:appendix_SLEFEs_width}.}
	The comparison of the area and number of vertices of the union and the convex hull is available in~\ref{sec:appendix_SLEFEs_smooth}. In \ref{sec:MinvoVsSLEFERunTime}, we compare MINVO and SLEFE in terms of runtime and simplicity of their implementation.

	\begin{figure*}[htb]
	\begin{centering}
		\includegraphics[width=0.9\textwidth]{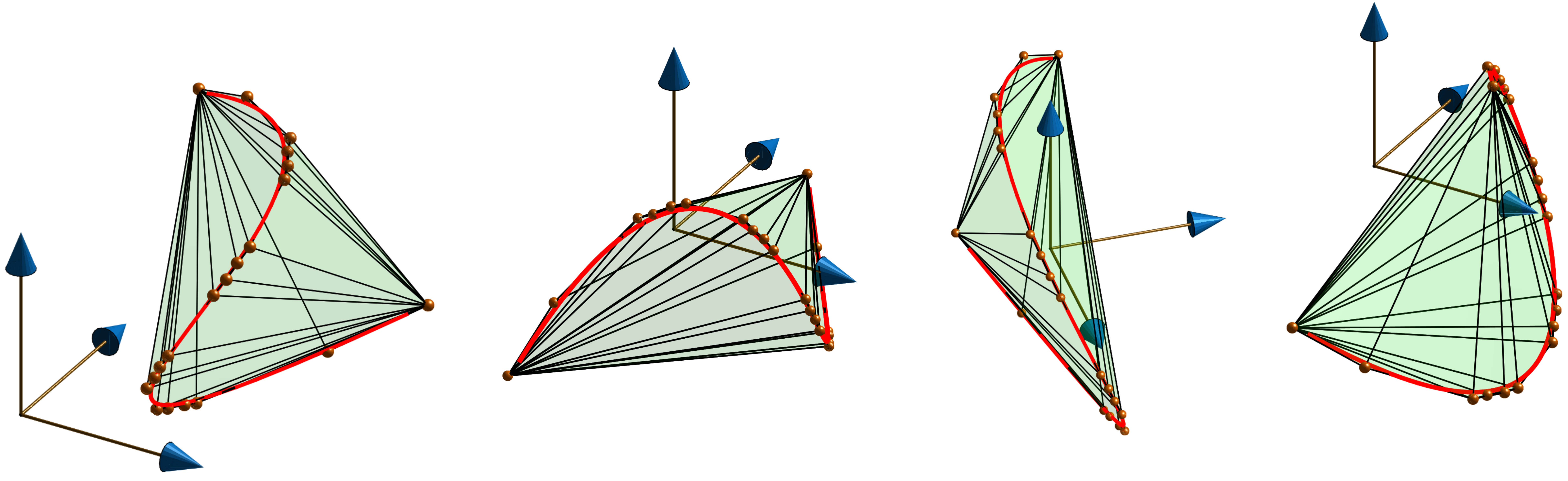}	
		\caption{Tighter polyhedral outer approximations for a curve $P\in\mathcal{P}^n$ can be obtained by splitting the curve into several \addmorerevision{subintervals}, calculating the MINVO $n$-simplexes that enclose each one of these \addmorerevision{subintervals} and then computing the convex hull of all these simplexes. In all these cases shown, the original curve $P\in\mathcal{P}^3$ is splitted into $5$ \addmorerevision{subintervals} \addmorerevision{(i.e., $s=5$)}, and the resulting convex hull is a polyhedron with $20$ \addrevision{vertices} that is $1.19$ times smaller than the smallest 3-simplex that encloses the whole curve (i.e., the simplex found by applying the MINVO basis to the whole curve).\label{fig:splitted_many_comparisons3d_poly_given}}
		\par\end{centering}
	\vskip 0ex
\end{figure*}

	\addrevision{Several conclusions can be drawn regarding the comparison between MINVO, B\'{e}zier, and SLEFE:
		\begin{itemize}
			\item \addmorerevision{Compared to SL\textsubscript{$2$}, MINVO achieves a smaller area for most of the $n$--$s$ combinations tested, and sometimes using only half of the vertices needed by SL\textsubscript{$2$}.~Compared to~SL\textsubscript{$h$} ($h\in[3,8]$), MINVO typically achieves a smaller area for the cases where either $s$ is small or the degree $n$ is high, and it usually does so using fewer number of vertices than~SL\textsubscript{$h$}. On the other hand, SL\textsubscript{$h$} tends to achieve a smaller area when either $s$ is large or the degree $n$ is small, but it usually requires more vertices than MINVO. Hence, MINVO is advantageous with respect to SL\textsubscript{$h$} in applications where having a small number of vertices is crucial.
			For example, a smaller number of vertices can substantially reduce the total computation time in algorithms that, after finding the enclosure, need to iterate through all of the vertices found to impose a constraint or perform a specific operation/check for each of them. \addextrarevision{If the number of vertices is not important for the specific application, then SL\textsubscript{$h$} ($h\ge4$) should be chosen, since it typically achieves a smaller area of the union and area of the convex hull.}}
			
			\item \addmorerevision{Compared to B\'{e}zier, MINVO also achieves a smaller \additionalrevision{union and hull} area for the cases where either $s$ is small or the degree $n$ is high, and, when $n\in[3,7]$, it does so using only up to $1.3$ times the number of vertices of B\'{e}zier.}
			
			\item \addextrarevision{In terms of the width of the enclosure (\ref{sec:appendix_SLEFEs_width}), SLEFE performs better than MINVO. The use of SLEFE is hence desirable in the cases where the width of the enclosure is more important for the specific application.}	
			
			\item For any of the techniques, \addmorerevision{operations like the union, the convex hull, or the outer boundaries computation for SLEFE (\cite{peters2004sleves}) are typically either not possible or computationally expensive in the applications where the curve itself is a decision variable of an optimization problem (as in, e.g.,~\cite{simon2009fitting, selinger2011efficient, tordesillas2020mader, tordesillas2021panther}). Instead, we can use the \textit{raw points} of the enclosure, which in 2D can be defined as the \textit{unique} control points of each subinterval (for MINVO and B\'{e}zier), and as the \textit{unique} vertices of each of the rectangles  generated per breakpoint 
			of each subinterval (for~SL\textsubscript{$h$})}. 
			The number of these \textit{raw points} for a general 2D curve is $ns+s$ for MINVO, $ns+1$ for B\'{e}zier, and  \addmorerevision{$4hs$ for~SL\textsubscript{$h$}. The comparison of these raw points is shown in Figure~\ref{fig:comparison_raw_points}. \conclusionRawPoints{}.    }

			\item As noted in \cite{peters2004sleves}, SLEFE depends \textit{pseudo-linearly} on the coefficients of the polynomial curve (i.e., linearly except for a min/max operation). In contrast, the MINVO or B\'{e}zier enclosures depend \textit{linearly} on the coefficients of the polynomial curve. This makes the MINVO and B\'{e}zier enclosures more suitable for time-critical optimization problems in which the coefficients of the curve are decision variables.	
		
	\end{itemize}}

	\section{Final Remarks}

	\subsection{Conversion between MINVO, Bernstein, and B-Spline}\label{subsec:conversion}
	
	Given a curve $P\in\mathcal{P}^n$, the control points (i.e., the \addrevision{vertices} of
	the $n$-simplex that encloses that curve) using a basis $\alpha$ can be obtained from the control points of a different basis $\beta$ ($\alpha,\beta\in\{\text{MV},\text{Be},\text{BS}\}$) using the formula
	\begin{equation} \label{eq:relationship_basis}
		\boldsymbol{V}_{\alpha}=\boldsymbol{P}\boldsymbol{A}_{\alpha}^{-1}=\boldsymbol{V}_{\beta}\boldsymbol{A}_{\beta}\boldsymbol{A}_{\alpha}^{-1} \addrevision{\, .}
	\end{equation}
	For instance, to obtain the Bernstein control points from the MINVO control points we can use $\boldsymbol{V}_{\text{Be}}=\boldsymbol{V}_{\text{MV}}\boldsymbol{A}_{\text{MV}}\boldsymbol{A}_{\text{Be}}^{-1}$.
	The matrices $\boldsymbol{A}_{\text{MV}}$ are the ones shown in Table
	\ref{tab:table_matrices}, while the matrices $\boldsymbol{A}_{\text{Be}}$
	and $\boldsymbol{A}_{\text{BS}}$ are available in \cite{qin2000general}.
	Note that all the matrices need to be expressed in the same interval
	($t\in[-1,1]$ in this paper)\addcad{, and that the inverses of these matrices can be easily precomputed offline}.

	\subsection{\add{Tighter volumes for Problem 1 via subdivision}}\label{subsec:tighter_volumes_subdivision}
	
	\add{\addrevision{As shown in Section~\ref{sec:SLEFEs}, and at} the expense of adding more \addrevision{vertices}, tighter} polyhedral solutions for Problem 1 can be obtained by \addmorerevision{dividing the polynomial curve into several subintervals} and then computing the convex hull of the MINVO \addmorerevision{enclosure of each subinterval}. \addmorerevision{To subdivide the curve, one can do it first in B\'{e}zier form (leveraging therefore the properties of De Casteljau's algorithm), and then compute the MINVO control points as linear functions of the B\'{e}zier control points of that subinterval using Eq.~\ref{eq:relationship_basis}. Alternatively, one can also tabulate (offline) the inverse of the matrices $\boldsymbol{A}_{\text{MV}}$, expressed in the subinterval desired, and then simply compute the MINVO control points of that subinterval as $\boldsymbol{V}_{\text{MV}}=\boldsymbol{P}\boldsymbol{A}_{\text{MV}}^{-1}$.}

	Several examples for $n=3$ are shown in \add{Figure}~\ref{fig:splitted_many_comparisons3d_poly_given}, where the original curve $P\in\mathcal{P}^3$ is split into $5$ \addmorerevision{subintervals} \addmorerevision{(i.e., $s=5$)}, and the resulting convex hull is a polyhedron with $20$ \addrevision{vertices} that is $1.19$ times smaller than the smallest 3-simplex that encloses the whole curve (i.e., the simplex found by applying the MINVO basis to the whole curve). 
	
	\add{Depending on the specific application, one might also be interested in obtaining a sequence of overlapping \addmorerevision{polyhedra} whose union (a nonconvex set in general) completely encloses the curve. This can be obtained by simply computing the MINVO \addmorerevision{enclosure} for every \addmorerevision{subinterval} of the curve.} %

	\subsection{When should each basis be used?}\label{sec:when_each_basis_used}
	\add{
		The Bernstein (B\'ezier) and B-Spline bases have many useful properties that are not shared by the MINVO basis. For example, a polynomial curve passes through the first and last B\'ezier control points, \addextrarevision{the derivative of a B\'ezier curve can be easily computed from the difference of the B\'ezier control points}, and the  B-Spline basis has built-in smoothness in curves with several segments. Hence, it may be desirable to use the B\'ezier or B-Spline control points
		to design and model the curve, and then convert the control points of every interval to the MINVO control points (using the simple \add{linear transformation given in} \addcad{Section~}\ref{subsec:conversion}) to perform collision/intersection checks, or to impose collision-avoidance constraints in an optimization problem \cite{herron1989polynomial,tordesillas2020mader,tordesillas2021panther}. This approach benefits from the properties of the Bernstein/B-Spline bases, while also exploiting the enclosures obtained by the MINVO basis. 
	}
	
	\section{Conclusions and Future Work}\label{sec:conclusions}
	
	This work derived and presented the MINVO basis. The key feature
	of this basis is that it finds the smallest $n$-simplex that encloses
	a given $n^\text{th}$\add{-degree} polynomial curve (Problem 1), and also finds the $n^\text{th}$\add{-degree} polynomial
	curve with largest convex hull enclosed in a given $n$-simplex (Problem 2). For $n=3$, the ratios of the improvement in the volume achieved by the MINVO basis with respect to the Bernstein and B-Spline bases are $2.36$ and $254.9$ respectively. When $n=7$, these improvement ratios increase to $902.7$ and $2.997\cdot10^{21}$ respectively. \add{Numerical global} optimality was proven for $n=1,2,3$, \add{numerical local optimality was proven for $n=4$, and high-quality feasible solutions \addmorerevision{for all}~$n\ge5$ were obtained. Finally, the} MINVO basis was also applied to polynomial curves with different $n$, $k$ and $m$ (achieving improvements ratios of up to $\approx 550$)\add{, and to some rational curves}. 
	
	The exciting results of this work naturally lead to the following
	questions and conjectures, that we leave as future work:
	\begin{itemize}
		\item Is the global optimum of Problem 4 the same as the global optimum
		of Problem 3? \add{I.e., are we losing optimality by imposing
			the specific structure on $\lambda_{i}(t)$? On a similar note, is it possible to obtain \addrevision{for any $n$} a bound on the distance between the objective value obtained by the model proposed in \addcad{Section~}\ref{sec:resultsnGT7}, and the global minimum of Problem~3?} 
		\item Does there exist a recursive formula to obtain the solution of Problem
		3 for a specific $n=q$ given the previous solutions for $n=1,\ldots,q-1$? Would this recursive
		formula allow to obtain the \textit{globally} optimal solutions \addmorerevision{for all}~\add{$n\in\mathbb{N}$} of Problem~3?
	\end{itemize}
	
	\add{Finally, the way polynomials are scaled (to impose $\boldsymbol{A}^T\boldsymbol{1}=\boldsymbol{e}$) in \addcad{Section~}\ref{sec:resultsnGT7} can suffer from numerical instabilities when the degree is very high ($n>30$). This is expected, since the monomial basis used to compute $\boldsymbol{A}$  is known to be numerically unstable \cite{blekherman2012semidefinite}. A more numerically-stable scaling, potentially avoiding the use of the monomial basis, could therefore be beneficial for higher degrees.}
	\section*{Acknowledgments}
	\add{The authors would like to thank Prof.~Gilbert Strang, Prof.~Luca Carlone, Dr.~Ashkan M. Jasour, Dr.~Kasra Khosoussi, \add{Dr.~David Rosen}, Juan José Madrigal, Parker Lusk, \addmorerevision{Dr.}~Kris Frey, and \addmorerevision{Dr.}~Maodong Pan} for helpful insights and
	discussions. \addmorerevision{The authors would also like to thank the anonymous reviewers, whose valuable feedback helped to improve the article.} Research funded in part by Boeing Research \& Technology.
	
	\appendix 
	\onecolumn
	\section{Volume of the Convex Hull of a Polynomial Curve\label{sec:AppVolume}}
	The volume \addrevision{of the convex hull of} any \add{$n^{\text{th}}$-degree} polynomial curve can be easily obtained
	using the result from \cite[Theorem~15.2]{karlin1953geometry}. In
	this work\footnote{Note that \cite{karlin1953geometry} uses the convention $t\in[0,1]$
		(instead of $t\in[-1,1]$), and therefore it uses the curve $\arraycolsep=1.4pt\left[\begin{array}{cccc}
			t & t^{2} & \cdots & t^{n}\end{array}\right]^{T}$. \addrevision{Note also that the convex hull of a moment curve is equal to a cyclic polytope \cite{ahmed2012volume,ziegler2012lectures} with infinitely many points evenly distributed along the curve.}}, it is shown that the volume of the convex hull of a curve $R$ with
	$\arraycolsep=1.4pt\boldsymbol{r}(t):=\left[\begin{array}{cccc}
		\frac{t+1}{2} & \left(\frac{t+1}{2}\right)^{2} & \cdots & \left(\frac{t+1}{2}\right)^{n}\end{array}\right]^{T}$ is given by 
	$$\text{vol}\left(\text{conv}\left(R\right)\right)=\prod_{j=1}^{n}\text{B}(j,j)=\prod_{j=1}^{n}\left(\frac{\left((j-1)!\right)^{2}}{(2j-1)!}\right)=\frac{1}{n!}\prod_{j=1}^{n}\left(\frac{j!(j-1)!}{(2j-1)!}\right)=\frac{1}{n!}\prod_{0\le i<j\le n}\left(\frac{j-i}{j+i}\right) \addrevision{\, ,}$$
	where $\text{B}(x,y)$ denotes the beta function. Let us now define
	$\tilde{\boldsymbol{t}}$ as $\arraycolsep=1.4pt\tilde{\boldsymbol{t}}:=\left[\begin{array}{cccc}
		t^{n} & t^{n-1} & \cdots & t\end{array}\right]^{T}$, $\boxtimes$ as any number \addmorerevision{in} $\mathbb{R}$ and write $\boldsymbol{r}(t)$ as:
	\[
	\boldsymbol{r}(t)=\underbrace{\left[\begin{array}{cccccc}
			0 & 0 & \cdots & 0 & \frac{1}{2} & \boxtimes\\
			0 & 0 & \cdots & \frac{1}{2^{2}} & \boxtimes & \boxtimes\\
			\vdots & \vdots & \ddots & \vdots & \vdots & \vdots\\
			0 & \frac{1}{2^{n-1}} & \cdots & \boxtimes & \boxtimes & \boxtimes\\
			\frac{1}{2^{n}} & \boxtimes & \cdots & \boxtimes & \boxtimes & \boxtimes
		\end{array}\right]}_{:=\boldsymbol{R}}\boldsymbol{t}=\boldsymbol{R}_{:,0:n-1}\tilde{\boldsymbol{t}}+\boldsymbol{R}_{:,n}
	\]
	
	Now, defining $\boldsymbol{Q}:=\boldsymbol{P}_{:,0:n-1}\boldsymbol{R}_{:,0:n-1}^{-1}$, note that $\left(\boldsymbol{p}(t)-\boldsymbol{P}_{:,n}\right)=\boldsymbol{P}_{:,0:n-1}\tilde{\boldsymbol{t}}=\boldsymbol{Q}\boldsymbol{R}_{:,0:n-1}\tilde{\boldsymbol{t}}=\boldsymbol{Q}\left(\boldsymbol{r}(t)-\boldsymbol{R}_{:,n}\right)=\boldsymbol{Q}\boldsymbol{r}(t)-\boldsymbol{Q}\boldsymbol{R}_{:,n}
	$. As the translation part does not affect the volume, we can write
			$$\text{vol}\left(\text{conv}\left(P\right)\right)=\text{vol}\left(\text{conv}\left(\left\{ \boldsymbol{p}(t)-\boldsymbol{P}_{:,n}\;|\;t\in[-1,1]\right\} \right)\right)=\text{vol}\left(\text{conv}\left(\left\{ \boldsymbol{Q}\boldsymbol{r}(t)\;|\;t\in[-1,1]\right\} \right)\right)=...$$
	\iffalse
	where 
	\[
	\text{conv}\left(\left\{ \boldsymbol{Q}\boldsymbol{r}(t)\;|\;t\in[-1,1]\right\} \right)=\left\{ \sum_{i\in\mathcal{I}}\alpha_{i}\boldsymbol{Q}\boldsymbol{r}(t_{i})\;|\;\sum_{i\in\mathcal{I}}\alpha_{i}=1,\alpha_{i}\ge0,t_{i}\in[-1,1],\forall i\in\mathcal{I}\right\} =\boldsymbol{Q}\text{conv}\left(R\right)
	\]
	Therefore:
	\fi
	\[
	...=\text{vol}\left(\boldsymbol{Q}\text{conv}\left(R\right)\right)=\text{abs}\left(\frac{\left|\boldsymbol{P}_{:,0:n-1}\right|}{\left|\boldsymbol{R}_{:,0:n-1}\right|}\right)\text{vol}\left(\text{conv}\left(R\right)\right) \addrevision{\, ,}
	\]
	where we have used the notation $\boldsymbol{Q}\text{conv}\left(R\right)$ to denote the set $\left\{ \boldsymbol{Q}\boldsymbol{x}|\boldsymbol{x}\in\text{conv}\left(R\right)\right\}$.
	
	As the determinant of $\boldsymbol{R}_{:,0:n-1}$ is $\left|\boldsymbol{R}_{:,0:n-1}\right|=\prod_{i=1}^{n}\frac{1}{2^{i}}=2^{\frac{-n(n+1)}{2}}$,
	we can conclude that:
	\[
	\boxed{\text{vol}\left(\text{conv}\left(P\right)\right)=\frac{\text{abs}\left(\left|\boldsymbol{P}_{:,0:n-1}\right|\right)}{n!}2^{\frac{n(n+1)}{2}}\prod_{0\le i<j\le n}\left(\frac{j-i}{j+i}\right)}
	\]
	\vskip-4.5ex
	\hfill $\square$
	
	\addrevision{ }

	\section{Invertibility of the matrix $\boldsymbol{A}$ \label{sec:AppInvertibility}}
	From \add{Eq.~}\ref{eq:PinS2}, we have that the $(n+1)\times(n+1)$ matrix $\boldsymbol{A}$ satisfies
	\[
	\left[\begin{array}{c}
		\boldsymbol{P}\\
		\boldsymbol{e}^{T}
	\end{array}\right]=\left[\begin{array}{c}
		\boldsymbol{V}\\
		\boldsymbol{1}^{T}
	\end{array}\right]\boldsymbol{A} \addrevision{\, .}
	\]
	
	As $\text{abs}\left(\left|\begin{array}{c}
		\boldsymbol{P}\\
		\boldsymbol{e}^{T}
	\end{array}\right|\right)=\text{abs}\left(\left|\boldsymbol{P}_{:,0:n-1}\right|\right)\neq0$, and $\arraycolsep=1.4pt\left|\begin{array}{c}
		\boldsymbol{V}\\
		\boldsymbol{1}^{T}
	\end{array}\right|=\left|\begin{array}{cc}
		\boldsymbol{V}^{T} & \boldsymbol{1}\end{array}\right|\neq0$ (see \addcad{Section~}\ref{sec:problems-definition}), we have that 
	$
	\text{rank}\left(\left[\begin{array}{c}
		\boldsymbol{P}\\
		\boldsymbol{e}^{T}
	\end{array}\right]\right)=\text{rank}\left(\left[\begin{array}{c}
		\boldsymbol{V}\\
		\boldsymbol{1}^{T}
	\end{array}\right]\right)=n+1
	$. Using now the fact that $\text{rank}\left(\boldsymbol{B}\boldsymbol{C}\right)\le\min\left(\text{rank}\left(\boldsymbol{B}\right),\text{rank}\left(\boldsymbol{C}\right)\right)$,
	we conclude that $\text{\text{rank}\ensuremath{\left(\boldsymbol{A}\right)}}=n+1$
	(i.e., $\boldsymbol{A}$ has full rank), and therefore $\boldsymbol{A}$ is invertible. \hfill $\square$

	\interfootnotelinepenalty=10000  %
	\section{Karush–Kuhn–Tucker conditions (for odd $n$)\label{sec:AppKKT}}
	In this \addcad{Appendix} we derive the KKT conditions for this problem:
	\[
	\fbox{\ensuremath{\begin{array}{c}
				\underset{\boldsymbol{A}\in\mathbb{R}^{(n+1)\times(n+1)}}{\min}\quad-\text{ln}\left(\left|\boldsymbol{A}^{T}\boldsymbol{A}\right|\right)\\
				\text{s.t.}\quad\;\boldsymbol{A}^{T}\boldsymbol{1}=\boldsymbol{e}\\
				\qquad\qquad\boldsymbol{A}\boldsymbol{t}\ge\boldsymbol{0}\quad\forall t\in[-1,1]
	\end{array}}}
	\]
	
	which is equivalent to Problem 3. For the sake of brevity, we present here the case $n$ odd (the case $n$ even can be easily obtained with small modifications). In the following, 
	$\boldsymbol{V}*\boldsymbol{W}$ will be
	the matrix resulting from the row-wise discrete convolution (i.e., $\left(\boldsymbol{V}*\boldsymbol{W}\right)_{i,:}=\boldsymbol{V}_{i,:}*\boldsymbol{W}_{i,:}$), and $\text{Top}(\boldsymbol{a},\boldsymbol{b})$
	will denote the Toeplitz matrix whose first column is $\boldsymbol{a}$
	and whose first row is $\boldsymbol{b}^{T}$. Let us also define: 
	\begin{equation*}
		\resizebox{1.0 \textwidth}{!}{
			$\arraycolsep=0.8pt\boldsymbol{R}_{\boldsymbol{G}}:=\text{Top}\left(\left[\begin{array}{c}
				1\\
				1\\
				\boldsymbol{0}_{n-1}
			\end{array}\right],\left[\begin{array}{c}
				1\\
				\boldsymbol{0}_{n-1}
			\end{array}\right]\right)=\left[\begin{array}{c}
				\boldsymbol{I}_{n}\\
				\boldsymbol{0}_{n}^{T}
			\end{array}\right]+\left[\begin{array}{c}
				\boldsymbol{0}_{n}^{T}\\
				\boldsymbol{I}_{n}
			\end{array}\right]\quad\arraycolsep=1.2pt\boldsymbol{R}_{\boldsymbol{H}}:=\text{Top}\left(\left[\begin{array}{c}
				-1\\
				1\\
				\boldsymbol{0}_{n-1}
			\end{array}\right],\left[\begin{array}{c}
				-1\\
				\boldsymbol{0}_{n-1}
			\end{array}\right]\right)=\left[\begin{array}{c}
				-\boldsymbol{I}_{n}\\
				\boldsymbol{0}_{n}^{T}
			\end{array}\right]+\left[\begin{array}{c}
				\boldsymbol{0}_{n}^{T}\\
				\boldsymbol{I}_{n}
			\end{array}\right]\quad\arraycolsep=1.2pt\boldsymbol{L}_{q}:=\left[\begin{array}{cc}
				\boldsymbol{0}^{T} & 0\\
				\boldsymbol{I}_{q-1} & \boldsymbol{0}
			\end{array}\right]_{q\times q}\, ,$
		} 
	\end{equation*}

	and the matrices $\boldsymbol{G}\in\mathbb{R}^{(n+1)\times\frac{n+1}{2}}$,
	$\boldsymbol{H}\in\mathbb{R}^{(n+1)\times\frac{n+1}{2}}$ and $\boldsymbol{\lambda}\in\mathbb{R}^{(n+1)}$.
	We know that
	\begin{equation*}
		\resizebox{1.0 \textwidth}{!}{
			$\left(\boldsymbol{A}\boldsymbol{t}\right)_{i}\ge0\quad\forall t\in[-1,1]\overset{(\text{a})}{\iff}\left(\boldsymbol{A}\boldsymbol{t}\right)_{i}=(t+1)g_{i}^{2}(t)+(1-t)h_{i}^{2}(t)\overset{(\text{b})}{\iff}\boldsymbol{A}_{i,:}=\left(\boldsymbol{G}_{i,:}*\boldsymbol{G}_{i,:}\right)\boldsymbol{R}_{G}^{T}+\left(\boldsymbol{H}_{i,:}*\boldsymbol{H}_{i,:}\right)\boldsymbol{R}_{H}^{T}\iff$
		}
	\end{equation*}
	\begin{equation}
		\text{...\ensuremath{\iff\boldsymbol{A}}=\ensuremath{\left[\begin{array}{c|c}
					\boldsymbol{G}*\boldsymbol{G} & \boldsymbol{H}*\boldsymbol{H}\end{array}\right]\left[\begin{array}{c}
					\boldsymbol{R}_{G}^{T}\\
					\boldsymbol{R}_{H}^{T}
				\end{array}\right]}} \addrevision{\, ,} \label{eq:defA}
	\end{equation}
	
	where $g_{i}(t)$ and $h_{i}(t)$ are polynomials of degree $\frac{n-1}{2}$.
	Note that (a) is given by the Markov–Luk\'{a}cs \addrevision{Theorem} (see \addcad{Section~}\ref{subsec:Equiv_formulation}). In (b) we
	have simply used the discrete convolution to multiply $g_{i}(t)$
	by itself, and the Toeplitz matrix $\boldsymbol{R}_{\boldsymbol{G}}$ to multiply the result
	by $(t+1)$ \cite{qin2000general}. An analogous reasoning applies for the term with $h_{i}$\footnote{Alternatively, we can also write:
		
		\[
		\left(\boldsymbol{A}_{i,:}\right)^{T}=\boldsymbol{R}_{\boldsymbol{G}}\text{Top}\left(\left[\begin{array}{c}
			\boldsymbol{G}_{i,:}\\
			\boldsymbol{0}_{\frac{n-1}{2}}
		\end{array}\right],\left[\begin{array}{c}
			\boldsymbol{G}_{i,0}\\
			\boldsymbol{0}_{\frac{n-1}{2}}
		\end{array}\right]\right)\boldsymbol{G}_{i,:}+\boldsymbol{R}_{\boldsymbol{H}}\text{Top}\left(\left[\begin{array}{c}
			\boldsymbol{H}_{i,:}\\
			\boldsymbol{0}_{\frac{n-1}{2}}
		\end{array}\right],\left[\begin{array}{c}
			\boldsymbol{H}_{i,0}\\
			\boldsymbol{0}_{\frac{n-1}{2}}
		\end{array}\right]\right)\boldsymbol{H}_{i,:}
		\]
	}. Using now $\boldsymbol{G}$ and $\boldsymbol{H}$ as the decision variables of the primal problem (where $\boldsymbol{A}$ is given by \add{Eq.~}\ref{eq:defA}), the Lagrangian is
	$$
	\ensuremath{\mathcal{L\ensuremath{=}}-\text{ln}(|\boldsymbol{A}^{T}\boldsymbol{A}|)+\boldsymbol{\lambda}^{T}(\boldsymbol{A}^{T}\boldsymbol{1}-\boldsymbol{e})} \addrevision{\, .}
	$$
	
	Differentiating the Lagrangian yields to
	\[
	\frac{\partial\mathcal{L}}{\partial\boldsymbol{G}_{ij}}=\text{tr}\left(-\underbrace{\frac{\partial\text{ln}\left(\left|\boldsymbol{A}^{T}\boldsymbol{A}\right|\right)}{\partial\boldsymbol{A}}}_{=2\boldsymbol{A}^{+}=2\boldsymbol{A}^{-1}}\boldsymbol{Q}_{\boldsymbol{G}_{ij}}\right)+\text{tr}\left(\boldsymbol{\lambda}\boldsymbol{1}^{T}\boldsymbol{Q}_{\boldsymbol{G}_{ij}}\right)=\text{tr}\left(\left(-2\boldsymbol{A}^{-1}+\boldsymbol{\lambda}\boldsymbol{1}^{T}\right)\boldsymbol{Q}_{\boldsymbol{G}_{ij}}\right)  \addrevision{\, ,}
	\]
	
	where
	\begin{equation}
		\arraycolsep=1.4pt\boldsymbol{Q}_{\boldsymbol{G}_{ij}}:=\frac{\partial\boldsymbol{A}}{\partial\boldsymbol{G}_{ij}}=2\ensuremath{\left(\boldsymbol{L}_{(n+1)}\right)^{i-1}\left[\begin{array}{c}
				\left[\begin{array}{cc}
					\boldsymbol{G}_{i,:} & \boldsymbol{0}^{T}\end{array}\right]\left(\boldsymbol{L}_{n}^{T}\right)^{j-1}\boldsymbol{R}_{\boldsymbol{G}}^{T}\\
				\boldsymbol{0}_{n\times(n+1)}
			\end{array}\right]_{(n+1)\times(n+1)}}   \addrevision{\, .}  \label{eq:eqQ}
	\end{equation}
	
	Same expression applies for $\boldsymbol{Q}_{\boldsymbol{H}_{ij}}:=\frac{\partial\boldsymbol{A}}{\partial\boldsymbol{H}_{ij}}$,
	but using $\boldsymbol{H}_{i,:}$ and $\boldsymbol{R}_{\boldsymbol{H}}^{T}$
	instead. The KKT equations can therefore be written as follows:
	
	\vspace{0.1cm}
	\noindent\fcolorbox{black}{white}{\begin{minipage}[t]{0.98\columnwidth}%
			
			\textbf{KKT equations: }Solve for $\boldsymbol{G},\boldsymbol{H},\boldsymbol{\lambda}$:
			\[
			\left\{ \begin{aligned}
				&\text{tr}\left(\left(-2\boldsymbol{A}^{-1}+\boldsymbol{\lambda}\boldsymbol{1}^{T}\right)\boldsymbol{Q}_{\boldsymbol{G}_{ij}}\right)=0\qquad\forall i\in\{0,...,n\}, \forall j\in\{0,...,\frac{n-1}{2}\}\\
				&\text{tr}\left(\left(-2\boldsymbol{A}^{-1}+\boldsymbol{\lambda}\boldsymbol{1}^{T}\right)\boldsymbol{Q}_{\boldsymbol{H}_{ij}}\right)=0\qquad\forall i\in\{0,...,n\}, \forall j\in\{0,...,\frac{n-1}{2}\}\\
				&\boldsymbol{A}^{T}\boldsymbol{1}=\boldsymbol{e} \kern 7.3cm
			\end{aligned}\right.
			\]
			where $\boldsymbol{A}$ is given by \add{Eq.~}\ref{eq:defA}. and $\boldsymbol{Q}_{\boldsymbol{G}_{ij}}$
			by \add{Eq.~}\ref{eq:eqQ}. $\boldsymbol{Q}_{\boldsymbol{H}_{ij}}$ is also given by \add{Eq.~}\ref{eq:eqQ}, but using $\boldsymbol{H}_{i,:}$ and $\boldsymbol{R}_{H}^{T}$ instead.
	\end{minipage}}

	\section{\addrevision{Plot of a polynomial $x(t)\in \mathbb{R}[t] $ over $t\in[-1,1]$\label{sec:ScalarPol}}}
	
\begin{figure}
		\begin{centering}
			\addrevisiongraphics{\includegraphics[width=1.0\columnwidth]{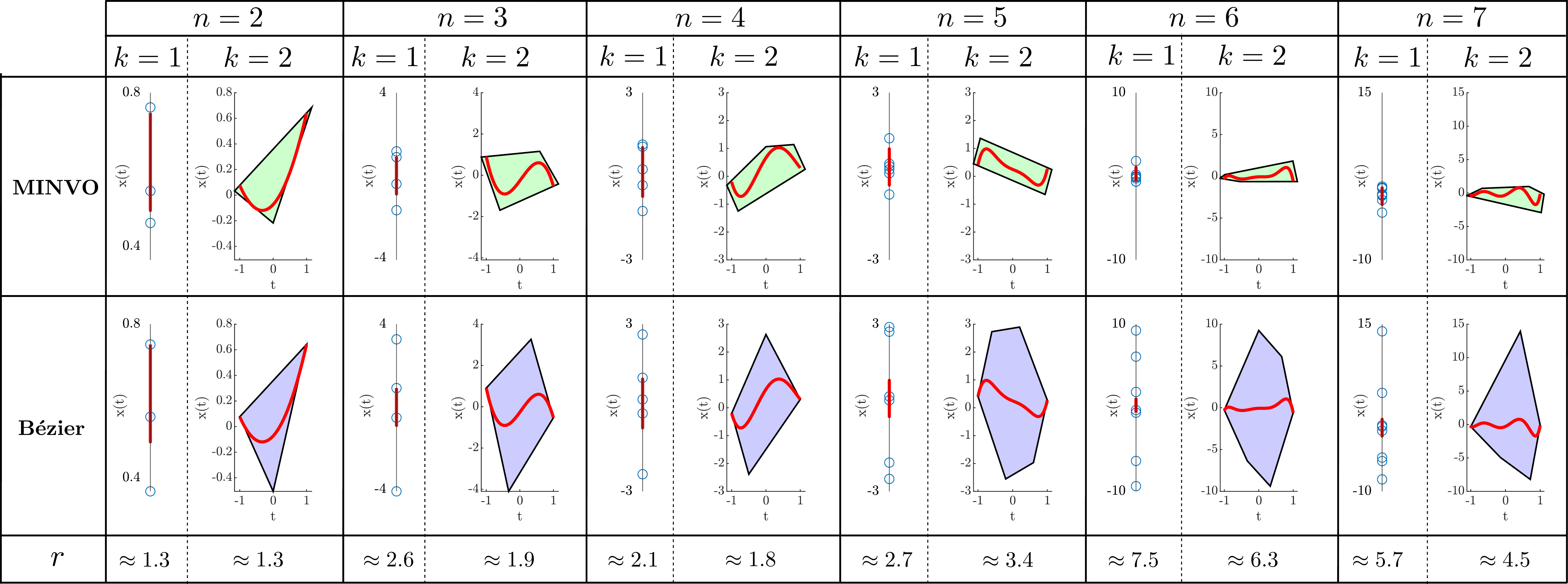}}
			\par\end{centering}
		\caption{\addrevision{Comparison of the convex hull of the MINVO and B\'{e}zier control points for $k=m\in\{1,2\}$ and different $n$. Letting $x(t)$ denote a polynomial of degree~$n$, the column with $k=1$ shows the plot of the curve $p(t)=x(t)$, and the blue circles denote the control points. The column with $k=2$ shows the plot of the curve  $\arraycolsep=1.6pt\boldsymbol{p}(t)=\left[\protect\begin{array}{cc}
				t & x(t)\protect\end{array}\right]^T$ (which corresponds to the graph of the polynomial $x(t)$ over $t\in[-1,1]$) and the convex hull of the control points.  In the last row, $r$ denotes the ratio of the longitudes $\frac{\text{Long}_\text{Be}}{\text{Long}_\text{MV}}$ (for $k=1$) or the areas $\frac{\text{Area}_\text{Be}}{\text{Area}_\text{MV}}$ (for $k=2$) of the convex hulls of the control points. The polynomials $x(t)$ pass through $n+1$ random points in the interval $[-1,1]$.}\label{fig:case_k1_diff_n}}
		\vskip 0ex
	\end{figure}

	\addrevision{This paper has focused on polynomial curves as defined in Section~\ref{sec:Notation-and-Definitions}. One particular case of such curves corresponds to the plot of a polynomial $x(t)\in \mathbb{R}[t] $ over $t\in[-1,1]$. Indeed, that plot simply corresponds to a curve that has $\arraycolsep=1.6pt\boldsymbol{p}(t)=[\begin{array}{cc}	t & x(t)\end{array}]^T$. Some examples of such curves and their associated convex hulls are shown in Figure~\ref{fig:case_k1_diff_n}. Here, we generate a polynomial $x(t)\in \mathbb{R}[t]$ passing through $n+1$ random points in $[-1,1]$, and then plot the control points for the case $k=1$ (i.e., $p(t)=x(t)$) and the convex hull of the control points for the case $k=2$ (i.e., $\arraycolsep=1.6pt\boldsymbol{p}(t)=[\begin{array}{cc}	t & x(t)\end{array}]^T$)}. \addrevision{Note that for the cases with $k=1$, a smaller convex hull can be obtained by simply reparametrizing the curve to a first-degree curve:  $p(t)=x_\text{min} + \frac{t+1}{2}(x_\text{max}-x_\text{min})$ (where $x_\text{min}:=\underset{t\in[-1,1]}{\text{min}}x(t)$ and $x_\text{max}:=\underset{t\in[-1,1]}{\text{max}}x(t)$), and then using the matrices $\boldsymbol{A}_\text{MV}$ and $\boldsymbol{A}_\text{Be}$ corresponding to $n=1$. This would give $x_\text{min}$ and $x_\text{max}$ as the control points of the curve.}
	
		\section{\addrevision{Application of the MINVO basis to surfaces \label{sec:AppSurfaces}}}
	
\begin{figure}
		\begin{centering}
			\addrevisiongraphics{\includegraphics[width=1.0\columnwidth]{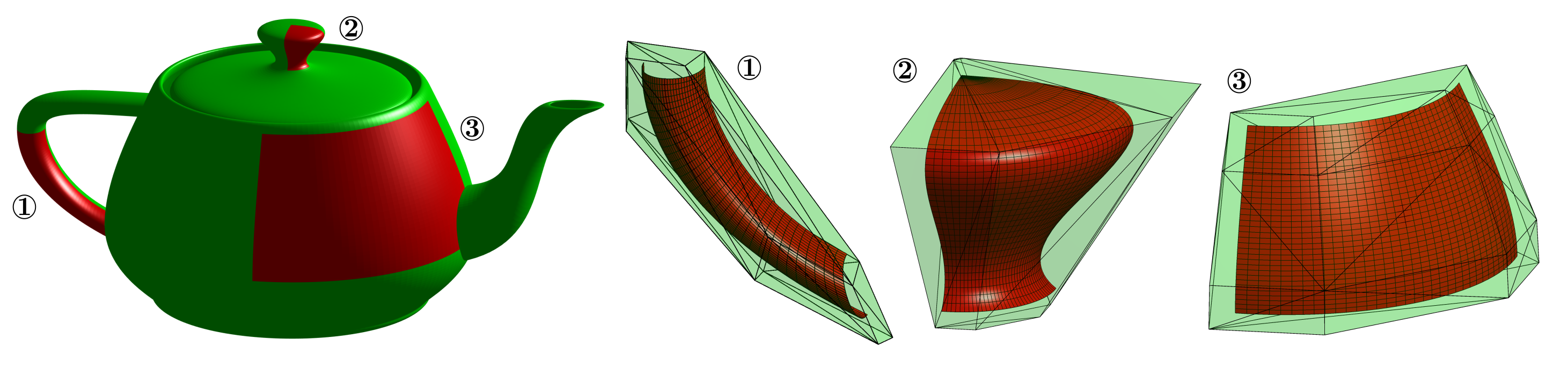}}
			\par\end{centering}
		\caption{\addrevision{MINVO basis applied to generate outer polyhedra that enclose three bicubic patches of the teapot.} \label{fig:teapot_all_only_minvo}}
		
		\vskip-2ex
	\end{figure}

	\addrevision{Similar to any other polynomial basis that is nonnegative and is a partition of unity (i.e., the polynomials in the basis sum up to one), the MINVO basis can be applied to generate polynomial surfaces of degree
		$(q,r)$ contained in the convex hull of its $(q+1)\cdot(r+1)$
		control points. Let us define $\arraycolsep=1.4pt\boldsymbol{u}_{q}:=\left[\begin{array}{ccc}
			u^{q} & u^{q-1}\cdots & 1\end{array}\right]^{T}$ and let $\boldsymbol{A}_{\text{MV},q}$ denote the matrix $\boldsymbol{A}_{\text{MV}}$
		for $n=q$. Analogous definitions apply for $v$, $\boldsymbol{v}_{r}$,
		and $\boldsymbol{A}_{\text{MV},r}$. Moreover, let $\boldsymbol{V}_{\text{MV},j}\in\mathbb{R}^{(q+1)\times(r+1)}$
		contain the coordinate $j\in\{x,y,z\}$ of the control points. The
		parametric equation of the surface will then be given by: 
		\[
		\mathbf{s}(u,v)=\left[\begin{array}{c}
			\left(\boldsymbol{A}_{\text{MV},q}\boldsymbol{u}_{q}\right)^{T}\boldsymbol{V}_{\text{MV},x}\\
			\left(\boldsymbol{A}_{\text{MV},q}\boldsymbol{u}_{q}\right)^{T}\boldsymbol{V}_{\text{MV},y}\\
			\left(\boldsymbol{A}_{\text{MV},q}\boldsymbol{u}_{q}\right)^{T}\boldsymbol{V}_{\text{MV},z}
		\end{array}\right]\boldsymbol{A}_{\text{MV},r}\boldsymbol{v}_{r}
		\]
		where $\mathbf{s}(u,v)\in\mathbb{R}^{3}$ and $u,v\in[-1,1]$. An example of the MINVO basis applied to generate different bicubic surfaces (i.e., $q=r=3$) is shown in \add{Figure}~\ref{fig:teapot_all_only_minvo}. Note however that, when applied to surfaces, the MINVO basis is not guaranteed to generate always a smaller enclosing polyhedron than, for example, the B\'{e}zier basis. An in-depth analysis of the comparison of the volumes produced by the B\'{e}zier and MINVO bases when applied to surfaces is out of the scope of this paper, and left as future work.}
	
\newpage

\FloatBarrier
\section{\addextrarevision{Definitions of the union, convex hull, and width of an enclosure (for 2D curves) \label{sec:appendix_definitions}}}
\addextrarevision{Let us define $I:=\{0,1,..,s-1\}$ and $J:=\{0,1,...,h-2\}$, where $s$ denotes the number of subdivisions used, and $h$ is the number of breakpoints per subinterval used in SLEFE. Moreover, let $\mathcal{E}$ denote an enclosure, $\text{conv\ensuremath{\left(\cdot\right)}}$ the convex hull of a set of vertices, $\text{vert}\left(\cdot\right)$ the vertices of an enclosure,  and $\text{lssbb}\left(\cdot\right)$ the length of the smallest side of the smallest arbitrarily-oriented bounding box of an enclosure. The union, convex hull, and width are then defined as follows:}

	\begin{center}
		\addextrarevisiontable{
		\begin{tabular}{|c|c|c|c|}
			\hline 
			\selectlanguage{american}%
			& \textbf{Union} & \textbf{Convex hull} & \textbf{Width}\tabularnewline
			\hline 
			\hline 
			\textbf{MV/Be} & $\bigcup\limits _{i\in I}\mathcal{E}_{i}$ & 
			$\text{conv}\left(\bigcup\limits _{i\in I}\text{vert\ensuremath{\left(\mathcal{E}_{i}\right)}}\right)$
			& $\text{max}\left(\bigcup\limits _{i\in I}\text{lssbb\ensuremath{\left(\mathcal{E}_{i}\right)}}\right)$
			\tabularnewline
			\hline 
			\textbf{SL} & $\ensuremath{\bigcup\limits _{i\in I,j\in J}\mathcal{E}_{ij}}$ & $\text{conv}\left(\bigcup\limits_{i\in I,j\in J}\text{vert}\left(\mathcal{E}_{ij}\right)\right)$
			& $\text{max}\left(\bigcup\limits_{i\in I,j\in J}\text{lssbb}\ensuremath{\left(\mathcal{E}_{ij}\right)}\right)$
			\tabularnewline
			\hline 
		\end{tabular}}
		\par\end{center}

\addextrarevision{Here, $\mathcal{E}_{i}$ (for MV and Be) is the enclosure of the subinterval $i$, while $\mathcal{E}_{ij}$ (for SL) is the enclosure of the segment $j$ of the subinterval $i$. Note that the area of the union is the sum of the individual areas minus the overlapping area, and it can be computed numerically using~\cite{matlabArea}. The area of the convex hull is the area of the smallest convex set that contains $\bigcup\limits _{i\in I}\text{vert\ensuremath{\left(\mathcal{E}_{i}\right)}}$ (for MV/Be) or $\bigcup\limits_{i\in I,j\in J}\text{vert}(\mathcal{E}_{ij})$ (for SL), and it can be computed numerically using~\cite{matlabConvHull}.}

\FloatBarrier

\section{\addextrarevision{Comparison with SLEFEs and B\'{e}zier: Width \label{sec:appendix_SLEFEs_width}}}

\addextrarevision{In this section, we compare the MINVO enclosure, B\'{e}zier enclosure, and SLEFE in terms of their widths. Given that the width is an important metric for some CAD applications, for this comparison we use data from different real CAD models.} 

\begin{figure}[htpb]
	\begin{centering}
		\addextrarevisiongraphics{\includegraphics[width=1.0\columnwidth]{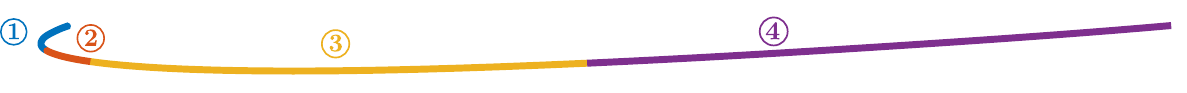}}
		\par\end{centering}
	\caption{\addextrarevision{2D trim curve of the model 10-23022015-110975 from \textcolor{blue}{\url{https://traceparts.com}}. This curve can be split into four B\'{e}zier curves, shown in different colors.} \label{fig:specific_curve_CAD_width}}
	
	\vskip-2ex
\end{figure}

\addextrarevision{We first use one of the 2D trim curves of the model 10-23022015-110975 from \textcolor{blue}{\url{https://traceparts.com}}. This curve, shown in Fig.~\ref{fig:specific_curve_CAD_width}, is a B-Spline of degree 8, which can be split into 4  B\'{e}zier curves. For each of these curves, the comparison between the widths of the enclosures is shown in Table~\ref{tab:comparison_width}.}

\begin{table*}[hbt!]
	\setlength\extrarowheight{0.2pt}
	\caption{\addextrarevision{Comparison between the width of the MINVO enclosure, B\'{e}zier enclosure, and SLEFE for the curve shown in Fig.~\ref{fig:specific_curve_CAD_width}. \notationSLh{}. We use $s=1$ for all the cases of this table.\label{tab:comparison_width}}  }
	\centering
		\begin{centering}
			\addextrarevisiontable{
				\begin{tabular}{c>{\centering}p{0.2\textwidth}>{\centering}p{0.2\textwidth}>{\centering}p{0.2\textwidth}>{\centering}p{0.2\textwidth}}
					\cmidrule{2-5} \cmidrule{3-5} \cmidrule{4-5} \cmidrule{5-5}

					& \textbf{Curve 1} & \textbf{Curve 2} & \textbf{Curve 3} & \textbf{Curve 4}\tabularnewline 
					\midrule
					\midrule 
					$\frac{\text{Width}_{\text{Be}}}{\text{Width}_{\text{MV}}}$
					& {1.05}           & {0.94}           & {0.92}           & {0.96}\tabularnewline           
					\midrule 
					$\frac{\text{Width}_{\text{SL}_{2}}}{\text{Width}_{\text{MV}}}$
					& {5.95}           & {1.43}           & {1.20}          & {3.92}\tabularnewline          
					\midrule 
					$\frac{\text{Width}_{\text{SL}_{3}}}{\text{Width}_{\text{MV}}}$
					& {2.20}          & {0.67}           & {1.24}          & {1.62}\tabularnewline           
					\midrule 
					$\frac{\text{Width}_{\text{SL}_{4}}}{\text{Width}_{\text{MV}}}$
					& {1.12 }          & {0.30}           & {0.49}           & {0.79}\tabularnewline           
					\midrule 
					$\frac{\text{Width}_{\text{SL}_{5}}}{\text{Width}_{\text{MV}}}$
					& {0.58}           & {0.19}           & {0.30}           & {0.47}\tabularnewline           
					\midrule 
					$\frac{\text{Width}_{\text{SL}_{6}}}{\text{Width}_{\text{MV}}}$
					& {0.37}           & {0.13}           & {0.20}           & {0.30}\tabularnewline          
					\midrule 
					$\frac{\text{Width}_{\text{SL}_{7}}}{\text{Width}_{\text{MV}}}$
					& {0.22 }          & {0.087}          & {0.14}          & {0.20}\tabularnewline          
					\midrule 
					$\frac{\text{Width}_{\text{SL}_{8}}}{\text{Width}_{\text{MV}}}$
					& {0.16}           & {0.067}          & {0.11}          & {0.15}\tabularnewline          
					\midrule 
					$\frac{\text{Width}_{\text{SL}_{9}}}{\text{Width}_{\text{MV}}}$
					& {0.13 }         & {0.054}          & {0.085}          & {0.12}\tabularnewline          
					\midrule 
					$\frac{\text{Width}_{\text{SL}_{10}}}{\text{Width}_{\text{MV}}}$
					& {0.097}          & {0.042}          & {0.067}          & {0.090}\tabularnewline          
					\bottomrule
			\end{tabular}}
			\par\end{centering}
\end{table*}

\begin{figure}[htpb]
	\begin{centering}
		\addextrarevisiongraphics{\includegraphics[width=0.8\columnwidth]{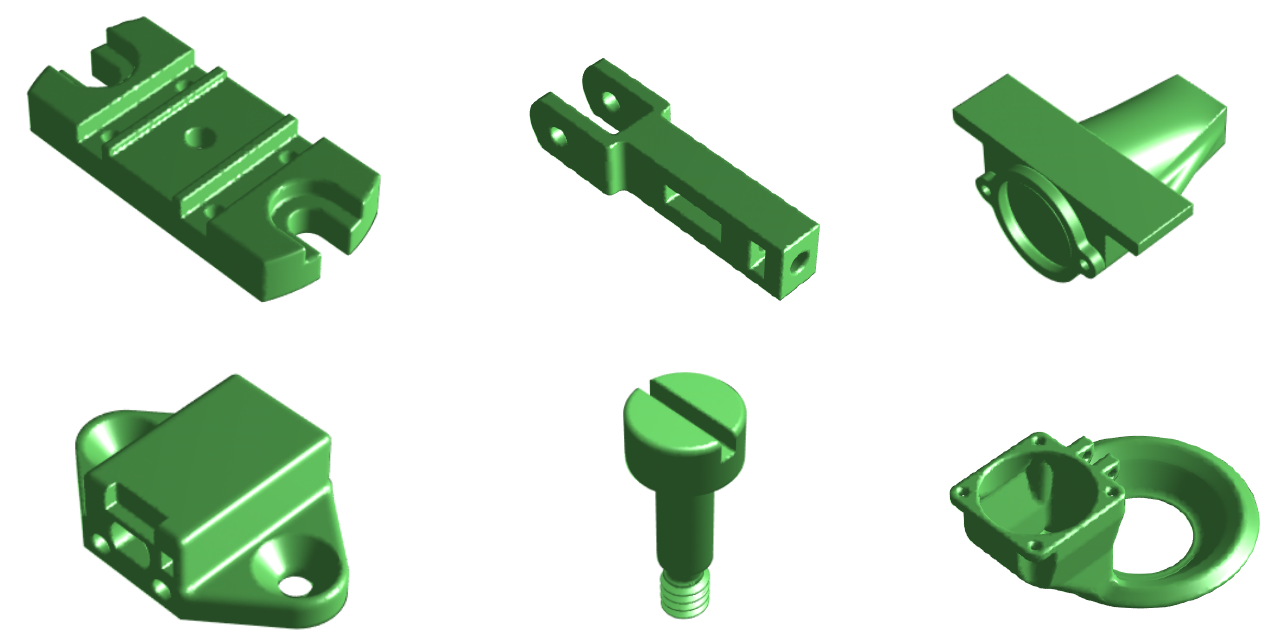}}
		\par\end{centering}
	\caption{\addextrarevision{Models taken from the ABC dataset~\cite{koch2019abc}.} \label{fig:parts_CAD}}
	
\end{figure}

\begin{figure}[htpb]
	\begin{centering}
		\addextrarevisiongraphics{\includegraphics[width=1.0\columnwidth]{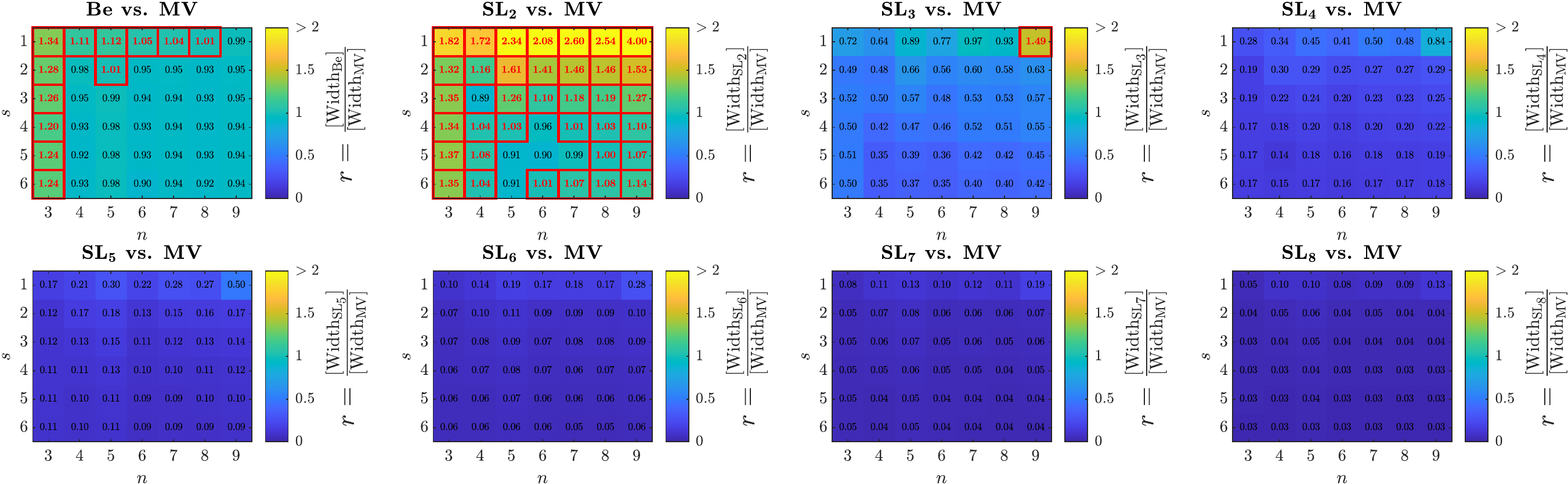}}
		\par\end{centering}
	\caption{\addextrarevision{Comparison between the width of the MINVO (MV) enclosure, B\'{e}zier (Be) enclosure, and SLEFE (SL) for \nthdegree{} 2D polynomial curves. Here, $s$ is the number of subintervals the curve is divided into, \notationSLh{}, and  $[\cdot]$ denotes the mean operator. For every $n$-$s$ combination, a total of 10 polynomial curves obtained from the models shown in Fig.~\ref{fig:parts_CAD} were used. The red squares denote the $n$-$s$ combination for which MINVO achieves a smaller width.} \label{fig:comparison_slefes_colored_matrices_CAD_width}}
	
\end{figure}

\addextrarevision{We then perform a similar analysis using six models taken from the ABC dataset~\cite{koch2019abc}, which is an extensive dataset of CAD models. These models are shown in Fig.~\ref{fig:parts_CAD}. We obtain the MINVO enclosure, B\'{e}zier enclosure, and SLEFE of the 2D (approximate) preimages of ten 3D curves of these models. These 2D curves have degrees ranging from 3 to 9. The results are shown in Fig.~\ref{fig:comparison_slefes_colored_matrices_CAD_width}.}

\addextrarevision{All these previous results (Table~\ref{tab:comparison_width} and Fig.~\ref{fig:comparison_slefes_colored_matrices_CAD_width}) allow us to conclude that SLEFE performs much better than MINVO in terms of the width of the enclosure. Hence, in applications where  it is crucial to have a small width of the enclosure, SLEFE should be preferred over MINVO.}

\newcommand{\captionNonSmoothPolSlefes}{Comparison between the MINVO \addmorerevision{(MV)} enclosure, the B\'{e}zier \addmorerevision{(Be)} enclosure, and the SLEFE \addmorerevision{(SL)} for \nthdegree{} 2D polynomial curves passing through $n+1$ random points in $[-1,1]^2$. \addmorerevision{\notationSLh{}, and $s$ is the number of subdivisions used. In all these plots, $h=n+1$ is used.} The notation used is $\text{Ve}=[a,b]$ (where $a$ is the number of vertices of the union and $b$ is the number of vertices of the convex hull) and $\text{Ar}=[c,d]$ (where $c$ is the area of the union of the enclosures and $d$ is the area of the convex hull of the enclosures). The black points are the vertices of the union.}

\newcommand{\captionSmoothPolSlefes}{Comparison between the MINVO \addmorerevision{(MV)} enclosure, the B\'{e}zier \addmorerevision{(Be)} enclosure, and the SLEFE \addmorerevision{(SL)} for \nthdegree{} 2D polynomial 
	\additionalrevision{curves obtained as described in~\ref{sec:appendix_SLEFEs_smooth}}.	\addmorerevision{\notationSLh{}, and $s$ is the number of subdivisions used. In all these plots, $h=n+1$ is used.} The notation used is $\text{Ve}=[a,b]$ (where $a$ is the number of vertices of the union and $b$ is the number of vertices of the convex hull) and $\text{Ar}=[c,d]$ (where $c$ is the area of the union of the enclosures and $d$ is the area of the convex hull of the enclosures). The black points are the vertices of the union. \addextrarevision{Note that this figure shows only some cases that have $h=n+1$, but the results in Fig.~\ref{fig:comparison_slefes_colored_matrices_smooth} include all the cases with different values of $h$, $n$, and $s$. }}

\clearpage %
\section{\addrevision{Comparison with SLEFEs and B\'{e}zier: \addextrarevision{Area and number of vertices} 
	}}\label{sec:appendix_SLEFEs_smooth}
\begin{figure*}[!ht]
	\begin{centering}
		\additionalrevisiongraphics{\includegraphics[width=0.9\textwidth]{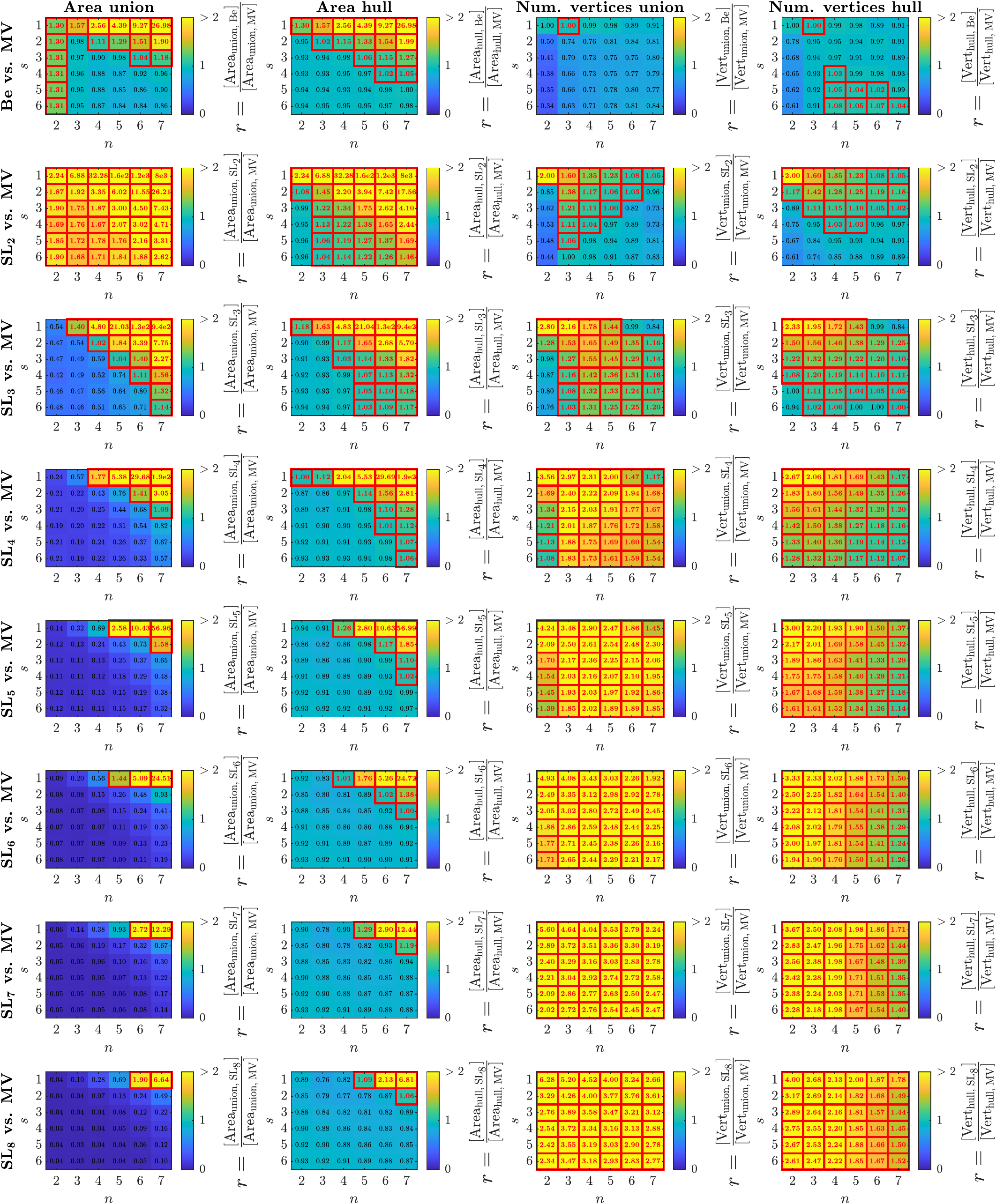}}
		\par\end{centering}
	\vskip-1ex
	\caption{Comparison between the MINVO (MV) enclosure, B\'{e}zier (Be) enclosure, and SLEFE (SL) for \nthdegree{} 2D polynomial curves. Here, $s$ is the number of subintervals the curve is divided into, \notationSLh{}, and  $[\cdot]$ denotes the mean operator. For every $n$-$s$ combination, a total of 100 polynomial \additionalrevision{curves obtained as described in~\ref{sec:appendix_SLEFEs_smooth}} were used. The red squares denote the $n$-$s$ combination for which MINVO achieves a smaller area (first two columns) or fewer number of vertices (last two columns). } \label{fig:comparison_slefes_colored_matrices_smooth}
	\vskip -2ex
\end{figure*}

\begin{figure}[!ht]
	\subfloat[$n=2$\label{fig:comparison_slefes_deg_2_smooth}]{%
		\additionalrevisiongraphics{\includegraphics[width=0.98\textwidth]{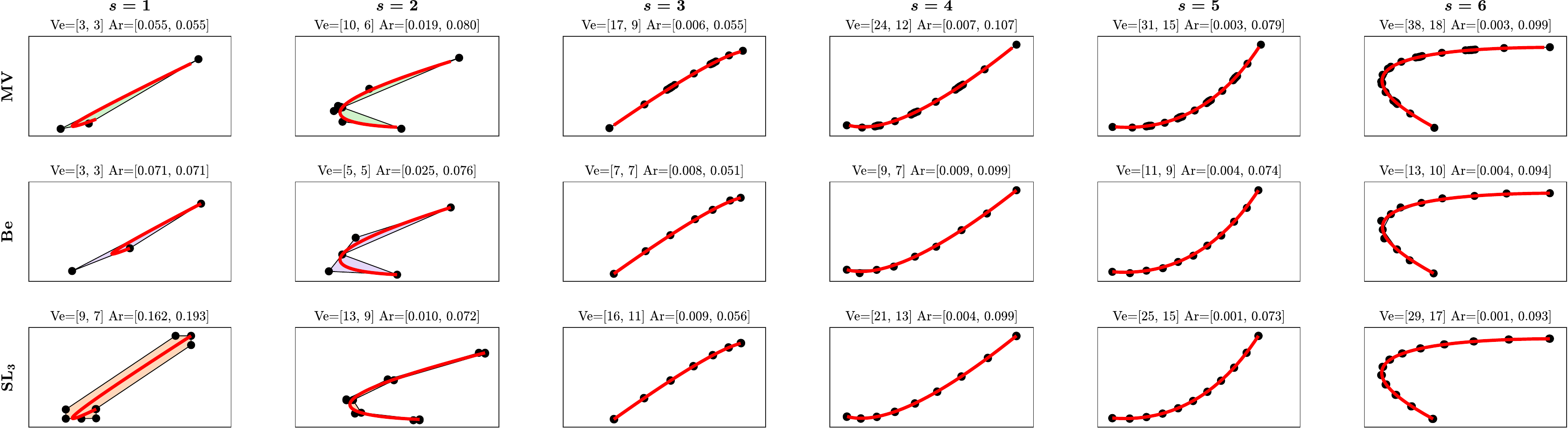}}
	}
	\vskip -3.0ex
	\hfill
	\subfloat[$n=3$\label{fig:comparison_slefes_deg_3_smooth}]{%
		\additionalrevisiongraphics{\includegraphics[width=0.98\textwidth]{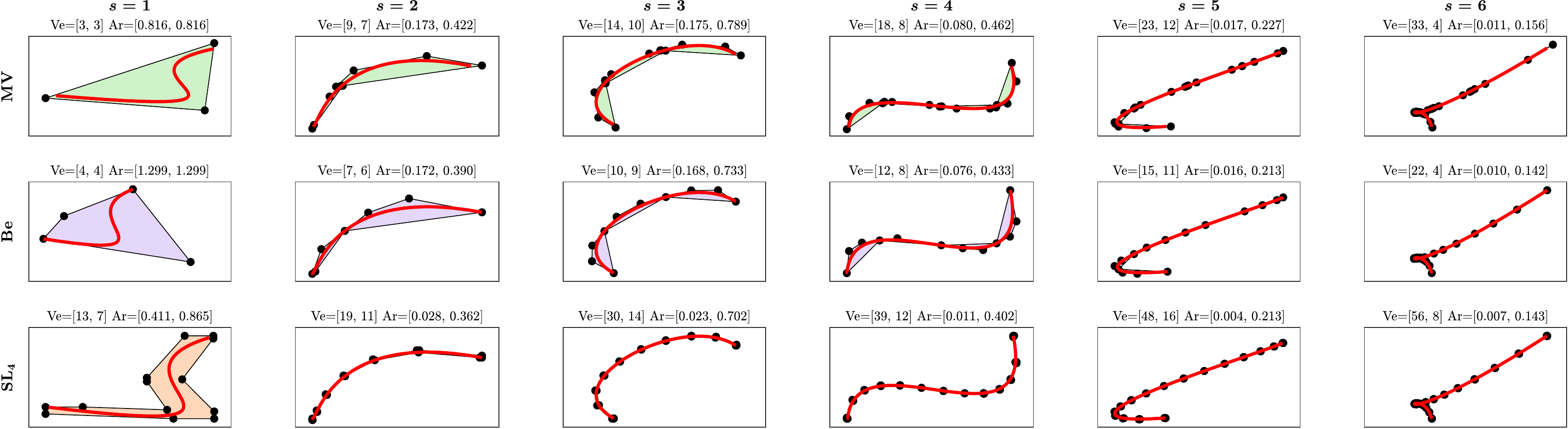}}
	}
	\vskip -3.0ex
	\hfill
	\subfloat[$n=4$\label{fig:comparison_slefes_deg_4_smooth}]{%
		\additionalrevisiongraphics{\includegraphics[width=0.98\textwidth]{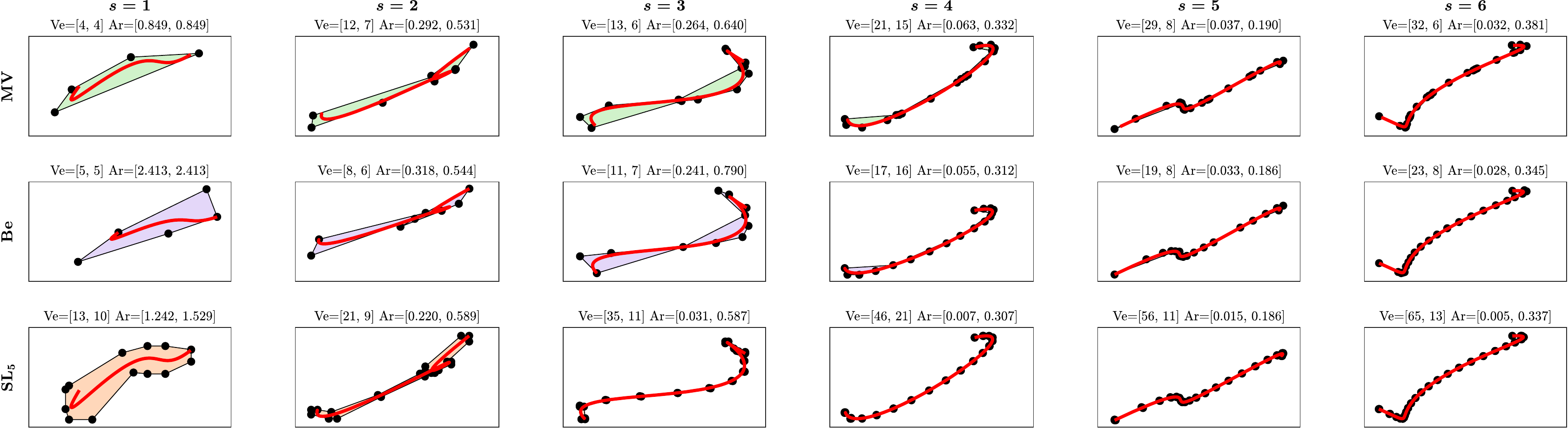}}
	}
	\vskip -3.0ex
	\hfill
	
	\subfloat[$n=5$\label{fig:comparison_slefes_deg_5_smooth}]{%
		\additionalrevisiongraphics{\includegraphics[width=0.98\textwidth]{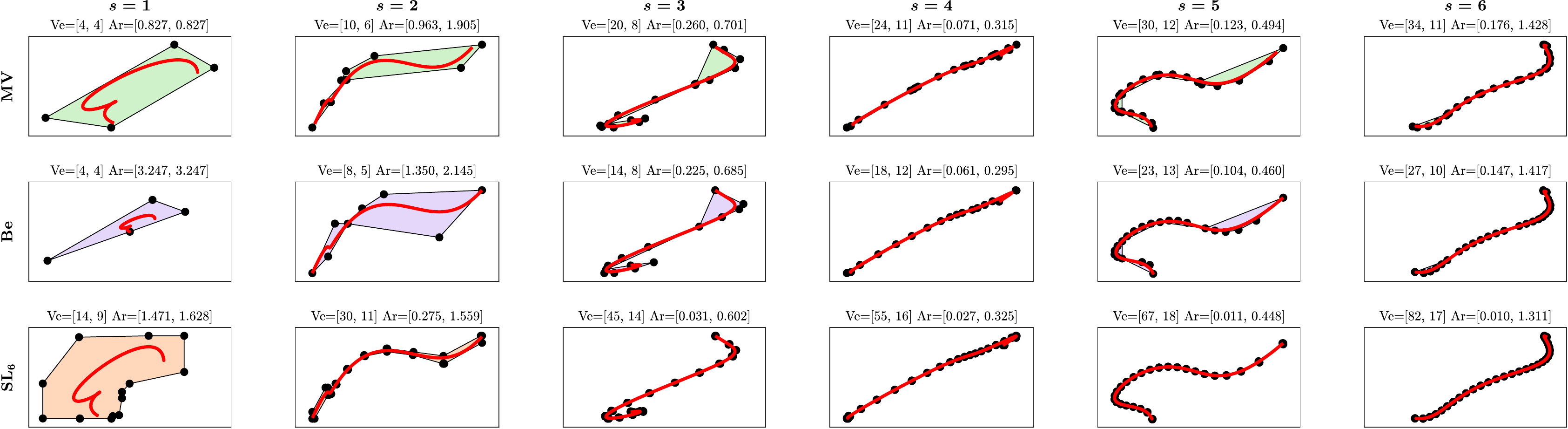}}
	}
	\vskip -2ex
	\caption{ \captionSmoothPolSlefes{} \label{fig:comparison_slefes_deg_2to5_smooth}}
\end{figure}

\begin{figure}[htpb]
	\subfloat[$n=6$\label{fig:comparison_slefes_deg_6_smooth}]{%
		\additionalrevisiongraphics{\includegraphics[width=0.98\textwidth]{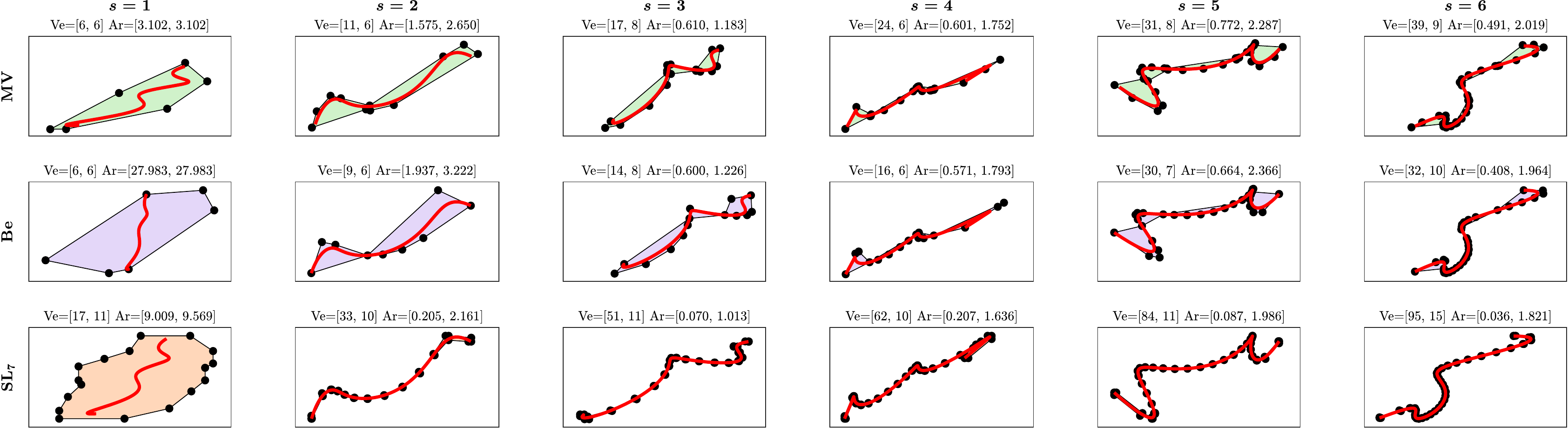}}
	}
	\vskip -2ex
	\hfill
	\subfloat[$n=7$\label{fig:comparison_slefes_deg_7_smooth}]{%
		\additionalrevisiongraphics{\includegraphics[width=0.98\textwidth]{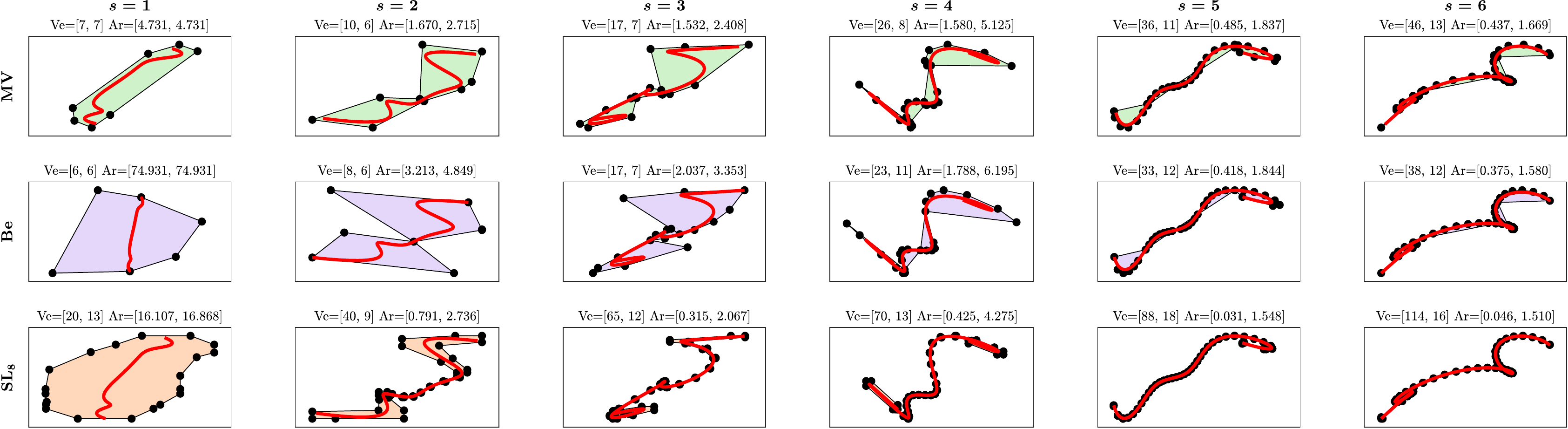}}
	}
	\vskip -2ex
	\caption{\captionSmoothPolSlefes{} \label{fig:comparison_slefes_deg_6to7_smooth}}
\end{figure}

	  \additionalrevision{The \nthdegree{} 2D polynomial curves used pass through $n+1$ points $\left\{\boldsymbol{x}_{0},\hdots,\boldsymbol{x}_{n}\right\}$ that satisfy the dynamical system $\boldsymbol{x}_{k+1}=\boldsymbol{x}_{k}+\text{rand}()-0.15\cdot\boldsymbol{1}$, where $\text{rand}()$ is a random vector in $[0,1]^2$ and $\boldsymbol{x}_{0}=\boldsymbol{0}$. \addextrarevision{Note that these curves are artificially generated, and do not correspond to real CAD data.}
	  	The results are shown in Figure~\ref{fig:comparison_slefes_colored_matrices_smooth}, and some examples are available in Figures~\ref{fig:comparison_slefes_deg_2to5_smooth}  and~\ref{fig:comparison_slefes_deg_6to7_smooth}.}

\FloatBarrier
\section{\addmorerevision{Comparison between MINVO and SLEFE in terms of runtime and simplicity in the implementation\label{sec:MinvoVsSLEFERunTime}}}

\addmorerevision{In Section~\ref{sec:SLEFEs} the MINVO enclosure and the SLEFE are compared in terms of enclosing area and number of vertices. The \textbf{enclosing area} is important in terms of conservativeness of the enclosure with respect to the curve, while the \textbf{number of vertices} can have a direct impact on the computation time (for example, in settings where the curve is a decision variable in an optimization problem and there is one constraint per vertex), and on the memory needed to store them.} 
	
\addmorerevision{There are also other applications where the \textbf{runtime to obtain the enclosure} is important. In this section, we compare the runtimes needed to obtain the SLEFE and the MINVO enclosure from a curve given in its B\'{e}zier form (i.e., $\boldsymbol{V}_\text{Be}$ is given):
\begin{itemize}
	\item  The MINVO enclosure is obtained by simply doing $\boldsymbol{V}_{\text{MV}}=\boldsymbol{V}_{\text{Be}}\boldsymbol{A}_{\text{Be}}\boldsymbol{A}_{\text{MV}}^{-1}$ (Eq.~\ref{eq:relationship_basis}), where the term $\boldsymbol{A}_{\text{Be}}\boldsymbol{A}_{\text{MV}}^{-1}$ is tabulated offline using the matrices available in Table~\ref{tab:table_matrices} and~\cite{qin2000general} for each degree $n$. If the curve were given instead in its monomial form (i.e., $\boldsymbol{P}$ is given), then the computation of the MINVO enclosure would also be a simple matrix multiplication: $\boldsymbol{V}_{\text{MV}}=\boldsymbol{P}\boldsymbol{A}_{\text{MV}}^{-1}$, where $\boldsymbol{A}_{\text{MV}}^{-1}$ is tabulated offline.
	\item The SLEFE enclosure is obtained as detailed in \cite[Section 3.3]{myles2005threading}, using the tabulated values given by the SubLiME package~\cite{SubLiME21slefe}. \additionalrevision{We optimize the speed of the SLEFE implementation leveraging vector and matrix operations. Note however that the SLEFE computation requires $\text{max}(\cdot)$ and $\text{min}(\cdot)$ operators, and hence it cannot be obtained as a single matrix multiplication.}
\end{itemize}}

\addmorerevision{The timing results are shown in Table \ref{tab:comparison_runtime_MINVO_slefe}. On average, the MINVO enclosure can be obtained \additionalrevision{$20.4$} times faster than the SLEFE enclosure. These timing results were obtained using Matlab\textsuperscript{\textregistered} R202\additionalrevision{1}b on an  AlienWare Aurora~r8 desktop running Ubuntu 18.04 and equipped with an Intel\textsuperscript{\textregistered} Core\textsuperscript{TM} i9-9900K CPU, 3.60GHz$\times$16 and 62.6~GiB. }   

\newcommand{\myHorizLine}{\cmidrule{2-8}}

\newcommand{\myratioSL}[1]{$\frac{\left[\text{ct}\left(\text{\text{SL}}_{#1}\right)\right]}{\left[\text{ct}(\text{MV})\right]}$}%

	\begin{table*}[hbt!]
	\setlength\extrarowheight{0.6pt}
	\caption{\addmorerevision{Computation times required to find the MINVO (MV) enclosure and the SLEFE (SL) for \nthdegree{}  polynomial curves.  \notationSLh{}, $\text{ct}(\cdot)$ denotes the computation time, and $[\cdot]$ denotes the mean operator. For each cell in this table, a total of 30 random polynomials passing through random points in $[-1,1]$ were used.  All these cases have $s=1$.\label{tab:comparison_runtime_MINVO_slefe}}  }
	\vskip-2ex
	\centering
	\noindent\resizebox{1.0\textwidth}{!}{%
		\begin{centering}
			\addmorerevisiontable{
			\begin{tabular}{>{\centering}p{0.04\textwidth}>{\centering}p{0.1\textwidth}>{\centering}m{0.1\textwidth}>{\centering}m{0.1\textwidth}>{\centering}m{0.1\textwidth}>{\centering}m{0.1\textwidth}>{\centering}m{0.1\textwidth}>{\centering}m{0.1\textwidth}>{\centering}m{0.1\textwidth}>{\centering}m{0.1\textwidth}>{\centering}m{0.1\textwidth}>{\centering}m{0.1\textwidth}>{\centering}m{0.1\textwidth}}
				\cmidrule{3-13} \cmidrule{4-13} \cmidrule{5-13} \cmidrule{6-13} \cmidrule{7-13} \cmidrule{8-13} \cmidrule{9-13} \cmidrule{10-13} \cmidrule{11-13} \cmidrule{12-13} \cmidrule{13-13} 
				&  & \multicolumn{11}{c}{\textbf{Degree n}}\tabularnewline
				\cmidrule{3-13} \cmidrule{4-13} \cmidrule{5-13} \cmidrule{6-13} \cmidrule{7-13} \cmidrule{8-13} \cmidrule{9-13} \cmidrule{10-13} \cmidrule{11-13} \cmidrule{12-13} \cmidrule{13-13} 
				&  & $\mathbf{2}$ &$\mathbf{3}$ & $\mathbf{4}$& $\mathbf{5}$ &$\mathbf{6}$ & $\mathbf{7}$\tabularnewline
				\midrule
				\midrule 
				\multirow{7}{0.04\textwidth}{\rotatebox[origin=c]{90}{\textbf{Ratios of comp. times\hspace{0.5cm}}}} 
& \myratioSL{2} & \additionalrevision{24.8} & \additionalrevision{24.0} & \additionalrevision{20.8} & \additionalrevision{18.3} & \additionalrevision{15.6} & \additionalrevision{13.3}\tabularnewline\myHorizLine{}  
& \myratioSL{3} & \additionalrevision{24.4} & \additionalrevision{22.8} & \additionalrevision{20.9} & \additionalrevision{18.3} & \additionalrevision{15.6} & \additionalrevision{14.4}\tabularnewline\myHorizLine{}  
& \myratioSL{4} & \additionalrevision{27.3} & \additionalrevision{24.9} & \additionalrevision{21.0} & \additionalrevision{18.9} & \additionalrevision{16.1} & \additionalrevision{15.1}\tabularnewline\myHorizLine{} 
& \myratioSL{5} & \additionalrevision{26.4} & \additionalrevision{23.3} & \additionalrevision{20.8} & \additionalrevision{18.6} & \additionalrevision{16.9} & \additionalrevision{15.6}\tabularnewline\myHorizLine{} 
& \myratioSL{6} & \additionalrevision{26.3} & \additionalrevision{25.0} & \additionalrevision{21.8} & \additionalrevision{19.2} & \additionalrevision{16.8} & \additionalrevision{15.1}\tabularnewline\myHorizLine{} 
& \myratioSL{7} & \additionalrevision{26.2} & \additionalrevision{24.2} & \additionalrevision{22.3} & \additionalrevision{19.0} & \additionalrevision{17.5} & \additionalrevision{15.6}\tabularnewline\myHorizLine{} 
& \myratioSL{8} & \additionalrevision{27.8} & \additionalrevision{25.9} & \additionalrevision{22.8} & \additionalrevision{19.8} & \additionalrevision{17.6} & \additionalrevision{15.6}\tabularnewline
				\bottomrule
			\end{tabular}}
			\par\end{centering}
	}
\end{table*}

\addmorerevision{Finally, another aspect that one may consider is the \textbf{simplicity} of the implementation. As detailed above, only a single matrix multiplication is required to obtain the MINVO enclosure, which translates into a simple one line of code in most of the modern programming languages. The SLEFE computation also has a simple implementation, in this case involving sums, multiplications, $\text{max}(\cdot)$, and $\text{min}(\cdot)$ operators. Code examples of how to use MINVO and SLEFE are available at \url{https://github.com/mit-acl/minvo} (for both MINVO and SLEFE) and~\cite{SubLiME21slefe} (for SLEFE).}

	\twocolumn
	\bibliographystyle{elsarticle-num}
	\bibliography{bibliography}

\end{document}